\documentclass[12pt]{article}
\usepackage{lscape,epsfig,amsmath,psfrag,styles/axodraw,styles/feynarts}
\usepackage[dvips]{color}
\usepackage{slashed}

\graphicspath{{figs/}}
\def\beq{\begin{equation}}
\def\eeq{\end{equation}}
\def\beqar{\begin{eqnarray}}
\def\eeqar{\end{eqnarray}}
\def\barr#1{\begin{array}{#1}}
\def\earr{\end{array}}
\def\bfi{\begin{figure}}
\def\efi{\end{figure}}
\def\btab{\begin{table}}
\def\etab{\end{table}}
\def\bce{\begin{center}}
\def\ece{\end{center}}

\def\text{\textstyle}

\def\al{\alpha}

\def\be{\beta}

\def\ga{\gamma}
\def\de{\delta}

\def\la{\lambda}
\def\si{\sigma}
\def\Si{\Sigma}
\def\Ga{\Gamma}
\def\De{\Delta}
\newcommand{\La}{\Lambda}

\newcommand{\TeV}{\unskip\,\mathrm{TeV}}

\newcommand{\cO}{{\cal O}}

\def\mathswitchr#1{\relax\ifmmode{\mathrm{#1}}\else$\mathrm{#1}$\fi}

\newcommand{\PW}{\mathswitchr W}

\def\mathswitch#1{\relax\ifmmode#1\else$#1$\fi}

\newcommand{\MW}{\mathswitch {M_W}}

\newcommand{\MZ}{\mathswitch {M_Z}}
\newcommand{\MH}{\mathswitch {M_H}}

\newcommand{\MHp}{M_{H^\pm}}

\newcommand{\Mt}{\mathswitch {m_t}}

\newcommand{\mh}{\mathswitch {m_h}}

\newcommand{\MA}{\mathswitch {M_A}}

\newcommand{\scrs}{\scriptscriptstyle}
\newcommand{\sw}{\mathswitch {s_{\scrs\PW}}}

\newcommand{\cw}{\mathswitch {c_{\scrs\PW}}}

\newcommand{\GF}{\mathswitch {G_\mu}}
\newcommand{\gf}{\GF}

\newcommand{\Tb}{\tan \beta\hspace{1mm}}

\newcommand{\CTb}{\cot \beta\hspace{1mm}}

\newcommand{\CQZb}{\cos^2 2\beta\hspace{1mm}}

\newcommand{\lag}{\mbox{${\cal L}$}}

\newcommand{\tchi}{\tilde\chi}

\newcommand{\mst}{m_{\tilde{t}}}

\newcommand{\mt}{\Mt}

\newcommand{\Sferm}{\tilde{f}}

\newcommand{\tsf}{\theta\kern-.20em_{\tilde{f}}}
\newcommand{\tsfp}{\theta\kern-.20em_{\tilde{f}\prime}}
\newcommand{\tsq}{\theta\kern-.15em_{\tilde{q}}}

\newcommand{\mf}{m_f}

\newcommand{\msusy}{M_{\mathrm{SUSY}}}

\newcommand{\sfl}{\tilde{f}_L}
\newcommand{\sfr}{\tilde{f}_R}

\newcommand{\lsim}
{\;\raisebox{-.3em}{$\stackrel{\displaystyle <}{\sim}$}\;}
\newcommand{\gsim}
{\;\raisebox{-.3em}{$\stackrel{\displaystyle >}{\sim}$}\;}

\newcommand{\wz}{\sqrt{2}}

\newcommand{\KL}{\left(}
\newcommand{\KR}{\right)}
\newcommand{\KKL}{\left[}
\newcommand{\KKR}{\right]}

\newcommand{\RA}{\Rightarrow}

\newcommand{\VL}{\left( \begin{array}{c}}
\newcommand{\VR}{\end{array} \right)}
\newcommand{\ML}{\left( \begin{array}{cc}}
\newcommand{\MLd}{\left( \begin{array}{ccc}}
\newcommand{\MLv}{\left( \begin{array}{cccc}}
\newcommand{\MR}{\end{array} \right)}

\newcommand{\hc}{\mbox {h.c.}}

\newcommand{\tev}{\,\, \mathrm{TeV}}
\newcommand{\gev}{\,\, \mathrm{GeV}}

\newcommand{\BC}{\begin{center}}
\newcommand{\EC}{\end{center}}
\newcommand{\BE}{\begin{equation}}
\newcommand{\EE}{\end{equation}}
\newcommand{\BEA}{\begin{eqnarray}}
\newcommand{\BEAnn}{\begin{eqnarray*}}
\newcommand{\EEA}{\end{eqnarray}}
\newcommand{\EEAnn}{\end{eqnarray*}}
\newcommand{\non}{\nonumber}
\newcommand{\id}{{\rm 1\kern-.12em
\rule{0.3pt}{1.5ex}\raisebox{0.0ex}{\rule{0.1em}{0.3pt}}}}

\def\als{\alpha_s}

\def\draftdate{\relax}
\def\mda{\relax}
\def\mua{\relax}
\def\mla{\relax}
\def\draft{
\def\thtystars{******************************}
\def\sixtystars{\thtystars\thtystars}
\typeout{}
\typeout{\sixtystars**}
\typeout{* Draft mode!
         For final version remove \protect\draft\space in source file
*}
\typeout{\sixtystars**}
\typeout{}
\def\draftdate{\today}
\def\mua{\marginpar[\boldmath\hfil$\uparrow$]%
                   {\boldmath$\uparrow$\hfil}%
                    \typeout{marginpar: $\uparrow$}\ignorespaces}
\def\mda{\marginpar[\boldmath\hfil$\downarrow$]%
                   {\boldmath$\downarrow$\hfil}%
                    \typeout{marginpar: $\downarrow$}\ignorespaces}
\def\mla{\marginpar[\boldmath\hfil$\rightarrow$]%
                   {\boldmath$\leftarrow $\hfil}%
                    \typeout{marginpar:
$\leftrightarrow$}\ignorespaces}
\def\Mua{\marginpar[\boldmath\hfil$\Uparrow$]%
                   {\boldmath$\Uparrow$\hfil}%
                    \typeout{marginpar: $\Uparrow$}\ignorespaces}
\def\Mda{\marginpar[\boldmath\hfil$\Downarrow$]%
                   {\boldmath$\Downarrow$\hfil}%
                    \typeout{marginpar: $\Downarrow$}\ignorespaces}
\def\Mla{\marginpar[\boldmath\hfil$\Rightarrow$]%
                   {\boldmath$\Leftarrow $\hfil}%
                    \typeout{marginpar:
$\Leftrightarrow$}\ignorespaces}
\overfullrule 5pt
\oddsidemargin -15mm
\marginparwidth 29mm
}

\definecolor{Blue}{named}{Blue}
\definecolor{Lblue}{cmyk}{1,0,0,0}
\definecolor{Red}{named}{Red}
\definecolor{Green}{named}{PineGreen}
\definecolor{Black}{named}{Black}
\definecolor{Magenta}{named}{Magenta}
\definecolor{Royal}{named}{RoyalBlue}
\definecolor{Orange}{named}{Orange}
\definecolor{Purple}{named}{Purple}
\definecolor{YOrange}{named}{YellowOrange}
\definecolor{Yellow}{named}{Yellow}
\definecolor{Mahogany}{named}{Mahogany}
\definecolor{Brown}{named}{Brown}
\definecolor{Gray}{named}{Gray}
\newcommand{\black}[1]{\color{Black}#1 \color{Black}}

\newcommand{\red}[1]{\color{Red}#1 \color{Black}}

\newcommand{\magenta}[1]{\color{Magenta}#1 \color{Black}}

\newcommand{\purple}[1]{\color{Purple}#1 \color{Black}}

\definecolor{GreenYellow}{named}{GreenYellow}

\definecolor{SkyBlue}{named}{SkyBlue}

\definecolor{Apricot}{named}{Apricot}

\def\ul#1{\underline{\vphantom{g}#1}}

\def\blackl#1{\black{\ul{#1}}}

\newcounter{chapter}
\evensidemargin \oddsidemargin
\allowdisplaybreaks

\definecolor{cyan2}{rgb}{0,0.6125,0.6125}
\definecolor{grey}{rgb}{0.8,0.8,0.8}
\definecolor{gold}{rgb}{1,0.839844,0}
\definecolor{sea}{rgb}{0.1, 0.1, 0.6}
\definecolor{mango}{rgb}{0.901961, 0.901961, 0.345098}
\definecolor{darkblue}{rgb}{0.,0.,1.0}
\definecolor{pink}{rgb}{1.,0.9,0.9}    
\definecolor{backgreen}{rgb}{0.8255,0.9869,0.8435}
\definecolor{foregreen}{rgb}{0.3,0.6333,0.3}
\definecolor{fondo}{rgb}{0.93,1.0,1.0}
\definecolor{fondo1}{rgb}{0.8,1.0,1.0}
\definecolor{darkgreen}{rgb}{0.0,0.7,0.}
\definecolor{darkred}{rgb}{1.0,0.,0.}
\definecolor{brown}{rgb}{.55,0.,0.}
\definecolor{mio}{rgb}{0.5,0.3,1.0}

\begin{document}


\def\thefootnote{\fnsymbol{footnote}}

\begin{flushright}
IFT-UAM/CSIC-14-001\\
FTUAM-14-1
\end{flushright}

\vspace{0.5cm}

\begin{center}

{\large\sc {\bf 
The Higgs System in and Beyond the Standard Model}}

\vspace{1cm}

{\sc
  Maria J.~Herrero$^{1}$%
\footnote{email: Maria.Herrero@uam.es}%
}

\vspace*{.7cm}

{\sl
$^1$Departamento de F\'isica Te\'orica and Instituto de F\'isica Te\'orica,
IFT-UAM/CSIC\\
Universidad Aut\'onoma de Madrid, Cantoblanco, Madrid, Spain

}

\end{center}

\vspace*{0.1cm}

\begin{abstract}
\noindent
After the discovery of the Higgs boson particle on the 4th of July of 2012 at the Large Hadron Collider, sited at the european CERN laboratory,  we are entering in a fascinating period for Particle Physics where both theorists and experimentalists are devoted to  
fully understand the features of this new particle and the possible 
consequences for High Energy Physics of the Higgs system both within and beyond the Standard Model of fundamental particle interactions. This paper is a summary of the lectures given at the third IDPASC school (Santiago de Compostela, Feb. 2013, Spain) addressed to PhD students, and contains a short introduction to the main basic aspects of the Higgs boson particle in and beyond the Standard Model. 

\end{abstract}

\def\thefootnote{\arabic{footnote}}
\setcounter{page}{0}
\setcounter{footnote}{0}

\newpage

\section{Introduction}
The Standard Model (SM)~\cite{Glashow:1961tr,GellMann:1964nj,Zweig:1964jf,Weinberg:1967tq,Salam:1968rm} describes with unprecedent precision ($0.1\%$) the properties of all known elementary particles, Leptons and Quarks, and their fundamental interactions, electromagnetic, strong and weak. Gravity is not included in the SM. The complete gauge symmetry group of the SM is $SU(3)_C \times SU(2)_L \times U(1)_Y$, with $SU(3)_C$ being the symmetry group of the strong interactions and $SU(2)_L \times U(1)_Y$ the symmetry group of the electroweak interactions. 

The different elementary particles described by the SM are collected in the following figure and include: 1)  the three fermion families with the three charged leptons, the electron $e$ the muon $\mu$ and the tau $\tau$, the three neutral leptons, i.e. the neutrinos $\nu_e$, $\nu_\mu$ and $\nu_\tau$, the three up-type quarks $u$ (up) , $c$ (charm) and $t$ (top), the three down-type quarks $d$ (down), $s$ (strange) and $b$ (bottom); 2) the force carriers: the photon $\gamma$, mediator of the electromagnetic interactions, the eight gluons $g_a$ ($a=1,..,8$), mediators of the strong interactions, and the three weak bosons, mediators of the weak interactions, the neutral $Z$ boson and the two charged $W^{\pm}$ bosons. 

\begin{figure}[ht!]
\begin{center}
\psfig{file= 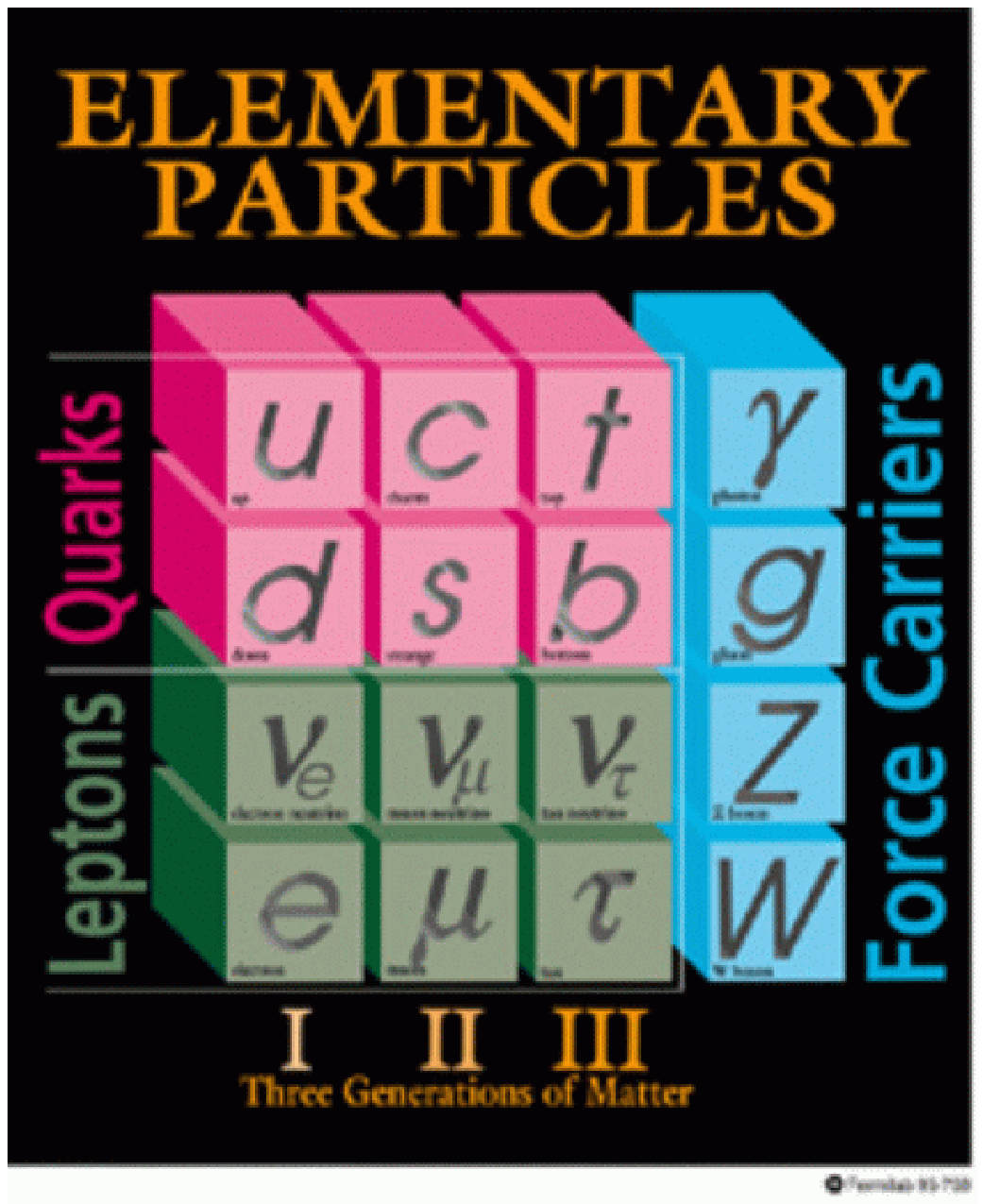,scale=0.5,clip=}\\
The particle content of the Standard Model
\end{center}
\end{figure} 
All the above particles have been experimentally seen and their properties have been measured  in many cases with very high precision~\cite{pdg}. The elementary particles that are the constituents of matter are of two types, leptons and quarks, and they are fermions with spin equal to 1/2. With respect to the mediators of the three interactions within the SM, i.e. the gauge particles, they are bosons with spin equal to 1, and their properties have also been tested in the experiments. One of the most important properties of these gauge bosons is their mass. 
The carriers of electromagnetic (photon) and strong interactions (gluons) are massless gauge bosons.
But the carriers of weak interactions, $W^{\pm}$ and $Z$,  are massive. The present measurements give: 
$M_W^{\rm exp}=80.385 \pm 0.015 \gev$ and $M_Z^{\rm exp}=91.1876 \pm 0.0021 \gev$. 
{}
{}

The fact that these weak bosons, $W^{\pm}$ and $Z$, have non zero masses leads to a problem in Gauge Quantum Field Theory which can be read as how to reconcile gauge invariance and massive gauge bosons. The SM, as any other Gauge Quantum Field Theory is built under the construction principle of gauge invariance where the exchanged field quanta with spin one defines the gauge particle and it must be massless in order to preserve this gauge invariance. Therefore, the observed weak boson masses must be explained in a different way. Within the SM, this way is the 
Higgs Mechanism~\cite{Higgs:1964ia,Higgs:1964pj,Higgs:1966ev,Englert:1964et,Guralnik:1964eu} that will be described in these lectures and that leads to the prediction of a new massive scalar particle the Higgs boson particle. The present consensus in the High Energy Physics Community points towards the interpretation that the recently discovered scalar particle at 
the LHC~\cite{Aad:2012tfa,Chatrchyan:2012ufa} with a mass between 125 and 126 GeV is indeed this Higgs boson, predicted in the SM. The most recent measurements by the ATLAS and CMS collaborations set $M_H^{\rm ATLAS}=125.5\pm 0.6 \gev$~\cite{Aad:2013wqa} and $M_H^{\rm CMS}=125.7\pm 0.4 \gev$~\cite{Chatrchyan:2013lba}, respectively. They also show that the most probable $J^P$ quantum numbers for this dicovered particle are $0^+$, and conclude that the measured Higgs couplings to the other SM particles are in agreement so far with the values predicted in the SM. 
But, although all these first LHC data are really encouraging,  there is still a long way to fully check the SM Higgs boson hypothesis. These commented properties above and many other features of this new scalar particle will be measured in the future with much higher precision than at present, and we hopefully will be able to disentangle finally which particle is really this one, the SM Higgs boson or something else. This is really a fascinating period in the History of Particle Physics.

 This paper is organized in two main blocks, corresponding to the two given lectures:

\begin{enumerate}
\item[]
{\bf  Lecture 1: The Higgs boson in the Standard Model}
\subitem \hspace{0.5cm} The building of the Electroweak Theory
\subitem \hspace{0.5cm} Electroweak Symmetry Breaking
\subitem \hspace{0.5cm} The spectra of the SM and the particle masses
\subitem \hspace{0.5cm} SM Higgs boson couplings
\subitem \hspace{0.5cm} SM Higgs boson decays and production at LHC
\subitem \hspace{0.5cm} Other interesting properties of the SM Higgs system
\item[]
{\bf Lecture 2: Some avenues beyond Standard Model Higgs}
\subitem \hspace{0.5cm} Motivations for looking beyond the Standard Model
\subitem \hspace{0.5cm} The hierarchy problem of the SM Higgs sector
\subitem \hspace{0.5cm} Two main avenues to solve the hierarchy problem
\subitem \hspace{0.5cm} Supersymmetry
\subitem \hspace{0.5cm} Compositness
\subitem \hspace{0.5cm} Electroweak Chiral Lagrangians
\end{enumerate} 


\section{The building of the Electroweak Theory}
In the following we shortly remind the basics of the gauge principle,  using QED as an illustrative example, and then apply it to the Electroweak Theory.

\blackl{The gauge principle:} \\[0.5em]
In order to get a Lagrangian that is invariant under local (gauge) transformations, massless gauge fields $A_\mu$ must be introduced with specific interactions with matter. The concrete prescription is provided by the covariant derivative.  Number of gauge bosons = Number of symmetries= Number of generators of the symmetry group.

In practice, one follows three steps: 1) Start with the Lagragian for propagating fermion fields without interactions,i.e., for free fields. 2) Replace the usual derivative by the covariant derivative. 3) Add the proper invariant kinetic terms for the gauge fields, such that they can propagate.

\blackl{QED as an example:}

Let use $\Psi$ to describe the field of a fermion with electric charge $Q$ (in units of $e$, the electron charge) and mass $m$. The associated free Lagrangian is:

$${\cal L}_{\rm free}=\bar \Psi( i{\not \! \partial} -m) \Psi$$

where, ${\not \! \partial} \equiv \partial_{\mu} \gamma^\mu$, $\gamma^\mu$= Dirac matrices.

The corresponding eq. of motion for $\Psi$ is the Dirac equation:

$$(i{\not \! \partial} -m) \Psi =0$$

Then we replace the normal derivative by the covariant derivative that includes the gauge field, here denoted by $A_\mu$ which defines the photon particle 
(${\gamma}$ in the figure),

$\partial_{\mu}\Psi \to D_{\mu}\Psi \equiv (\partial_{\mu}-ieQ A_\mu)\Psi$ ; 

$F_{\mu \nu}=\partial_{\mu}A_\nu-\partial_{\nu}A_\mu$ 

With this covariant derivative we then build the QED Lagrangian:
$$\RA{\cal L}_{\rm QED}=\bar \Psi( i \slashed{D} -m) \Psi - \frac{1}{4}F_{\mu \nu}F^{\mu \nu}$$
where $F_{\mu \nu}=\partial_{\mu}A_\nu-\partial_{\nu}A_\mu$ is the electromagnetic field stress tensor. 

\BC
\setlength{\unitlength}{1pt}
\begin{picture}(350, 100)
\ArrowLine(60,10)(150,30)
\ArrowLine(150,30)(185,95)
\Vertex(150,30){3}
\Photon(150,30)(300,5){3}{8.5}
\put(70,17){$\Psi$}
\put(156,74){$\Psi$}
\put(225,32){{$\ga$}}
\GCirc(300,5){15}{.5}
\put(325,0){nucleus}
\end{picture}
\EC

One can check easily that ${\cal L}_{\rm QED}$ is invariant under $U(1)$ Gauge transformations, with one single generator give by $Q$: 
$$\Psi \to e^{ieQ \theta(x)}\Psi \,\,\,;\,\, A_\mu \to A_\mu -\frac{1}{e} \partial_\mu \theta(x)$$
Notice that a mass term for the photon of the type 
$m^2 A^\mu A_\mu$ is not $U(1)$ gauge invariant, therefore in QED the gauge invariance principle implies that the photon is massless, which is in total agreement with data.
However, this is not the case of the W and Z electroweak gauge bosons and the immediate questions arise: why are they massive? How do they get their masses?...


\blackl{The gauge invariance in the Electroweak Theory:}\\[1em]

The Electroweak Theory (EW) refers to the part of the SM that describes together  the electromagnetic and weak interactions within the same framework of a Gauge Quantum Field Theory based on the 
gauge principle invariance of the electroweak interactions.

The gauge symmetry group of the Electroweak Theory is $SU(2)_L \times U(1)_Y$, with  4 generators. $SU(2)_L$ is the weak isospin group  which is non abelian, and has 
3 generators $T_{1,2,3}=\sigma_{1,2,3}/2$, with $\sigma_{1,2,3}$ being the Pauli matrices.
$U(1)_Y$ is the weak hypercharge group which is abelian and has 1 generator $Y/2$.
The electromagnetic group appears as a subgroup of the electroweak group, $U(1)_{\rm em} \subset SU(2)_L \times U(1)_Y$; and the corresponding generator is a combination of the third component of the weak isospin and the weak hypercharge, $Q=T_3+Y/2$.

The elementary particles of the SM, Quarks and Leptons, transform as:\\[0.5em]
1) Under  $SU(2)_L$: $\Psi_L \to e^{i\frac{\vec{\sigma}}{2}\vec{\theta}(x)}\Psi_L$, doublets ; $\Psi_R \to \Psi_R$, singlets \\[0.5em]
2) Under  $U(1)_Y$:
$\Psi \to e^{i\frac{Y}{2}\beta(x)}\Psi$. \\[0.5em] 
Where $\Psi_L=(1-\gamma_5)/2$ and $\Psi_R=(1+\gamma_5)/2$ refer to the two possible chiral projections, for left and right handed chiralities of the fermion $\Psi$, respectively.
 
The corresponding quantum numbers for the first generation of quarks and leptons are collected in 
the tables.

\vspace{2cm}
\begin{minipage}[h]{8cm}
\begin{center}
\begin{tabular}{crrrr}
\hline
Lepton & $T$ & $T_3$ & $Q$ & $Y$\\
\hline
$\nu_L$& $\frac{1}{2}$ & $\frac{1}{2}$ & $0$ & $-1$ \\
$e_L$ & $\frac{1}{2}$ & $-\frac{1}{2}$ & $-1$ & $-1$\\
$e_R$ & $0$ & $0$ & $-1$ & $-2$ \\
\hline
\end{tabular}
\end{center}
\end{minipage}
\begin{minipage}[h]{8cm}
\begin{center}
\begin{tabular}{crrrr}
\hline
Quark & $T$ & $T_3$ & $Q$ & $Y$\\
\hline
$u_L$& $\frac{1}{2}$ & $\frac{1}{2}$ & $\frac{2}{3}$
&$\frac{1}{3}$  \\
$d_L$ & $\frac{1}{2}$ & $-\frac{1}{2}$ & $-\frac{1}{3}$ &
$\frac{1}{3}$\\
$u_R$ & $0$ & $0$ & $\frac{2}{3}$ & $\frac{4}{3}$ \\
$d_R$ & $0$ & $0$ & $-\frac{1}{3}$ & $-\frac{2}{3}$\\
\hline
\end{tabular}
\end{center}
\end{minipage} 

\vspace{1cm}
\blackl{The particle content and the Lagrangian of the Electroweak Theory}\\[1em]

The particle content of the SM is summarized schematically in the following:

\vspace{0.5cm}
{\small 
\begin{minipage}[h]{8cm}
\begin{center}
{\bf Matter particles} \\[1.5em]
 $1^{st}$ family:  
$\left(\begin{array}{c}
 \nu_e \\
  e^- 
\end{array} \right)_L$, $e^-_R$,  
 $\left(\begin{array}{c}
 u \\
 d  
\end{array} \right)_L$, $u_R$, $d_R$ 
\end{center} 
 
\begin{center}
 $2^{nd}$ family:  
$\left(\begin{array}{c}
 \nu_{\mu} \\
  \mu^- 
\end{array} \right)_L$, $\mu^-_R$,  
 $\left(\begin{array}{c}
 c \\
 s  
\end{array} \right)_L$, $c_R$, $s_R$ 
\end{center} 
 
\begin{center}
 $3^{rd}$ family:  
$\left(\begin{array}{c}
 \nu_{\tau} \\
  \tau^- 
\end{array} \right)_L$, $\tau^-_R$,  
 $\left(\begin{array}{c}
 t \\
 b  
\end{array} \right)_L$, $t_R$, $b_R$ 
\end{center} 
\end{minipage}
\begin{minipage}{.2cm}\textcolor{black}{\rule{.05cm}{7cm}}
\end{minipage}
\begin{minipage}[h]{8cm}
\BC
{\bf Gauge particles}\\[1em]

$SU(2)_L$: 3 generators $T_i$, 3 gauge bosons $W_i^\mu$\\[0.3em]
$U(1)_Y$: 1 generator $\frac{Y}{2}$, 1 gauge boson $B^\mu$ \\[0.3em]
$W^i_{\mu \nu}=\partial_\mu W^i_\nu -\partial_\nu W^i_\mu + 
g \epsilon^{ijk}W^j_\mu W^k_\nu$\\[0.3em]
$B_{\mu \nu} =\partial_\mu B_\nu -\partial_\nu B_\mu$ \\[0.5em]

{\bf Physical EW bosons} \\[1em]

$W_\mu^{\pm}=\frac{1}{\sqrt{2}}(W_\mu^1\mp i W_\mu^2)$ \\[0.3em]
$Z_\mu=\cos \theta_W W_\mu^3-\sin \theta_W B_\mu$ \\[0.3em]
$A_\mu=\sin \theta_W W_\mu^3+\cos \theta_W B_\mu$ 
\EC
\end{minipage}

\vspace{0.5cm}
The electroweak interactions are introduced via the gauge principle, as in the previous example, by the replacement in the free Lagrangian of the normal derivative by the corresponding covariant derivative:  
$$\partial_\mu \Psi \to D_\mu \Psi =(\partial_\mu -ig \vec{T}.\vec{W}_\mu -i g' \frac{Y}{2} B_\mu)\Psi$$
where $g$ is the $SU(2)_L$ gauge coupling and $g'$ is the $U(1)_Y$ gauge coupling. The relation between the electromagnetic coupling $e$ and  these two couplings $g$ and $g'$ is a consequence of $U(1)_{\rm em}$ being a subgroup of $SU(2)_L\times U(1)_Y$: 
 $$g=\frac{e}{\sin \theta_W} \,\,\,;\,\,g'=\frac{e}{\cos \theta_W}$$ 
where $\theta_W$ is the weak angle that defines the physical neutral gauge bosons $Z_\mu$ and $A_\mu$ in terms of the EW interaction eigenstates $W_\mu^3$ and $B_\mu$.

The Lagrangian of the Electroweak Theory is then given by:
$${\cal L}_{\rm EW}= \sum_{\Psi}i \overline{\Psi}\gamma^{\mu}D_{\mu}\Psi 
-\frac{1}{4}W^i_{\mu \nu}W^{\mu\nu}_i -\frac{1}{4}B_{\mu \nu}B^{\mu \nu}$$
where the sum runs over all the fermions of the SM: quarks and leptons.

This Lagrangian ${\cal L}_{\rm EW}$ is invariant under $SU(2)_L\times U(1)_Y$ gauge 
transformations. However, notice that it doesnt content any mass term for any of the SM fields.
It can be easily checked that a mass term like $m \overline{\Psi}\Psi =m(\overline{\Psi}_L \Psi_R +\overline{\Psi}_R \Psi_L)$ and a mass term like $M_W^2 W_\mu W^\mu$ are not $SU(2)_L$ invariant. Therefore ${\cal L}_{\rm EW}$ does not describe properly yet the masses for the fermions nor the weak gauge bosons and a new piece in the SM Lagrangian must be introduced to generate the particle masses which is directly related with some sort of breaking of the Electroweak symmetry. Then the full SM Lagrangian will be finally built from the two terms:  
$${\cal L}_{\rm SM}={\cal L}_{\rm EW}+{\cal L}_{\rm EWSB}$$
where ${\cal L}_{\rm EWSB}$ refers to the Lagrangian for the Electroweak Symmetry Breaking
that will be described next.
\section{Electroweak Symmetry Breaking}
In this section we shortly summarize the basics of the Electroweak Symmetry Breaking in the SM and  set the steps to follow for the building of ${\cal L}_{\rm EWSB}$.

The most relevant aspects of the Electroweak Symmetry Breaking can be organized in three main points:

\begin{enumerate}
 
\black{\item}
The Phenomenon of Spontaneous Symmetry Breaking \\[-1.5em]

\black{\item}
Spontaneous Symmetry Breaking: the Goldstone Theorem\\[-1.5em]

\black{\item} 
Electroweak Symmetry Breaking: the Higgs Mechanism\\[-1.5em]

\end{enumerate}
These are important to understand separately and will be commented in the following. 
 
\vspace{1cm} 
\blackl{The Phenomenon of Spontaneous Symmetry Breaking }\\[1em]
{\bf A simple definition:}\\[0.2em]
{\it A physical system has a symmetry that is spontaneously
broken if the interactions governing the dynamics of the system possess
such a symmetry but the ground state of this system does not}.\\[0.2em]
{\bf A simple example:} \\[0.2em]
Let us consider an infinitely extended ferromagnet at temperature $T$ close to the Curie temperature $T_C$. The system is described by an infinite set of elementary spins and their interactions (given by the Lagrangian) are rotational invariant. The ground state of this system presents two different situations depending 
on the value of $T$ being above or below the Curie temperature. These two situations are schematically described below:
\\[0.2em]
\begin{minipage}[h]{9cm}
Situation I: $T>T_C$ \\
\BC
\includegraphics[width=5cm]{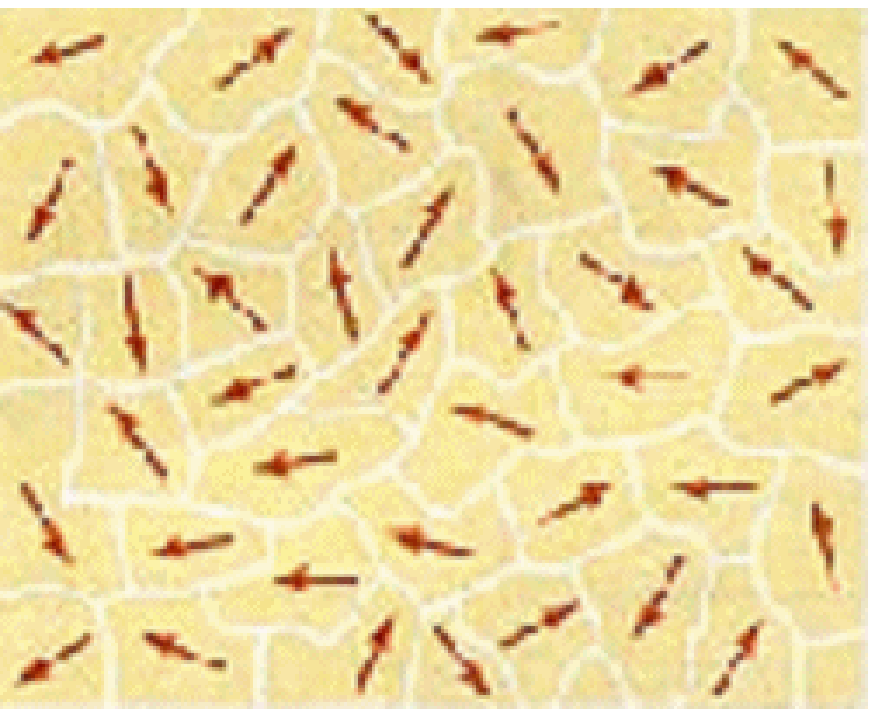} 
\EC
the spins are randomly oriented\\
the ground state is rotationaly invariant\\
the average Magnetization (order parameter)\\
vanishes, $\vec{M}_{\rm average}=0$ 
\end{minipage}
\begin{minipage}[h]{9cm}
Situation II: $T<T_C$ 
\BC
\includegraphics[width=5cm]{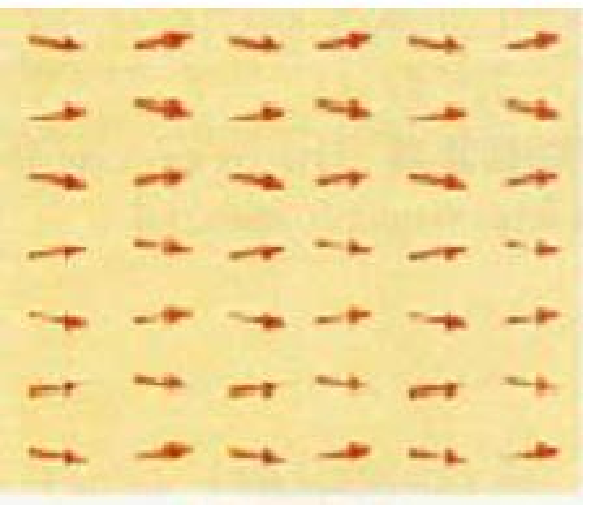}
\EC
the spins are oriented to some particular\\ (and arbitrary) direction\\
the ground state is not rotationaly invariant\\
$\vec{M}_{\rm average} \neq 0$ ({\it Spontaneous Magnetization})\\
$\exists$ infinite possible ground states, \\
but the system chooses a particular one.
\end{minipage}

\vspace{1cm}
With regard the mathematical description of this behavior of the ground state in the extended ferromagnet example, there is a very simple theoretical framework that describes successfully this phenomenon of spontaneous symmetry breaking, the Theory of Ginzburg-Landau.

\vspace{0.5cm}
{\bf The Theory of Ginzburg-Landau (1950)}\\[1em]
 In this theory, for $T$ near $T_C$, the free energy density
$u(\vec{M})$ for small $\vec{M}$  is given by:
$$ 
u(\vec{M}) = (\partial_i \vec{M})(\partial_i \vec{M})+V(\vec{M})\;;\;
i=1,2,3 \,\,$$ 
where the potential is:
$$V(\vec{M})=\alpha_1(T-T_C)(\vec{M}.\vec{M})+\alpha_2(\vec{M}.\vec{M})^2
\;;\;\alpha_1,\alpha_2>0 $$
 Notice that in the drawings to simplify we have chosen a two dimensional (instead of three) magnetization vector $\vec{M}=(M_X,M_Y)$.
 
The magnetization of the ground state is obtained from the condition of
extremum:
\begin{equation}
\frac{\delta V(\vec{M})}{\delta M_i}=0\Rightarrow
\vec{M}.\left [ \alpha_1(T-T_C)+2\alpha_2(\vec{M}.\vec{M})\right ]=0 
\non
\end{equation}

This leads to two solutions for $\vec{M}$, depending on the value of $T$ which correspond respectively to the previous described situations I and II. These two qualitative different solutions describe the so-called symmetric and non symmetric phases of the system.\\[1cm]
\begin{minipage}[h]{8cm}
Situation I: $T>T_C$, Symmetric phase\\
\BC
\includegraphics[width=5cm]{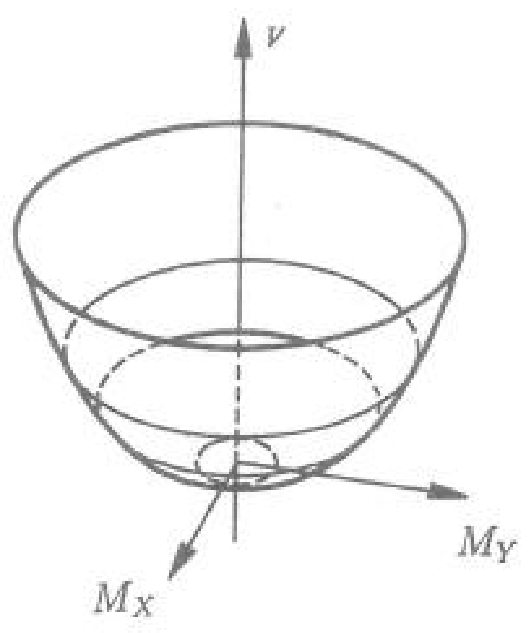} \\
\EC
Unique minimum at $\vec{M}=0$ and $V(0)=0$ 
\end{minipage}
\begin{minipage}[h]{10cm}
Situation II: $T<T_C$, Non symmetric phase\\
\BC
\includegraphics[width=6cm]{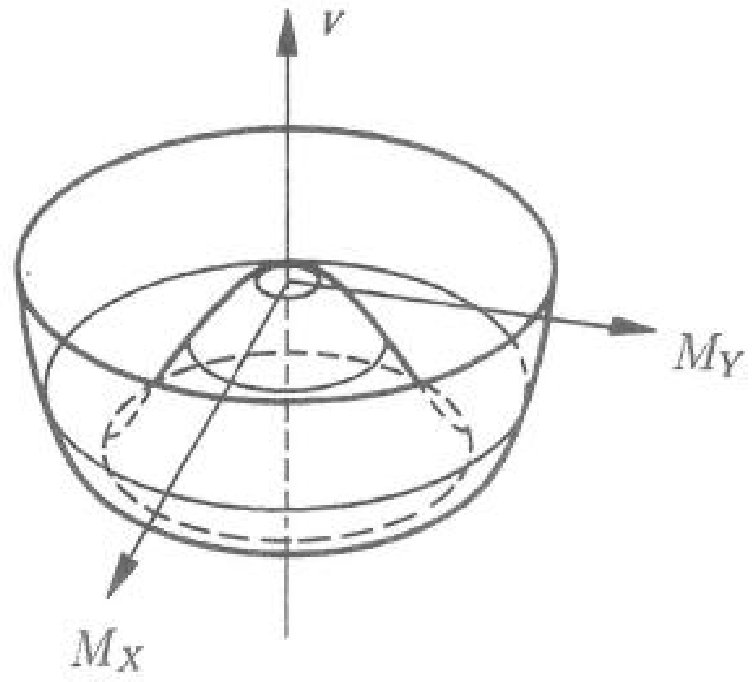} \\
\EC
$\vec{M}=0$ is a local maximum \\
$\exists$ infinite degenerate minima all having same $|\vec{M}|$ \\
$ \alpha_1(T-T_C)+2 \alpha_2(\vec{M}.\vec{M})=0 \Rightarrow
|\vec{M}|=\sqrt{\frac{\alpha_1(T_C-T)}{2\alpha_2}}$ \\
The choice of a particular minimum (direction) \\
is what generates the spontaneous breaking.
\end{minipage}

\vspace{1cm} 
\blackl{Goldstone Theorem (Nambu,Goldstone,1960-1962)}\\[0.5em]

The Goldstone Theorem applies to Quantum Field Theories (QFT) with Spontaneous Symmetry Breaking (SSB).\\[0.5em]
{\bf SSB stated in simple words:}\\[0.5em]
{\it In QFT, a system is said to
have
a symmetry that is spontaneously broken if the Lagrangian describing the
dynamics of the system is invariant under this symmetry
transformation, but the vacuum of the theory is not. The vacuum
$|0>$ is the state where the Hamiltonian expectation value $<0|H|0>$ is
minimum.} \\[0.5em]
{\bf Goldtone Theorem stated in simple words:}\\[0.5em]
{\it If a QFT has a global  symmetry of the Lagrangian which is not a
symmetry of the vacuum $\RA$ there must exist one massless
boson, scalar or pseudoscalar, associated to each
generator which does not annihilate the vacuum and having its same
quantum numbers. These modes are referred to as Nambu-Goldstone bosons
or simply as Goldstone bosons (GBs).}\\[0.5em]
Notice that:\\[0.1em]
$ U|0> = |0> \; \mbox{with} \; U= \exp(i\epsilon^a Q^a) \RA Q^a|0>=0\,\,\,  \forall \,\,a $\\[0.1em]
and:\\[0.1em]
$ U|0> \neq |0> \; \mbox{with} \; U= \exp(i\epsilon^a Q^a) \RA \exists \,\,Q^a 
\,\,/\,\,  Q^a|0> \neq 0$ 

\vspace{0.5cm} 
\blackl{QCD as an example}

\vspace{0.5cm}
One illustrative example of the phenomenon of SSB and the consequences of the Goldstone Theorem is provided by the well known case of QCD with two flavors, $u,d$, where there is a global symmetry, the chiral symmetry, that is known to be spontaneously broken. We comment briefly on this next. 

Let us start with the QCD Lagrangian, given in terms of quarks, $q_i\equiv q (i=1,2,3)$, and gluons, $g_\mu^a \equiv A_\mu^a (a=1,..8)$, by:  
 \begin{equation}
{\cal L}_{\rm QCD}=-\frac{1}{2} TrG^{\mu\nu} G_{\mu\nu}+\sum_{u,d}
(i\bar{q}\gamma^{\mu}D_{\mu}q-m_q\bar{q}q) \non
\end{equation}
where,
\begin{eqnarray}
  G_{\mu\nu}&=&\partial_{\mu}A_{\nu}-\partial_{\nu}A_{\mu}-ig_s
  \left[A_{\mu},A_{\nu}\right]\nonumber \\
  D_{\mu}q &=& (\partial_{\mu}-ig_s A_{\mu})q \nonumber \\
 A_{\mu}&=&\sum_{a=1}^8 \frac{1}{2}A_{\mu}^a\lambda_a \non
\end{eqnarray} 
The generators of the $SU(3)_C$ color group are the eight $\lambda_a/2$ matrices, with $\lambda_a$ being the well known $3 \times 3$
Gell-Mann matrices. $g_s$ is the strong coupling constant, and $m_q$ is the mass of the quark $q$. 

In addition to the $SU(3)_C$ gauge symmetry of QCD, that is the responsible for the strong interactions among quarks and gluons, ${\cal L}_{\rm QCD}$ has an extra global symmetry for the case of massless quarks, $m_{u,d}=0$:  
\black{$$SU(2)_L \times SU(2)_R \,\,\equiv \,\,\mbox{Chiral Symmetry}$$}
defined by:
\[ \Psi_L \rightarrow \Psi_L'=U_L \Psi_L =\exp(i\alpha_L^a Q_L^a)\Psi_L\; ;\;Q_{L}^{1,2,3} \,\,\,\mbox{generators of}\,\,\, SU(2)_{L}\]
\[ \Psi_R \rightarrow \Psi_R'=U_R \Psi_R =\exp(i\alpha_R^a Q_R^a)\Psi_R\; ;\;Q_{R}^{1,2,3} \,\,\,\mbox{generators of}\,\,\, SU(2)_{R} \]
where,
\[ \Psi = \left ( \begin{array}{c}u \\ d \end{array} \right ) \;;\;
\Psi_L=\frac{1}{2}(1-\gamma_5)\Psi\;;\;
\Psi_R=\frac{1}{2}(1+\gamma_5)\Psi \]
When the $m_{u,d}\neq 0$ terms are included into ${\cal L}_{\rm QCD}$ then the chiral symmetry is explicitly broken, but not much since these quark masses are small. Then, the chiral symmetry is not an exact global symmetry but it is a very good approximate symmetry of QCD.

On the other hand, it happens that this chiral symmetry is not a symmetry of the QCD vacuum, therefore it must be a spontaneously broken symmetry.
Indeed, this  chiral symmetry is spontaneously broken down to the
isospin symmetry, given by the subgroup $SU(2)_V$ of the chiral group, $SU(2)_L \times SU(2)_R$:
\[ SU(2)_L \times SU(2)_R = SU(2)_V \times SU(2)_A \rightarrow SU(2)_V 
\,\,;\,\, SU(2)_V=SU(2)_{R+L}\,\,;\,\,SU(2)_A=SU(2)_{R-L}\] \\[.3em]
The SSB phenomenon occurs here because ${\cal L}_{\rm QCD}$ is invariant under $SU(2)_L \times SU(2)_R$ but the 
QCD vacuum is NOT fully $SU(2)_L \times SU(2)_R$ invariant. It  
is only invariant under the subgroup $SU(2)_V \subset SU(2)_L \times SU(2)_R$. Schematically we write this SSB as: 
$$ SU(2)_L \times SU(2)_R \to SU(2)_V$$
But, how do we know from experiment that the QCD vacuum is not $SU(2)_L \times SU(2)_R$ symmetric?. The demonstration of this fact goes by starting with the 'negative'  hypothesis, i.e assuming a QCD symmetric vacuum, and ending in an acceptable conclusion. \\[.5em]
Let us assume that  $|0>$ is chiral invariant $\Rightarrow$
 \[  U_L|0>=|0>\;;\;U_R|0>=|0>
  \Rightarrow Q_L^a|0>=0\;;\;
Q_R^a|0>=0 \]

Let $|\Psi>$ be an eigenstate of the
Hamiltonian and parity operator such that:
\[ H|\Psi>=E|\Psi>\;;\;P|\Psi>=|\Psi> \]
Then, from the two previous assumptions, one finds a new eigenstate $|\Psi'>$ of the Hamiltonian with the same eigenvalue as $|\Psi>$ but with opposite parity: 
\[
\exists
|\Psi'>= \frac{1}{\sqrt{2}}(Q_R^a-Q_L^a)|\Psi>\; / \; H|\Psi'>=E|\Psi'>
\;;\;P|\Psi'>=-|\Psi'> \]
But, it turns out that {\it there are not such parity doublets in the
hadronic spectrum} $\RA$ $SU(2)_A$ is NOT a symmetry of the vacuum, or equivalently, $Q_A^a|0>\neq 0 (a=1,2,3)$. $\RA$ {\it chiral symmetry must be
spontaneously broken to the reduced symmetry of the vacuum, $SU(2)_V$}. 

Now, according to Goldstone Theorem, and as a consequence of the previous breaking, 
 there must exist one massless
 Goldstone boson, scalar or pseudoscalar, associated to each
generator which does not annihilate the vacuum and having its same
quantum numbers. More specifically, the spontaneous breaking of the chiral symmetry in QCD, implies the existence of three massless Goldstone bosons:
$$ SU(2)_L \times SU(2)_R \to SU(2)_V\,\,\,;\,\,\,\mbox{with} ,\,\,Q_A^a|0>\neq 0 (a=1,2,3)$$ 

\black{$\RA \,\,\,\,\exists$  3 massless GBs, pseudoscalars, $\pi^a(x)\;\; a=1,2,3$.}\\[.7em] 
A very important feature for the phenomenology of low energy QCD is that these three GBs are identified with the physical pions. More specifically, their combinations: $\pi^+=(\pi^1-i\pi^2)/\sqrt{2}$, $\pi^-=(\pi^1+i\pi^2)/\sqrt{2}$ and $\pi^0=\pi^3$.

Since, in Nature, $m_\pi \neq 0$ $\RA$ chiral symmetry is explicitly broken, and the pions are pseudo-GB. But the important outcome is that the hierarchy $m_\pi<<m_{\rm hadrons}$ is explained.

The dynamics of pion interactions is well described by the so-called  Chiral Lagrangian of QCD and the associated Effective Quantum Field Theory called Chiral Perturbation Theory (ChPT). We will come back to the subject of Chiral Lagrangians in the next lecture where we will comment on some applications o these type of effective Lagrangians for beyond the Standard Model Physics.

\vspace{0.5cm}
\blackl{The Higgs Mechanism:}\\[.7em]
The Goldstone Theorem is for theories with spontaneously broken global
symmetries but does not hold for gauge theories. When a spontaneous
symmetry breaking takes place in a gauge theory, the so-called Higgs
Mechanism operates. As will be seen in the following, the Higgs Mechanism when applied to the case of the SM leads to the prediction of a new scalar particle, the so-called Higgs boson particle, whose experimental discovery by the collaborations ATLAS and CMS at the Large Hadron Collider (LHC), placed at the laboratory CERN, close to Geneva,  was anounced in an international open web-conference on the 4th of July 2012. Recently, on the 8th of October 2013, Peter Higgs and Francois Englert have received the Physics Nobel Prize 2013 for the proposal of this mass generation mechanism and for the prediction of the Higgs boson particle.

In the historical development of the guiding ideas that ended up with the final Higgs Mechanism there were indeed many authors involved, including: 
Higgs (1964); Englert, Brout (1964); Guralnik, Hagen, Kibble (1964). Many of these contributions were inspired in previous works within Solid State Physics, including those by Anderson (1963). See also the works by Schwinger (1962) where the generation of mass for gauge fields was already mentioned. See also the BCS Theory of Superconductivity, the existence of Cooper pairs and the absence of massless GBs in presence of electromagnetic interactions which can be found in the works by  Nambu (1960). \\[.7em]   
{\bf How to generate mass for gauge bosons in gauge theories (in simple words):}\\[.7em]  
{\it When a spontaneous
symmetry breaking takes place in a gauge theory the would-be Goldstone bosons associated to the global symmetry
breaking do not manifest
explicitly in the physical spectrum but instead they 'combine' with the
massless gauge bosons and as result, once the spectrum of the theory is
built up on the non-symmetric vacuum, there appear massive vector
bosons. The number of vector bosons that acquire a mass is
precisely equal to the number of these would-be-Goldstone bosons, which in turn are equal to the number of symmetries that the vacuum has lost}. 

Before going to the SM case, we illustrate first the Higgs Mechanism with one very simple example.

\blackl{An illustrative example:  $U(1)$ gauge symmetry breaking:}\\[.7em]
 Consider the simplest case of a gauge theory based in a $U(1)$ gauge symmetry, with one complex scalar  
 $\Phi= \frac{1}{\sqrt{2}}(\Phi_1+i\Phi_2)$, one gauge boson $A_\mu$, and a potential of Ginzburg-Landau type.
 The Lagrangian for this $U(1)$ gauge theory is:\\[.7em]
${\cal L}= (D_\mu \Phi)^{\dagger}(D^\mu \Phi) - \frac{1}{4}F_{\mu \nu}F^{\mu \nu}- V(\Phi),$ 
\\[.3em]
with $D_\mu \Phi= (\partial_\mu-ig A_\mu)\Phi \,\,\,;\,\,\,
F_{\mu\nu}=\partial_\mu A_\nu - \partial_\nu A_\mu $, and
$V(\Phi)=\mu^2 \Phi^{\dagger} \Phi + \lambda (\Phi^{\dagger} \Phi)^2 \,\,\,;\,\,\,\lambda >0$.\\[.3em]
${\cal L}$ is invariant under $U(1)$ gauge transformations given by:\\[.3em]
$\Phi \to e^{-i\alpha (x)}\Phi \,\,\,;\,\,\,D_\mu \Phi \to e^{-i\alpha (x)} D_\mu \Phi \,\,\,;\,\,\,e^{-i\alpha (x)}\subset U(1)$\\[.3em]
$A_\mu \to A_\mu - \frac{1}{g} {\partial}_{\mu}\alpha (x)$\\[.3em] 
It is interesting to compare $V(\Phi)$ with the previous ferromagnet case:\\[.3em]
$V(\vec{M})=\alpha_1(T-T_C)(\vec{M}.\vec{M})+\alpha_2(\vec{M}.\vec{M})^2
\;;\;\alpha_1,\alpha_2>0$\\[.3em]
All said previously applies now with the replacements: $(M_X,M_Y) \to \frac{1}{\sqrt{2}}(\Phi_1+i\Phi_2)$\\[.3em]
$\alpha_1(T-T_C) \to \mu^2$;  $\alpha_2 \to \lambda$;   
$\vec{M}_{\rm ground \,\,state} \to <0|\Phi|0> \equiv <\Phi> $ 
 
Then, in this case one similarly finds two different situations, but now corresponding to 
having either  $\mu^2>0$ or  $\mu^2<0$. These two situations describe the symmetric phase and the non-symmetric phase of the $U(1)$ gauge theory, as summarized schematically in the following:  
\begin{minipage}[h]{8cm}
Situation I: $\mu^2>0$, Symmetric phase\\
\BC
\includegraphics[width=5cm]{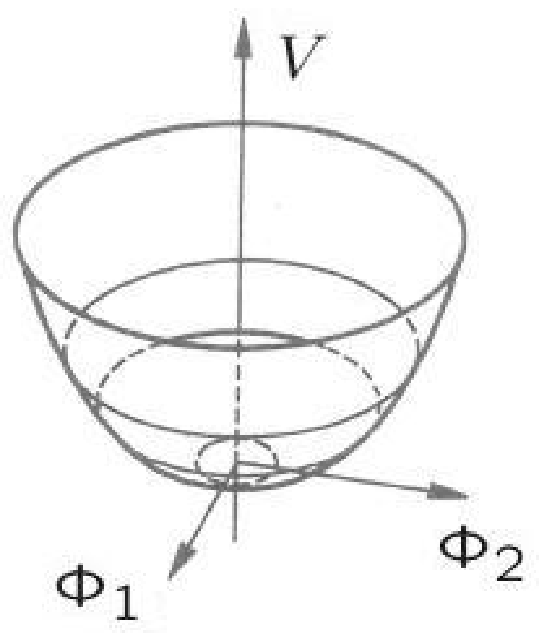} \\
\EC
Unique vacuum (minimum) at $<\Phi>=0$ \\
and $V(\Phi)=0$ at $<\Phi>=0$\\
The  {vacuum IS invariant} under $U(1)$
\end{minipage}
\begin{minipage}[h]{10cm}
Situation II: $\mu^2<0$, Non symmetric phase\\
\BC
\includegraphics[width=6cm]{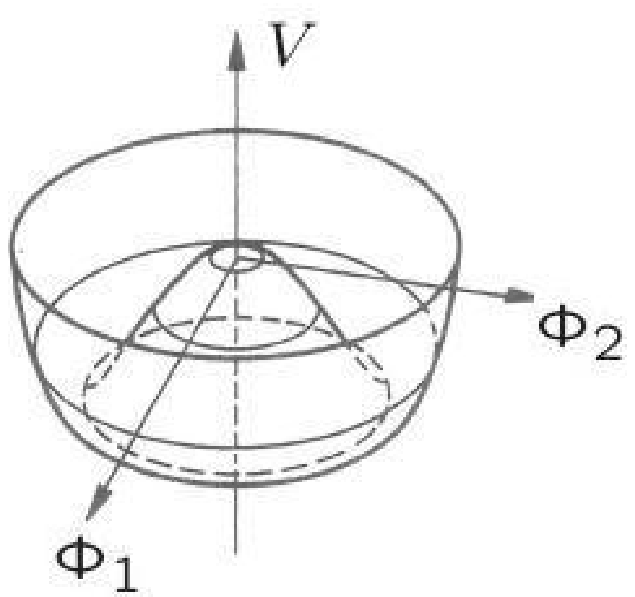} \\
\EC
$<\Phi>=0$ is a local maximum \\
$\exists$ infinite degenerate vacua (minima) all having same \\
 $|<\Phi>|$ but different 
 complex phases:\\
$|<\Phi>|=\sqrt{\frac{-\mu^2}{2\lambda}}\equiv \frac{v}{\sqrt{2}} \neq 0\,\,$;
$\arg<\Phi>$ arbitrary \\
A particular {vacuum IS NOT invariant} under U(1)\\
\end{minipage}
At this point, there are two important features that are worth to emphasize: \\
1){\it the choice of a particular vacuum (complex phase) 
is what generates the spontaneous breaking of $U(1)$}, \\
2)  
{\it the building of the spectra on top of this non-invariant vacuum (minimum) is what generates the gauge boson mass}.\\ 
The first point is clear, since once a particular complex phase has been chosen to describe the vacuum, this is not any more invariant under a $U(1)$ transformation which precisely rotates this phase and would change the starting vacuum into another one with a different phase.
Let us see now the second point in more detail.
 
Let us choose first a particular vacuum configuration, for instance, let us take a real one:\\
$|<\Phi>|=\sqrt{\frac{-\mu^2}{2\lambda}}\neq 0\,\,$;
$\arg<\Phi>=0$ $\RA$ $<\Phi_1>=\sqrt{\frac{-\mu^2}{\lambda}}=v$, $<\Phi_2>=0$ \\
Then, we change coordinates to new fields ($\equiv$ shifting the origen):\\
$\Phi'_1 \equiv \Phi_1-v$; $\Phi'_2 \equiv \Phi_2$ such that $<\Phi'_1>=0$;
$<\Phi'_2>=0$ \\
Next, write everything in terms of these new $\Phi'_{1,2}$ fields:
$$(D_\mu \Phi)^\dagger (D_\mu \Phi)=\left((\partial_\mu 
+i g A_\mu) \frac{1}{\sqrt{2}}(\Phi_1- i\Phi_2)\right)
\left((\partial_\mu 
-i g A_\mu) \frac{1}{\sqrt{2}}(\Phi_1+ i\Phi_2)\right)=....$$
$$
\frac{1}{2}(\partial_\mu\Phi'_1+g A_\mu \Phi'_2)^2
+\frac{1}{2}(\partial_\mu\Phi'_2-g A_\mu \Phi'_1)^2
-g v A^\mu(\partial_\mu \Phi'_2+g A_\mu \Phi'_1)
+ {\frac{1}{2}g^2v^2 A_\mu A^\mu }$$
And we see that a mass term for $A_\mu$} has appeared, i.e. the last term above. But it is not the physical basis yet since
there is a (nonphysical) mixing term $\sim gv A^\mu \partial_\mu \Phi'_2$ which 'combines' the gauge boson and the scalar fields.  In order to find the physical states this mixing term has to be removed.
It is convenient then to first  
  choose some proper coordinates: for instance, let us take
 'polar' coordinates to describe 'small oscillations' around vacuum configuration:
$$\Phi(x)=\frac{1}{\sqrt{2}}(v+\eta(x))e^{i\frac{\xi(x)}{v}}$$ 
Second, let us choose the proper gauge, i.e., make a gauge transformation to the unitary gauge (by fixing the gauge parameter to $\alpha(x)=\frac{\xi(x)}{v}$) where the unwanted mixing terms do not appear:
$$\Phi(x) \to e^{-i\frac{\xi(x)}{v}}\Phi(x)=\frac{1}{\sqrt{2}}(v+\eta(x))$$
$$A_\mu(x) \to A_\mu (x)-\frac{1}{gv} \partial_\mu \xi (x) \equiv B_\mu(x)$$ 
Finally, we write the Lagrangian in terms of the new fields $B_\mu$ and $\eta$:  
$${\cal L}=\frac{1}{2}(\partial_\mu \eta)^2+{\mu^2 \eta^2}
-\frac{1}{4}(\partial_\mu B_\nu -\partial_\nu B_\mu)^2 + {\frac{1}{2}(gv)^2B_\mu B^\mu} +\phantom{\RA}\phantom{\RA} $$
$$ \phantom{\RA} \hspace{2.6cm} \frac{1 }{2}g^2 B_\mu B^\mu \eta (2 v+ \eta)
-\lambda v \eta^3 - \frac{1}{4} \lambda \eta^4 $$
And we see clearly that these new fields, which are now physical, describe a massive gauge boson particle $B_\mu$ with spin 1 and mass $M_{B_\mu}=gv$, and a massive scalar particle $\eta$ with spin 0 and mass $m_\eta=\sqrt{2}|\mu|$. Notice also that the would-be-Goldstone boson in this example is the $\xi$ field and it has dissapeared from the spectrum. There is one symmetry of the Lagrangian that is not preserved by the vacuum and as a consequence there is one gauge boson getting mass. There is also one remaining scalar particle in the physical spectrum, the $\eta$ particle that is the Higgs particle of this example.

\vspace{0.5cm}
\blackl{The 'nice' properties of the Higgs Mechanism:}\\[.7em]
Here we collect some of the general properties of the Higgs Mechanism:\\[.7em]
${\star}$ The gauge symmetry of the interactions (i.e, of ${\cal L}$) is preserved\\[0.5em]
${\star}$ The renormalizability of the massless gauge theories is preserved\\[0.5em]
${\star}$ The total number of polarization degrees is preserved\\
For instance, in the previous $U(1)$ case:\\
Before SSB: total polarization degrees = 4 = (2 of $A_\mu$)+(2 of $\Phi$)\\
After SSB: total polarization degrees = 4 = (3 of $B_\mu$)+(1 of $\eta$)\\[0.5em]  
${\star}$ The nonphysical fields (i.e. the would-be-GBs) have dissapeared from the spectrum. In the previous $U(1)$ case: $\xi(x)$\\[0.5em]
${\star}$ The number of gauge bosons getting a mass = number of would-be-GBs= number of symmetries of ${\cal L}$ that are not symmetries of the vacuum. In the previous $U(1)$ case, this number is 1.\\[0.5em]
${\star}$ The would-be-GBs combine with the massless gauge bosons to give them a mass. This 'combine' in the U(1) example occurs indeed due to the mixing term $\sim gv A^\mu \partial_\mu \Phi'_2$\\[0.5em]
${\star}${Comment:} The Higgs mechanism does not necessarily imply the existence of a Higgs particle. It appears JUST  when required by the polarization degrees preservation property.   
 
\vspace{0.5cm} 
\blackl{The Higgs Mechanism applied to the Standard Model:}\\[.2em]
We want to generate masses for 3 gauge fields: $Z$, $W^+$ and $W^-$, 
but we want to keep the photon $\gamma$ massless.\\
{\bf Strategy:}
Introduce ({\it ad hoc}) a new scalar field, $\Phi$, and a potential of Ginzburg-Landau type, $V(\Phi)$ that make the job. Then one requires the following properties to this $\Phi$:\\
$\RA$ It must provide the 3 needed polarization degrees to play the role of the would-be-GBs \\
$\RA$ It must have non-zero $SU(2)_L \times U(1)_Y$ quantum numbers, such that the vacuum is not invariant under the complete symmetry, but just invariant under the subgroup $U(1)_{\rm em}$.\\
$\RA$ The field component in $\Phi$ acquiring a vev must be electrically neutral
to preserve  $U(1)_{\rm em}$.

Within the SM  these $\Phi$ and $V(\Phi)$ are chosen to be the simplest ones fulfilling all the above requirements:
$$\mbox{The SM introduces one complex scalar SU(2) doublet:} \quad \black{\Phi = \VL \phi^+ \\ \phi^0 \VR},$$
with particular $SU(2)_L \times U(1)_Y$ quantum numbers given by:
$$T(\Phi)=\frac{1}{2}\,\,,\,\, Y(\Phi)=1,$$
and with a potential defined as:  
$$
\black{
V(\Phi) = \mu^2 \Phi^{\dagger} \Phi + \la
(\Phi^{\dagger} \Phi)^2, \quad \la > 0 },
$$
that is a copy of the  Ginzburg-Landau one with the replacements: 
$\alpha_1(T-T_C) \to \mu^2$;  $\alpha_2 \to \lambda$ and    
$\vec{M} \to \Phi$.

The two phases in the SM case, are reached, as in the previous example, by setting the sign o 
$\mu^2$, either to $\mu^2>0$ if we want to place the SM vacuum in the symmetric phase, or to  
$\mu^2<0$ if we want the SM vacuum  to be in the non-symmetric phase. These two phases are schematically described below: 

\begin{minipage}[h]{10cm}
\vspace{0.5cm}
$\mu^2>0$: Symmetric phase:\\ 
 $V(\Phi)$ has a unique minimum at $<0|\Phi|0>=0$.\\
 The SM vacuum is $SU(2)_L \times U(1)_Y$ invariant.\\[1em]
 $\mu^2<0$: Non symmetric phase:\\ 
 $\exists$ infinite degenerate minima at:\\
 $|<0|\Phi|0>|=\left(
\begin{array}{c} 0 \\
 \frac{v}{\sqrt{2}} \end{array}\right)\;\;;\;\; {\rm with} \;\; v\equiv\sqrt{\frac{\mu^2}{\lambda}}\\
 \;\;\;\;\; {\rm and}\;\;\;{\rm  arbitrary}\;\;
  {\rm arg}\;\Phi .$\\
 It is the choice of a particular ${\rm arg}\;\Phi$ \\
 what produces the breaking.
%
\end{minipage}
\begin{minipage}[h]{9cm}
\vspace{0.3cm}
\black{$\mu^2 > 0: SU(2)_L \times U(1)_Y $} \\[0.5em]
\black{$\mu^2 < 0: SU(2)_L \times U(1)_Y \to U(1)_{\rm em}$} \\[0.5em]
\includegraphics[height=5cm,width=6cm,angle=0]{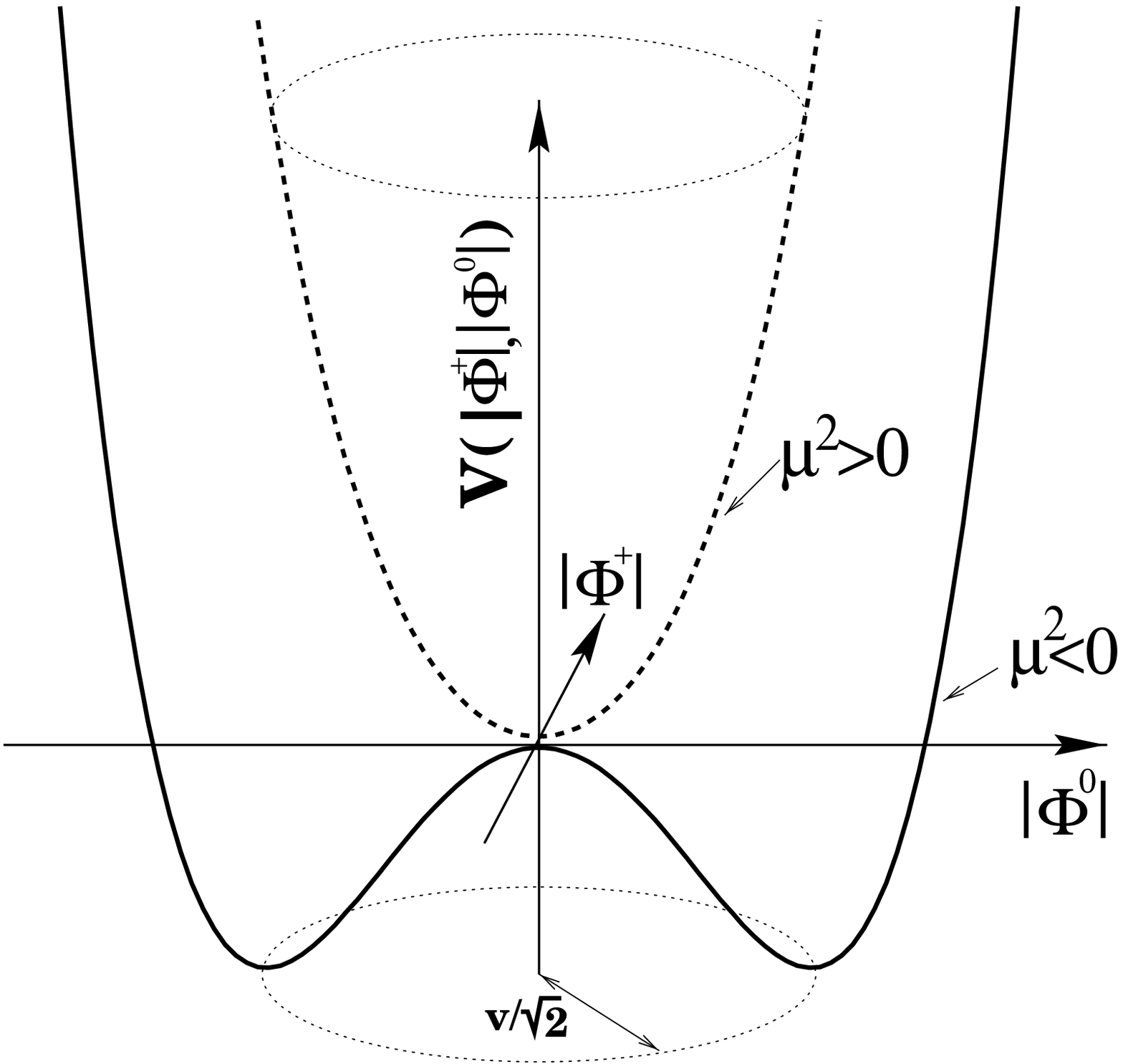}
\end{minipage}

\vspace{0.5cm}
Summarizing the outcome of this spontaneous EW symmetry breaking: for $\mu^2 < 0$ the SM vacuum is not $SU(2)_L \times U(1)_Y$ invariant
but it is just $U(1)_{\rm em}$ invariant.\\ 
This SSB is usually represented by the sequence: full EW symmetry $\to$ vacuum symmetry, namely,
$SU(2)_L \times U(1)_Y \to U(1)_{\rm em}$.

\section{The spectra of the SM and the particle masses}
In order to get the proper gauge boson and fermion masses by means of the Higgs Mechanism the EW symmetry breaking Lagrangian  ${\cal L}_{\rm EWSB}$ has to be properly defined.
Within the SM, ${\cal L}_{\rm EWSB}$ is built by including the previous $V(\Phi)$, the proper covariant derivatives of $\Phi$ and the Yukawa interactions of $\Phi$ with fermions. Specifically, one defines: 
\begin{eqnarray}
 {\cal L}_{\rm EWSB}&=&{\cal L}_{\rm SBS}+{\cal L}_{\rm YW}, \non  
 \end{eqnarray}
 where the Symmetry Breaking Sector Lagrangian (SBS) is given by:
 \begin{eqnarray}
{\cal L}_{\rm SBS}& =& (D_\mu \Phi)^\dagger (D^\mu \Phi)-V(\Phi), \non  
\end{eqnarray}
with 
\begin{eqnarray}
V(\Phi) & =& \mu^2\Phi^\dagger \Phi + \lambda (\Phi^\dagger \Phi)^2, \non 
\end{eqnarray}
and 
\begin{eqnarray}
 D_\mu \Phi &  = &  ( \partial_\mu - \frac{1}{2} i g
\vec{\tau}\cdot\vec{W}_\mu
-\frac{1}{2} i g' B_\mu) \Phi, \nonumber
\end{eqnarray}
The Yukawa Lagrangian (YW) is given in terms of the $\lambda_q$ Yukawa couplings by:
\begin{eqnarray}
{\cal L}_{\rm YW}&=& \lambda_e\bar{l}_L\Phi e_R+
\lambda_u\bar{q}\widetilde{\Phi}u_R+\lambda_d\bar{q}_L\Phi d_R+h.c.+
2^{\rm nd}\; {\rm and}\; 3^{\rm rd}\;{\rm families},\non
\end{eqnarray}
with
\begin{eqnarray}
l_L& = & \left ( \begin{array}{c}\nu_L \\ e_L \end{array}\right )\;;\;\;
q_L =  \left (\begin{array}{c}u_L \\ d_L \end{array}\right ) \nonumber
\\
\Phi & = & \left(
\begin{array}{c}\phi^+ \\
 \phi_0 \end{array}\right)\;;\;\;
\widetilde{\Phi}=i\tau_2\Phi^*=\left(\begin{array}{c}
\phi^*_0\\-\phi^-\end{array}\right)\nonumber
 \end{eqnarray}

It is a simple exercise to check that the \black{${\cal L}_{\rm EWSB}$ above is gauge $SU(2)_L \times U(1)_Y$ invariant.}


Once ${\cal L}_{\rm EWSB}$ is defined, then one follows the following steps:\\[1em]
 1) Fix a particular non-symmetric vacuum. For instance: 
  \[<0|\Phi|0>=\left( \begin{array}{c} 0\\ \frac{v}{\sqrt{2}}
  \end{array}\right )\;\;;\;\; {\rm arg} \Phi =0\]
 2) Perform 'small
oscillations' around this vacuum: 
\[\Phi(x)=\exp{\left ( i\frac{\vec{\xi}(x)\vec{\tau}}{v}\right ) }
\left(\begin{array}{c}
0\\ \frac{v+H(x)}{\sqrt{2}}\end{array}\right ) \]
where $\vec{\xi}(x)=({\xi}_1(x),{\xi}_2(x),{\xi}_3(x))$ and $H(x)$ are 'small' fields.\\[1em]
3) To eliminate the nonphysical (would-be-GBs) fields $\vec{\xi}$  make the
 following gauge transformation (i.e. go to the unitary gauge):\vspace{-0.6cm}
\begin{eqnarray}
\Phi'&=&U(\xi)\Phi=\left(\begin{array}{c}0\\ \frac{v+H}
{\sqrt{2}}\end{array}\right)\;;\;
U(\xi)=\exp\left(-i
\frac{\vec{\xi}\vec{\tau}}{v}\right) \nonumber\\
l_L'&=&U(\xi)l_L\;;\;e_R'=e_R\;;\;q_L'=U(\xi)q_L
\;;\;u_R'=u_R\;;\;d_R'=d_R \nonumber\\
\left(\frac{\vec{\tau}\cdot\vec{W}'_{\mu}}{2}\right)&=&
U(\xi)\left(\frac{\vec{\tau}\cdot\vec{W}_{\mu}}{2}\right) U^{-1}(\xi)
-\frac{i}{g}(\partial_{\mu}U(\xi))U^{-1}(\xi)\;;\;B'_{\mu}=B_{\mu} \non
\end{eqnarray}

4) Rotate the weak eigenstates  to the mass
eigenstates:  
\begin{eqnarray}
W^\pm_\mu & = & \frac{W'^1_\mu \mp i W'^2_\mu}{\sqrt 2}\;;\;
\nonumber\\[2mm] Z_\mu & = & \cos \theta_W\; W'^3_\mu - \sin \theta_W\; B'_\mu \;;\;
\nonumber \\[2mm] A_\mu & = & \sin \theta_W\; W'^3_\mu + \cos \theta_W\; B'_\mu\;;\;
\nonumber
\end{eqnarray}
where the weak angle $\theta_W$, defining the physical electroweak neutral gauge bosons, gives also the relations between the electromagnetic and the weak couplings:  
$$g=\frac{e}{\sin \theta_W}\;;\;
g'=\frac{e}{\cos \theta_W}$$  
5) Read the (tree level) particle masses from the proper terms in ${\cal L}_{\rm EWSB}$:
\begin{eqnarray}
(D_{\mu}\Phi')^{\dagger}(D^{\mu}\Phi')&=&\left(\frac{g^2v^2}{4}\right)
W^+_{\mu}W^{\mu -}+\frac{1}{2}\left(\frac{(g^2+g'^2)v^2}{4}\right)
Z_{\mu}Z^{\mu}+... \nonumber \\
V(\Phi')&=&\mu^2H^2+...\nonumber \\
{\cal L}_{\rm YW}&=&
-\left(\lambda_e \frac{v}{\sqrt{2}}\right)\bar{e}'_L e'_R-
\left(\lambda_u \frac{v}{\sqrt{2}}\right)\bar{u}'_L u'_R-
\left(\lambda_d \frac{v}{\sqrt{2}}\right)\bar{d}'_L d'_R+ h.c.+...
\nonumber
\end{eqnarray}
And, from these expressions above, one finally gets the tree level particle masses:
  
\begin{eqnarray}
 M_W&=&\frac{gv}{2}\;;\;M_Z=\frac{\sqrt{g^2+g'^2}v}{2}\;;\; M_H=\sqrt{2}|\mu|\;;\;\nonumber \\
m_e&=&\lambda_e\frac{v}{\sqrt{2}}\;;\;
m_u=\lambda_u\frac{v}{\sqrt{2}}\;;\;
m_d=\lambda_d\frac{v}{\sqrt{2}}\;;...
\nonumber 
\end{eqnarray} 
And, by construction, the photon and the neutrinos are got massless within the SM.

The first immediate conclusion, after the building of the SM spectra on top of the non-symmetric vacuum, is that one finds three massive weak gauge bosons, $W^+$, $W^-$ and $Z$ and one physical scalar massive boson with positive parity, the $H$ particle. This $0^+$ particle is named the Higgs boson particle of the SM. 

Notice also that, as expected, the number of bosonic degrees of freedom is preserved:
\begin{center}
before SSB =12 (4x2gauge+4scalar); after SSB = 12 (3x3gauge+1x2gauge+1scalar).
\end{center}
The second conclusions from the above expressions is that all the SM particle masses, as predicted from the Higgs Mechanism, are given in terms of the parameter $v$ with energy dimension, and whose relation with the input parameters $\mu$ and $\lambda$ in the potential is given by:
$$v=\sqrt{\frac{-\mu^2}{\lambda}}\,\,\,\,\,,\,\,\,\mbox{with} \,\,\,\mu^2<0\,\,\,\,\,,\,\,\,\mbox{and}\,\,\,\lambda>0.$$
Notice that both $\mu$ and $\lambda$ are unknown parameters of the model. Therefore, the predicted tree level Higgs boson mass above, $M_H=\sqrt{2}|\mu|$ and the Higgs self interactions given by $\lambda$ are unknown quantities within the SM.
In contrast, it is worth recalling that the value of the parameter $v$, i.e. the vacuum expectation value of the $\Phi$ field, was known from the experiments long time ago,
indeed before the discovery of $W^{\pm}$ and $Z$.
It was obtained from physical observables, well known from experiment.
For instance, for the muon decay width $\Gamma(\mu^- \to \nu_\mu {\bar \nu}_e e^-)$ it was known the prediction from the V-A Theory (Feynman, Gell-Mann 1958) in terms of the Fermi constant $G_F$ and the muon mass $m_\mu$ given by:
$$ \frac{1}{\tau_\mu}= \Gamma(\mu^- \to \nu_\mu {\bar \nu}_e e^-) \simeq
\frac{G_F^2 m_\mu^5}{192 \pi^3}$$
which provides a rather good prediction for the muon life time:
$$ \tau_\mu =2.2 \times 10^{-6} s \,\,\,\,\,,\,\,\,\mbox{for} \,\,\,G_F= 1.167\times 10^{-5}  \gev^{-2}\,\,\,\,\,,\,\,\,\mbox{and} \,\,\,m_\mu=0.10566 \gev.$$
On the other hand, within the SM, the muon decay proceeds via an intermediate virtual $W$ exchange:
Therefore, by matching the above $\Gamma$ to the prediction in the SM one gets:
\BC 
$\frac{G_F}{\sqrt 2}=\frac{g^2}{8M_W^2}=\frac{1}{2v^2}\,\,\,\RA$  $v= 246 \gev$
\EC
Finally, by using this $v$, the experimental value for $\sin^2\theta_W \simeq 0.23$ from {\it e.g.} DIS data and 
$g=e/\sin \theta_W$, with $e$ set by the fine structure constant, $\alpha=e^2/(4\pi)$, one gets the tree level mass values:
\BC 
$\RA M_W^{\rm tree} \simeq 78 \gev\,,\,M_Z^{\rm tree} \simeq 89 \gev$.
\EC
Thus, these values were known much before the weak bosons $W^{\pm}$ and $Z$ were discovered at CERN in 1983 and, indeed, they were pretty close to the experimental measured values!!. 

At present there are much more precise predictions of $M_W$ and $M_Z$, beyond tree level, i.e., with radiative corrections included, and there are also much more precise measurements of these masses, and the agreement between theory and experiment is excellent. These predictions of the gauge boson masses are probably one of the greatest successes of the SM.

In contrast to the gauge boson masses, the values of the fermion masses were not predicted from the Yukawa couplings, since these later were not extracted previously from other physical observables. In fact, in the fermion sector the predictions were rather the other way around: namely, the Yukawa couplings were extracted from the experimental measurements of the fermion masses.
For instance, for $m_e\simeq 0.5 \times 10^{-3} \gev$ one gets $\lambda_e \simeq 3 \times 10^{-6}$, for 
$m_t \simeq 173 \gev$ one gets $\lambda_t\simeq 1$, and similarly for the other fermions. One of the greatest mysteries nowadays in particle physics, not explained by the SM, is to understand the origin of such a widely spread fermion masses, or equivalently, the origin of the large hierarchy among the various fermion Yukawa couplings, which vary from extremely small values for leptons, more specifically those of the first generation and obviously all the tiny neutrino Yukawa couplings, to quite sizable values in the quark sector, specially that of the top quark being close to one!!!. 
Whatever explains this must be beyond SM physics.

\section{SM Higgs boson couplings}
In order to get the SM Higgs boson couplings to gauge bosons and to fermions one has to work out the interaction Lagrangian terms from the previous ${\cal L}_{\rm SBS}$ and ${\cal L}_{\rm YW}$ and express them in terms of the physical basis. Instead of writing the final interaction Lagrangian, we prefer here to express the Higgs interaction terms by the corresponding Feynman rules and the corresponding Higgs boson couplings. These are collected in the following drawings.

The most remarkable feature of the Higgs couplings is that they grow with the mass of the particle that is coupled to. Thus, the Higgs coupling to fermions $f$ are larger for larger $m_f$, the Higgs couplings 
to the gauge bosons $W$ and $Z$ go respectively with $M_W$ and $M_Z$, and the Higgs self-couplings, both the triple and the quartic, are more intense for heavier $M_H$.  
A very simple exercise is to use the present experimentally measured value at LHC of $M_H$ to get an estimate of the value of this self-coupling. Thus, by assuming $M_H \simeq 125 \gev$ and using the SM tree level relation: 
$$ \lambda=\frac{g^2M_H^2}{8M_W^2},$$
one gets, 
$\lambda \simeq 0.12$, which is indeed a small coupling if we compare it with either the electromagnetic coupling, $e \simeq 0.3$ and the weak coupling, $g \simeq 0.63$.

\begin{minipage}[h]{7cm}
\begin{center}
\vspace{2cm}
\begin{picture}(550, 80)
\DashLine(120,100)(120,40){5}
\Photon(120,40)(180,0){3}{4.5}
\Photon(60,0)(120,40){3}{4.5}
\put(185,-10){$W_\nu^-$}
\put(20,-10){$W_\mu^+$}
\put(100,95){$H$}
\put(150,40){$igM_Wg_{\mu\nu}$}
\end{picture}
\end{center}


\begin{center}
\begin{picture}(550, 80)
\DashLine(120,100)(120,40){5}
\Photon(120,40)(180,0){3}{4.5}
\Photon(60,0)(120,40){3}{4.5}
\put(185,-10){$Z_\nu$}
\put(35,-10){$Z_\mu$}
\put(100,95){$H$}
\put(150,40){$i\dfrac{g}{c_W}M_Zg_{\mu\nu}$}
\end{picture}
\end{center}

\bigskip

\begin{picture}(550, 80)
\DashLine(120,100)(120,40){5}
\ArrowLine(120,40)(180,0)
\ArrowLine(60,0)(120,40)
\put(185,-10){$f$}
\put(40,-10){$f$}
\put(100,95){$H$}
\put(140,40){$-i\dfrac{g}{2}\dfrac{m_f}{M_W}=-i\dfrac{\lambda_f}{\sqrt{2}}$}
\end{picture}

\bigskip

\begin{picture}(550, 80)
\DashLine(120,100)(120,40){5}
\DashLine(120,40)(180,0){5}
\DashLine(60,0)(120,40){5}
\put(185,-10){$H$}
\put(37,-10){$H$}
\put(100,95){$H$}
\put(140,40){$-ig\dfrac{3}{2}\dfrac{M_H^2}{M_W}=-i6\lambda v$}
\end{picture}

\bigskip
\end{minipage}
\begin{minipage}[h]{15cm}

\bigskip
\begin{picture}(550, 80)
\DashLine(60,80)(120,40){5}
\DashLine(120,40)(60,0){5}
\Photon(120,40)(180,80){3}{4.5}
\Photon(180,0)(120,40){3}{4.5}
\put(185,-10){$W_\mu^+$}
\put(185,75){$W_\nu^-$}
\put(40,-10){$H$}
\put(40,75){$H$}
\put(180,40){$i\dfrac{g^2}{2}g_{\mu\nu}$}
\end{picture}

\bigskip
\bigskip

\begin{picture}(550, 80)
\DashLine(60,80)(120,40){5}
\DashLine(120,40)(60,0){5}
\Photon(120,40)(180,80){3}{4.5}
\Photon(180,0)(120,40){3}{4.5}
\put(185,-10){$Z_\mu$}
\put(185,75){$Z_\nu$}
\put(40,-10){$H$}
\put(40,75){$H$}
\put(180,40){$i\dfrac{g^2}{2c_W^2}g_{\mu\nu}$}
\end{picture}

\bigskip
\bigskip

\begin{picture}(550, 80)
\DashLine(60,80)(120,40){5}
\DashLine(120,40)(60,0){5}
\DashLine(120,40)(180,80){5}
\DashLine(180,0)(120,40){5}
\put(185,-10){$H$}
\put(185,75){$H$}
\put(40,-10){$H$}
\put(40,75){$H$}
\put(150,40){$-ig^2\dfrac{3}{4}\dfrac{M_H^2}{M_W^2}=-i6\lambda$}
\end{picture}

\end{minipage}\\[5em]

Regarding the comparison of the SM predictions for the Higgs boson couplings to fermions and bosons with the experimental data from LHC, there seems to be a good agreement up to now, although the statistical significance of this agreement is still not very high. In the next plot we see that when comparing the value of the measured Higgs couplings to a particle $P$, $\lambda_P$,  versus the corresponding mass of the particle  $P$ that the Higgs is coupled to, $m_P$, one also finds a good agreement (black dashed line is best fit to data and the dotted black lines are 68\% CL ranges) with the SM prediction (red solid line). This is clearly signaling that the Higgs particle that has been discovered at LHC has couplings to the fundamental fermions and gauge bosons that are proportional to their mass, as in the SM.\\
\begin{center}
\includegraphics[width=10cm]{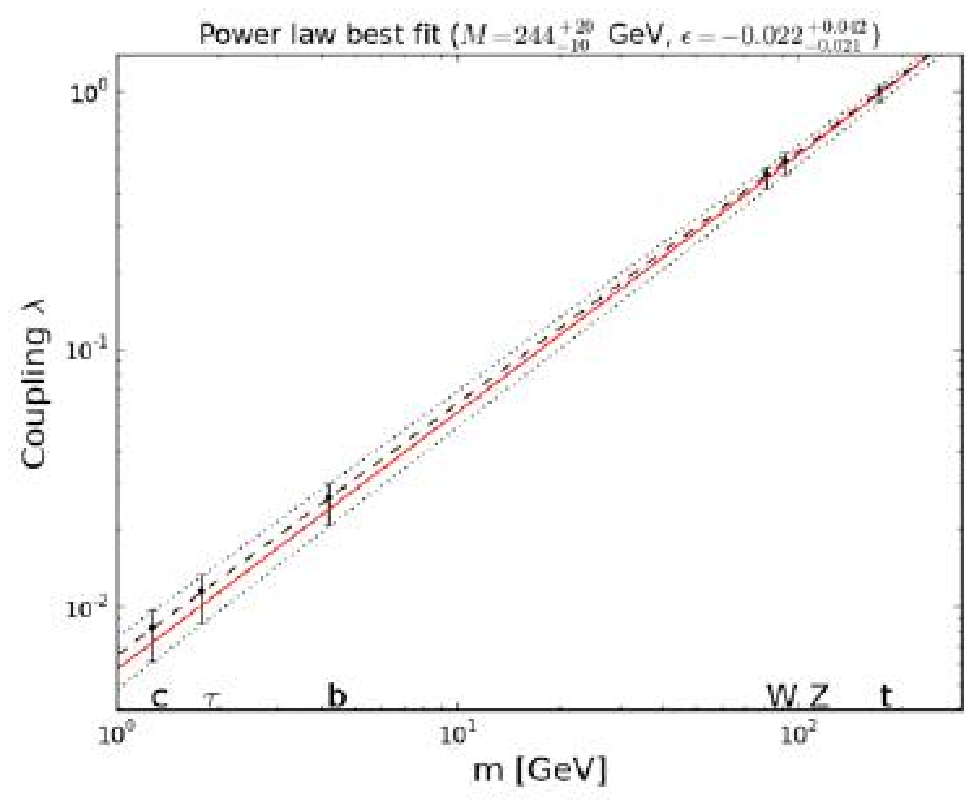}\\
(Plot taken from \cite{Ellis:2013lra})
\end{center}

\section{SM Higgs boson decays and production at LHC}

\vspace{1cm}
\blackl{1.) Higgs decay into fermions} 

\vspace{1cm}
The Higgs boson coupling to fermion $f$ can be written in terms of the Fermi  constant $G_F$ as:
$$
{g_{f\bar f H} = \KKL \wz\,G_F \KKR^{1/2} m_f}
$$
The Higgs boson partial decay width to fermions can then be expressed as:
$$
{\Ga(H \to f \bar f)} = 
  K \frac{G_F\,{\MH}}{4\wz\,\pi} {m_f^2(\MH^2)} \,
  \KL 1 - 4 \frac{\mf^2}{\MH^2} \KR^{3/2} 
$$
where $K=N_c=$ number of colors if $f$ is a quark, or $K=1$ if $f$ is a charged lepton.

Notice that the previous expression has the same functional form as the tree level partial decay width, except that the dominant radiative corrections are included into the value of the running fermion mass. Thus, for instance, the bulk of {QCD corrections} for decays to quarks are mapped into
$$
m_q^2(\mbox{pole}) \to {m_q^2(\MH^2)}
$$
A simple numerical estimate of the previous partial decay width shows that the dominant decay process is : $H \to b \bar b$.

%


\vspace{1cm}
\blackl{2.) Decay to heavy gauge bosons ($V = W, Z$)}

\vspace{1cm}
The relevant Higgs boson coupling here is the coupling to a couple of gauge bosons $VV$, with 
$V=W$ or $V=Z$, that can be written as:
$$
{g_{VVH} = 2\,\KKL \wz\, G_F \KKR^{1/2} M_V^2}
$$

The on-shell decay width ($\MH > 2 M_V$) at the tree level can be easily computed and gives:
$$
{\Ga(H \to VV)} = \de_V \frac{G_F {\MH^3}}
                                  {16\,\wz\,\pi}
   \KL 1 - 4 \frac{M_V^2}{\MH^2} + 12 \frac{M_V^4}{\MH^4} \KR \;
   \KL 1 - 4 \frac{M_V^2}{\MH^2} \KR^{1/2} 
$$
with $\de_{W,Z} = 2, 1$

 

\vspace{1cm} 
\blackl{3.) Higgs boson decay to massless gauge bosons ($gg$, $\ga\ga$)}

\vspace{1cm}
The Higgs decays to two gluons and two two photons do not exit at the tree level, and the first non vanishing contributions within the SM appear at the one loop level. These are very relevant decays precisely because of this fact, and any deviation from these SM predictions coming from new physics will be noticed most probably in these kind of decay channels.

The Higgs decay width into two gluons can be estimated with the dominant one-loop diagram where the top quark is propagated through the triangle, as in the figure below:

\vspace{1cm}
\setlength{\unitlength}{1pt}
\begin{picture}(500, 100)
\Vertex(200,90){3}
\ArrowLine(200,90)(200,10)
\put(215,45){{$t$}}
\Vertex(200,10){3}
\ArrowLine(200,10)(140,50)
\put(145,10){{$t$}}
\Vertex(140,50){3}
\ArrowLine(140,50)(200,90)
\put(145,77){{$t$}}
\put(285,86){{$g$}}
\Gluon(270,90)(200,90){5}{5}
\put(285,6){{$g$}}
\Gluon(270,10)(200,10){5}{5}
\DashLine(140,50)(70,50){5}
\put(45,45){{$H$}}
\end{picture}

\vspace{1cm}
Besides, to account for the huge higher order QCD radiative corrections one has to correct the one-loop result by using the running strong coupling constant and some extra terms summarized into the one with $C$ below. 
$$
{\Ga(H \to gg)} \; = \; 
  \frac{\gf\,\als^2(\MH^2) \, {\MH^3}}
       {36 \, \wz \, \pi^3}
  \KKL 1 + {C} \frac{\als(\mu)}{\pi} \KKR
$$
$$
{C} = \frac{215}{12} - \frac{23}{6} \log \KL \frac{\mu^2}{\MH^2} \KR
          + \cO(\als)
$$

In the case of Higgs boson decays to a couple of photons, there are two relevant one-loop diagrams, the triangular one with top quarks and the triangular one with $W$ gauge bosons propagating through the triangle, as given below:
  
\vspace{1cm}
\setlength{\unitlength}{1pt}
\mbox{}\hspace{-3cm}
\begin{picture}(500, 100)
\Vertex(200,90){3}
\ArrowLine(200,90)(200,10)
\put(215,45){{$t$}}
\Vertex(200,10){3}
\ArrowLine(200,10)(140,50)
\put(145,10){{$t$}}
\Vertex(140,50){3}
\ArrowLine(140,50)(200,90)
\put(145,77){{$t$}}
\put(285,86){{$\ga$}}
\Photon(270,90)(200,90){5}{5}
\put(285,6){{$\ga$}}
\Photon(270,10)(200,10){5}{5}
\DashLine(140,50)(70,50){5}
\put(45,45){{$H$}}
\mbox{}\hspace{-2cm}
\Vertex(500,90){3}
\Photon(500,90)(500,10){5}{5}
\put(515,45){{$W$}}
\Vertex(500,10){3}
\Photon(500,10)(440,50){5}{5}
\put(450,10){{$W$}}
\Vertex(440,50){3}
\Photon(440,50)(500,90){5}{5}
\put(450,77){{$W$}}
\put(585,86){{$\ga$}}
\Photon(570,90)(500,90){5}{5}
\put(585,6){{$\ga$}}
\Photon(570,10)(500,10){5}{5}
\DashLine(440,50)(370,50){5}
\put(345,45){{$H$}}
\end{picture}

\vspace{1cm} 
In this case, the computation of these two diagrams is sufficient to approximately describe this decay. The resulting partial decay width can be written as:

$$
{\Ga(H \to \ga\ga)} \; = \; 
  \frac{G_F \, \al^2 \, {\MH^3}}
       {128 \, \wz \, \pi^3} 
  \Big| \frac{4}{3} e_t^2 - 7 \Big|^2
$$
where the first term is from the top quark loop and the second one from the $W$~boson loop, and 
$e_t$ is the top quark electric charge.


\vspace{1cm}
\blackl{Summary of the branching ratios for the SM Higgs boson decays}\\[1em]
The SM predictions for the branching ratios for all the Higgs boson decays as a function of the Higgs mass are collected in the plot below. The total uncertainties in these predictions are also included. 
\BC
\includegraphics[height=8cm]{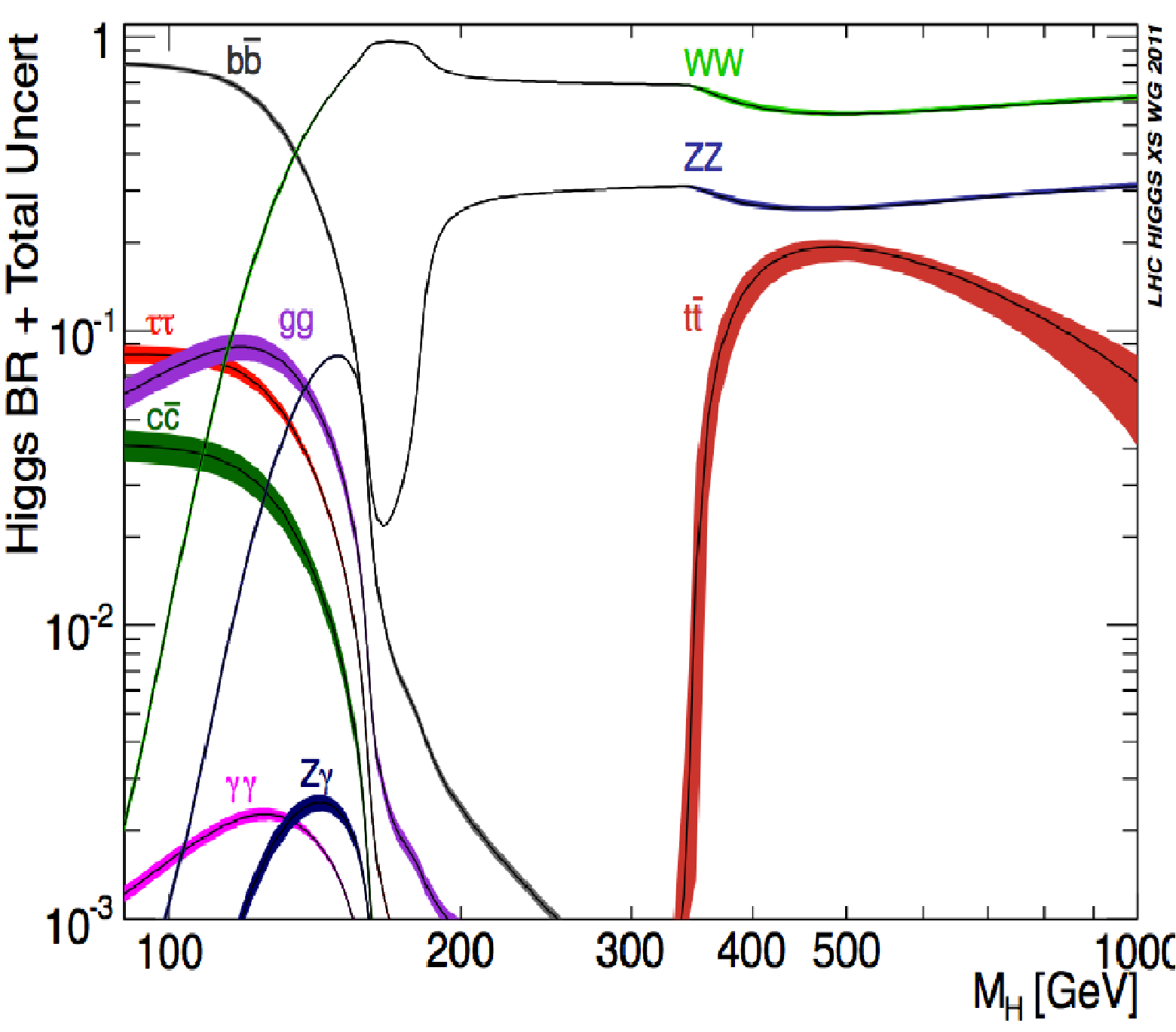}
\EC
At low values of $M_H$ the decay to $b{\bar b}$ dominates and for high $M_H$ the decays to weak gauge bosons dominate. In the region of most interest, namely with $M_H$ close to $126 \gev$, most of the channels enter into play and, in particular, the $H \to \ga \ga$ channel, even though it has a small branching ratio of about $2 \times 10 ^{-3}$, is indeed one of the most relevant channels at LHC due to the experimental feasibility to detect photons. The other most relevant channels proceed via $ZZ$ and these are the decays to four fermions: $H \to ZZ \to f_1 {\bar f}_1 f_2 {\bar f}_2$. The cases where these fermions are either a muon or an electron, i.e, $H \to \mu^+ \mu^- \mu^+ \mu^-$, $H \to e^+ e^- e^+ e^-$, and $H \to \mu^+ \mu^- e^+e^-$ (the so-called golden-channels) have received much attention in the recent years since these leptons are well measured at LHC. But overall, noways all the channels are being studied at LHC and ATLAS and CMS provide indeed measurements for most of these branching ratios and the couplings involved. 

\vspace{1cm}
\blackl{ The total width of the SM Higgs boson}\\

The prediction of total Higgs width as a function of the Higgs mass within the SM is shown in the plot below.  
\BC
\includegraphics[height=10cm]{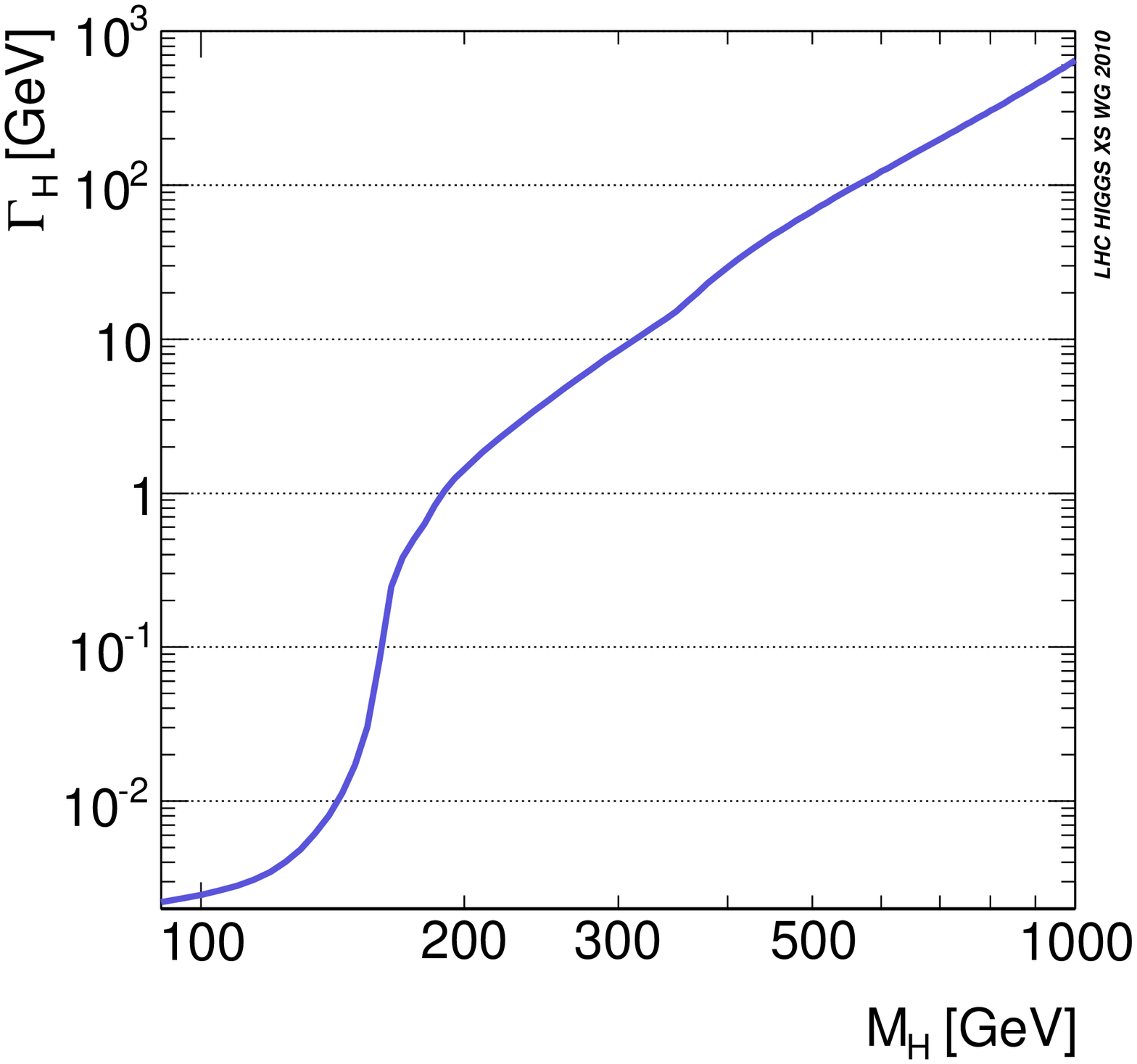}
\EC

The total width grows with $M_H$, and reaches large values at large $M_H$. In fact, $\Gamma_H$ gets comparable with the mass itself in the extreme case of $M_H$ close to $1000 \gev$. However, for $M_H$ close to $126 \gev$  the total width is very narrow, below $10^{-2} \gev$, and therefore it is very difficult to measure. At present there is not an experimental measurement of the total Higgs width.

\newpage

\vspace{0.5cm} 
\blackl{Higgs boson production at the LHC}

\vspace{0.5cm}
The most important SM Higgs production channels at the LHC are:\\[1em]
1) gluon fusion: $gg \to H$\\[.5em]
2) weak boson fusion (WBF): $q \bar q \to q' \bar q' H$\\[.5em]
3) $W$ boson associated production:
$q \bar q' \to WH$\\[0.5em]
4) $Z$ boson associated production:
$q \bar q \to ZH$\\[0.5em]
5) top quark associated production:
$gg, q \bar q \to t \bar t H$.

The relevant Feynman diagrams for the two first channels are shown below: 
\begin{center}

\mbox{}\hspace{2cm}
\setlength{\unitlength}{1pt}
\begin{picture}(400, 140)
\put(-80,120){{Gluon-Fusion:}}
\Vertex(000,90){3}
\ArrowLine(000,90)(000,10)
\put(-15,45){{$t$}}
\Vertex(000,10){3}
\ArrowLine(000,10)(060,50)
\put(030,12){{$t$}}
\Vertex(060,50){3}
\ArrowLine(060,50)(000,90)
\put(030,77){{$t$}}
\put(-85,86){{$g$}}
\Gluon(-70,90)(000,90){5}{5}
\put(-85,6){{$g$}}
\Gluon(-70,10)(000,10){5}{5}
\DashLine(060,50)(130,50){5}
\put(135,45){{$H$}}
\hspace{-1cm}
\put(200,120){{WBF:}}
\ArrowLine(310,05)(230,05)
\ArrowLine(390,-25)(310,05)
\ArrowLine(230,095)(310,095)
\ArrowLine(310,095)(390,125)
\Photon(310,05)(340,50){3}{4.5}
\Photon(310,095)(340,50){3}{4.5}
\DashLine(340,50)(410,50){5}
\Vertex(310,05){3}
\Vertex(310,095){3}
\Vertex(340,50){3}
\put(205,00){{$q$}}
\put(205,090){{$q$}}
\put(395,-30){{$q'$}}
\put(395,120){{$q'$}}
\put(335,15){{$W$}}
\put(335,070){{$W$}}
\put(415,45){{$H$}}
\end{picture}
\end{center}

\vspace{1cm}
The predictions within the SM  for the Higgs boson cross sections at LHC with $\sqrt{s}= 8$ TeV in the various channels as a function of the Higgs boson mass are collected in the figure below:
 
\BC 
\includegraphics[width=9cm]{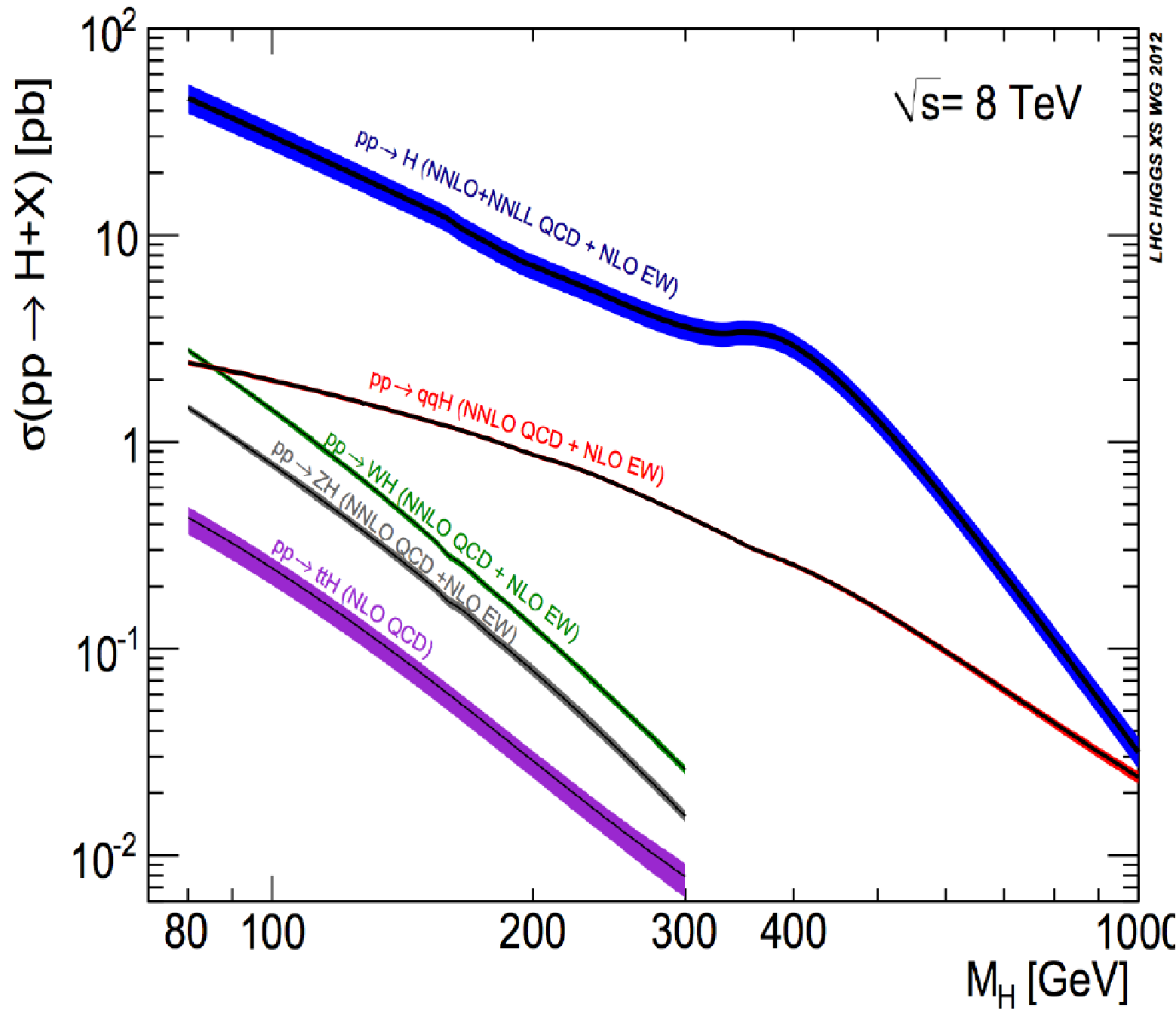}
\EC 
 
\section{Other interesting properties of the SM Higgs system}
There are several properties of the SM Higgs system that are worth to keep in mind, since they may give us some clue in the future studies of the fundamental physics underling the EW symmetry breaking sector. We select here three: 1) the role of the Higgs boson in the scattering of longitudinal weak gauge bosons, 2) the Equivalence Theorem, and 3) the theoretical limits on the value on the Higgs mass.
We will comment shortly on these next. \\[0.7em]
 \blackl{Higgs boson role in scattering of longitudinal $W$ and $Z$~bosons:}
\\[0.8em]
It is interesting to recall that when one computes the scattering amplitude for longitudinal gauge bosons, $W_L$ and/or $Z_L$ without including the diagrams with a Higgs boson, one gets a resulting amplitude that does not preserve unitarity at high energies. For instance, computing the three diagrams below for $W_L W_L \to W_L W_L$ scattering:
\BC
\setlength{\unitlength}{1pt}
\begin{picture}(530, 80):
%
\Photon(0,80)(60,40){3}{4.5}
\Photon(60,40)(0,0){3}{4.5}
\Vertex(60,40){3}
\Photon(60,40)(120,40){3}{4.5}
\Vertex(120,40){3}
\Photon(120,40)(180,80){3}{4.5}
\Photon(180,0)(120,40){3}{4.5}
\put(70,20){{$\ga$, $Z$}}
\put(200,40){$+$}
\Photon(240,80)(300,60){3}{4.5}
\Photon(300,60)(300,20){3}{4.5}
\Photon(300,20)(240,0){3}{4.5}
\Vertex(300,60){3}
\Vertex(300,20){3}
\Photon(300,60)(360,80){3}{4.5}
\Photon(300,20)(360,0){3}{4.5}
\put(310,40){{$\ga$, $Z$}}
\put(370,40){$+$}
\end{picture}

\begin{picture}(530, 80):
\Photon(300,80)(360,40){3}{4.5}
\Photon(300,0)(360,40){3}{4.5}
\Vertex(360,40){3}
\Photon(360,40)(420,80){3}{4.5}
\Photon(360,40)(420,0){3}{4.5}
\end{picture}
\EC
one gets, for large $E$, the following behavior:
$$ T_V= {- g^2 \frac{E^2}{\MW^2} + ...}$$ 
which implies a {\it violation of unitarity} at high energies.\\[0.5em]
However, if one considers, in addition, the contribution of a scalar particle $H$ with 
couplings proportional to the mass, given by the two extra diagrams below:
\BC
\setlength{\unitlength}{1pt}
\begin{picture}(550, 80)
\Photon(0,80)(60,40){3}{4.5}
\Photon(60,40)(0,0){3}{4.5}
\Vertex(60,40){3}
\DashLine(60,40)(120,40){5}
\Vertex(120,40){3}
\Photon(120,40)(180,80){3}{4.5}
\Photon(180,0)(120,40){3}{4.5}
\put(80,20){{$H$}}
\put(210,40){$+$}
\Photon(240,80)(300,60){3}{4.5}
\DashLine(300,60)(300,20){5}
\Photon(300,20)(240,0){3}{4.5}
\Vertex(300,60){3}
\Vertex(300,20){3}
\Photon(300,60)(360,80){3}{4.5}
\Photon(300,20)(360,0){3}{4.5}
\put(310,35){{$H$}}
\end{picture}
\EC
whose behavior at high energies is given by:
$$ T_S = {g^2_{WWH} \frac{E^2}{\MW^4} +...}$$
%
Then the final result for the total amplitude is: 
$$
{T_{\mathrm{tot}}} = {T_V } + {T_S }
= \frac{E^2}{\MW^4} \KL {g_{WWH}^2} - {g^2 \, \MW^2} \KR + \ldots
$$
 and the potential terms with bad high-energy behavior happens to cancel for
$$
{g_{WWH}} {=} {g \, \MW}
$$
which is exactly the value of the SM Higgs coupling to the $W$ gauge bosons. In conclusion, the SM Higgs particle repairs the bad high energy behavior of the longitudinal weak gauge bosons and the resulting total amplitudes are unitary at all energies. However, it is also convenient to keep in mind that this is not a unique solution and the SM Higgs system could be replaced by something else (new Higgs particles, new gauge bosons, new resonances, etc) which could effectively play this same role.

 
\vspace{1cm}
\blackl{Comparing the WW scattering with the would-be-GB scattering}
\\[0.5em]
An interesting result is provided by the so-called Equivalence Theorem that relates the scattering of massive gauge bosons and the scattering of GBs\\[1em]
{\bf Equivalence Theorem} (Cornwall {\it et al} 1974, Lee {\it et al} 1977):\\[1em]
{\it The scattering amplitudes of longitudinal gauge bosons} $V_L$
($V=W^{\pm},Z$), {\it at high energies}, $\sqrt{s}>>M_V$, {\it are
equivalent
to the scattering amplitudes of their corresponding would-be Goldstone
bosons} $w$
\begin{equation}
|T(V_L^1V_L^2...V_L^N \rightarrow V_L^1V_L^2...V_L^{N'})|\approx
|T(w_1w_2...w_N\rightarrow w_1w_2...w_{N'})|
\non
\end{equation} 
For instance, instead of using the unitary gauge one can use the more general Feynman rules of $R_\xi$ gauges and get the following relations: 1) \\[-1.2em] \begin{equation}
T(W_L^+W_L^-\rightarrow W_L^+W_L^-)=
T(w^+w^-\rightarrow w^+w^-)+O(\frac{M^2}{s}),\,\ \mbox{for}\, \sqrt{s}>>M_W,M_Z
\non
\end{equation}
and, 2) for $M_H>>M_{W,Z}$: 
\begin{eqnarray}
\Gamma(H\rightarrow W_L^+W_L^-)&=&\Gamma(H\rightarrow
w^+w^-)+O(\frac{M_W}{M_H})\nonumber \\
\Gamma(H\rightarrow Z_LZ_L)&=&\Gamma(H\rightarrow
zz)+O(\frac{M_Z}{M_H})\non
\end{eqnarray} 
These results above and some others provided by the Equivalence Theorem, apart of being of practical use, may give us also some clue in the future to further understand the fundamental dynamics underlying the EW symmetry breaking sector. \\[1em] 

\blackl{Theoretical limits on the Higgs mass}\\[0.3em] 
Next we present the three most popular Higgs mass limits from theory: I) The upper Higgs mass
limit from unitarity, II) The upper Higgs mass limit from triviality, and III) The Lower Higgs mass limit from vacuum stability. \\[0.3em] 

\blackl{I: Upper Higgs mass bound from unitarity}\\[0.3em] 
 
Let us study here the behavior of the scattering amplitude of longitudinal gauge bosons with respect to the value of $M_H$. The complete tree level result is given by:
\begin{eqnarray}
T(W_L^+W_L^-\rightarrow W_L^+W_L^-)&=&
-\frac{1}{v^2}\{-s-t+\frac{s^2}{s-M_H^2}+\frac{t^2}{t-M_H^2}
      +2M_Z^2+
\nonumber \\ & &
      \frac{2M_Z^2s}{t-M_Z^2}+
\frac{2t}{s}(M_Z^2-4M_W^2)-\frac{8s_W^2M_W^2M_Z^2s}{t(t-M_Z^2)}\}
\non 
\end{eqnarray}

Next we decompose $T$ in partial waves $a_J$ defined by:

\begin{equation}
T(s,\cos\theta)=16\pi\sum_{J=0}^{\infty}(2J+1)a_J(s)P_J(\cos\theta)
\,\,\,,\,\,P_J=\mbox{Legendre\,\,\,polynomials}
\non
\end{equation}

One can then compute the cross-section in terms of these partial waves:
\begin{equation}
\sigma_{\rm tot}\simeq \sigma_{\rm el}=\frac{16\pi}{s}\sum_{J=0}^{\infty}(2J+1)|a_J(s)|^2
\non
\end{equation}
On the other hand if we require $\sigma$ to fulfill the Optical Theorem (this OT is a consequence of unitarity $T^\dagger T=TT^\dagger =1$):
\begin{equation}
\sigma_{\rm tot}(1+2\rightarrow {\rm anything})=\frac{1}{s}
 {\rm Im}\;T(s,\cos\theta=1)
 \non
\end{equation}\\[-1em]
when applied to $\sigma_{\rm el}$, the OT can then be written in terms of the partial waves as:
\begin{equation}
|a_J(s)|^2={\rm Im}\;a_J(s)\;;\;\forall J \,\,\,\RA \,\,\,
|a_J|^2\leq 1\;;\;0\leq {\rm Im}\;a_J\leq 1\;;\;|{\rm Re}\; a_J|\leq \frac{1}{2}
\;;\; \forall J 
\non
\end{equation}
For instance, when applied to the lowest partial wave, defined by:
\begin{equation}
a_0(W_L^+W_L^-\rightarrow W_L^+W_L^-)=\frac{1}{32\pi}\int_{-1}^1
T(s,\cos\theta)d(\cos\theta) 
\non
\end{equation}
one finds an expression for $|Re\;a_0|$ valid in the high energy limit, $\sqrt{s}>>M_H,M_W$:
\begin{equation}
|a_0| \stackrel{s>>M_H^2,M_V^2}{\longrightarrow} \frac{M_H^2}{8\pi v^2}
\non
\end{equation}
Therefore, the corresponding unitary bound for $a_0$ leads to an upper bound for $M_H$, which in this particular case is:
$$|{\rm Re}\; a_0| \leq \frac{1}{2} \RA M_H < 860 \gev$$
This unitary bound can be improved if higher order corrections beyond tree level are included, and also by considering other possible channels. But the size of the final upper bound remains close to this. \\[1em]

\blackl{Upper Higgs mass bound from triviality}\\[1em]
Let us first consider the running of the Higgs self-coupling at the one-loop level, whose dominant contributions are given by the three diagrams below:
\begin{center}
\setlength{\unitlength}{1.5pt}
\begin{picture}(130,110) 
\DashLine(0,100)(50,50){5}
\DashLine(0,0)(50,50){5}
\DashLine(100,100)(50,50){5}
\DashLine(100,0)(50,50){5}
\put(-15,70){{$H$}}
\put(-15,-5){{$H$}}
\put(70,-5){{$H$}}
\put(70,70){{$H$}}
\put(30,45){{$\la$}}
\end{picture}
\begin{picture}(150,90)
\DashLine(0,100)(50,50){5}
\DashLine(0,0)(50,50){5}
\DashLine(150,100)(100,50){5}
\DashLine(150,0)(100,50){5}
\DashCArc(75,50)(25,0,360){3}
\put(-15,70){{$H$}}
\put(-15,-5){{$H$}}
\put(45,60){{$H$}}
\put(100,-5){{$H$}}
\put(100,70){{$H$}}
\end{picture}
\begin{picture}(110,110)
\DashLine(0,0)(50,50){5}
\DashLine(0,150)(50,100){5}
\DashLine(100,100)(150,150){5}
\DashLine(100,50)(150,0){5}
\ArrowLine(50,50)(100,50)
\ArrowLine(100,50)(100,100)
\ArrowLine(100,100)(50,100)
\ArrowLine(50,100)(50,50)
\put(-15,90){{$H$}}
\put(-15,-5){{$H$}}
\put(45,75){{$t$}}
\put(100,-5){{$H$}}
\put(100,90){{$H$}}
\end{picture} 
\EC
\setlength{\unitlength}{1pt}
  The relevant renormalization group equation (RGE) for the self-coupling $\lambda$ is:
$$
\frac{d\,{\la}}{d\,{t}} \; = \; \frac{3}{16\,\pi^2}
\KKL 4{\la^2} + 2{\la}g_t^2 - g_t^4 + 
     \frac{1}{16} \KL 2 g_2^4 + (g_2^2 + g_1^2)^2 \KR \KKR~, \quad
{t} = \log\KL\frac{Q^2}{v^2}\KR
$$
Notice that we use here a different notation than before: the top Yukawa coupling is $g_t$, and the SM gauge couplings are $g_1$, $g_2$ and $g_3$ respectively. 
 
 The so-called 'Triviality Problem' arises when $\lambda$ is large and it is related to the existence of a pole, named the Landau pole, in the solution to the previous RGE. For large $\lambda$, one can neglect the contributions from  $g_t$, $g_1$, $g_2$ and $g_3$ in the RGE and keep just the dominant contribution from $\lambda$, leading to a simple solution for the running coupling constant ${\la}({Q})$ in terms of the bare coupling constant $\lambda_0$:
\BEA
\frac{d\,{\la}}{d\,{t}} \; &=& \; \frac{3}{4\,\pi^2}
\KKL {\la^2} \KKR \non \\
\RA \qquad {\la}({Q}) \;& = &\; 
    \frac{\la_0}
         {1 - \frac{3 \la_0 }{2\,\pi^2} 
              \log\KL\frac{{Q}}{\La}\KR} \;\;\;;\;\;\lambda_0 \equiv \lambda(\La)	      
	      \non
\EEA 
where the presence of the Landau pole is manifest.
Now, by taking the $\La \to\infty$ limit, while fixing $\lambda_0$ to a finite value, one finds that the effective coupling $\la(Q)\to 0$  and in consequence the theory is trivial, i.e. non-interacting. The only way out from this is to assume the existence of a finite physical cut-off $\La_{\rm phys}$ such that $\la(Q) \neq 0$ all the way up to this cut-off. Then, by defining the renormalized Higgs mass in terms of $\lambda(v)$ as: 
\begin{equation}
M_H^2 = 2 \lambda(v)v^2 \,\,\,\,\mbox{with}\,\,\,\, 
 \lambda(v)=\frac{\lambda_0}{1-\frac{3}{2\pi^2}\lambda_0
\log(\frac{v}{\Lambda_{\rm phys}})},
\non
\end{equation} 
one finds that for decreasing (increasing) $\La_{\rm phys}$ $\RA$   $M_H$ increases (decreases) and indeed they may cross. This crossing point where $M_H(\La_{\rm phys})\simeq \La_{\rm phys}$ is what gives the upper bound to $M_H$. Clearly, this is a cut-off dependent bound.\\[1em]

\blackl{Lower Higgs mass bound from vacuum stability}\\[1em]
The problem of vacuum instability arises for small or negative {$\la$}. It can be understood either from the behavior of the effective potential or from the behaviour of the solution to the RGE for $\lambda$.

In few words, the behavior of the effective potential is as follows. The minimum of the effective potential (including loop corrections) changes with $\la(Q)$ and, a too small or negative $\la(Q)$ may change the true vacuum:  from wanted stable vacuum $V(v) < V(0)$ to the unwanted unstable vacuum with $V(v) > V(0)$ $\RA$ in which case the electroweak symmetry breaking does not take place. Indeed the situation can be even worse, since it can lead to an effective potential that is not even bounded from below!!. 
In summary, by requiring vacuum stability, namely by imposing $V(v) < V(0)$,  one then gets a lower bound on $\la(v)$ and in consequence also on
$M_H$. This lower bound on $M_H$ is also cut-off dependent.

One can alternatively solve the one loop RGE in the small $\lambda$ regime by, for instance, keeping just the dominant terms:
\BEA
\frac{d\,{\la}}{d\,{t}} \; &=& \; \frac{3}{16\,\pi^2}
\KKL - g_t^4 + \frac{1}{16} \KL 2 g_2^4 + (g_2^2 + g_1^2)^2 \KR \KKR \non \\
\RA \qquad {\la}({Q^2}) \; &=& \; {\la}(v^2)+
  \frac{3}{16\,\pi^2}
\KKL - g_t^4 + \frac{1}{16} \KL 2 g_2^4 + (g_2^2 + g_1^2)^2 \KR \KKR
  \log\KL\frac{{Q^2}}{v^2}\KR \non
\EEA
and then require a positive $\lambda(\Lambda)$. Thus, one gets a lower limit on $M_H$ that again depends on $\Lambda$:
$$
 {\la(\La) > 0}\; \RA \;
{\MH^2 > \frac{3v^2}{8\,\pi^2}
\KKL  g_t^4 - \frac{1}{16} \KL 2 g_2^4 + (g_2^2 + g_1^2)^2 \KR \KKR
  \log\KL\frac{{\La^2}}{v^2}\KR 
 }
$$
 
\newpage
\blackl{Both $M_H$ limits, upper and lower, combined}
\BC
 
 \mbox{}\hspace{5cm}\red{$\mt = 174 \gev$}\\[-1em]
 \includegraphics[height=10cm,width=10cm]{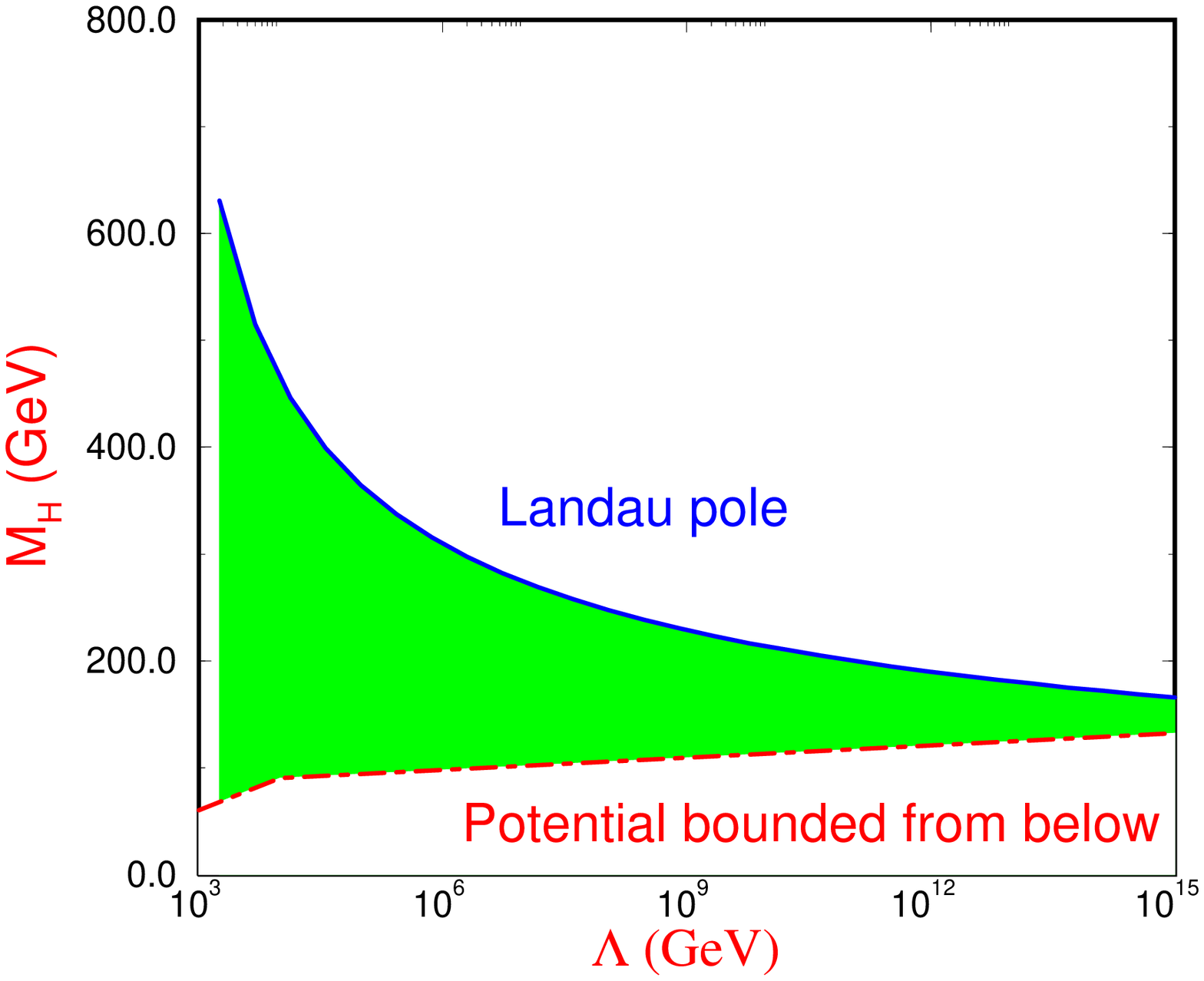}
\EC

\vspace{-0.5cm}
In the figure above the two limits on $M_H$ previously commented are plotted together as a function of the cut-off $\La$, which is interpreted to be the scale up to which the SM is valid. The upper blue line is the upper limit from avoiding the Landau pole/Triviality  and the lower dot-dashed red line is the lower limit from avoiding the vacuum instability and by requiring the potential to be bounded from below. The shaded green region in between the two lines is the allowed area for the SM value of $M_H$. For instance, if one requires the SM to be valid up to the scale of the Grand Unification Theory, then the Higgs mass should be within the following approximate interval:
   $$\mbox{For}\hspace{0.2cm}\La = M_{\rm GUT} \RA 130 \gev \lsim \MH \lsim 180 \gev$$
and, as can be seen in the above figure, the allowed interval gets narrower for larger $\Lambda$.

Recent computations of the stability lower bound include a NNLO analysis of the Higgs potential and realistic error estimates.\\
The condition for absolute stability up to the Planck scale gives the following bound (see for instance, Degrassi {\it et al} 2012): 
$$M_H(\gev) > 129.4 + 1.4 \left(\frac{m_t(\gev) -173.1}{0.7}\right)
-0.5\left(\frac{\alpha_s(M_Z)-0.1184}{0.0007}\right)\pm 1.0_{\rm th} $$
$$\RA M_H > 129.4 \pm 1.8 \gev$$
>From this lower bound one then may conclude that vacuum stability of the SM up to the Planck scale is excluded at $2  \sigma (98\% CL)$ for $M_H<126 \gev$ !!!. This is a quite remarkable result, given the present experimental meassurement which is precisely pretty close to 126 GeV. 
 
\vspace{1cm}
\blackl{Higgs mass limits from radiative corrections}\\[1em]

Another interesting Higgs mass limits can be extracted from the contributions of the Higgs particle, via radiative corrections, to the electro-weak precision observables (EWPO). 
For instance, the Higgs particle can propagate into the loops that contribute to the $e^+ e^- \to \mu^+ \mu^-$ scattering, and correct the tree level prediction for the observables associated to this proccess by an amount whose size depend, among other parameters, on the value of $M_H$. One example of a one-loop diagram where the Higgs enters in a relevant way is the one shown in the figure below, where the Higgs and the $Z$ bosons propagate inside the loop correcting the intermediate $Z$ boson propagator. A comparison between the prediction for a EWPO from the SM at a given order in perturbation theory and for a given $M_H$ value with the experimental measurement for this EWPO allows to set an allowed interval on $M_H$ (or equivalently a preferred by data $M_H$ window) and also set exclusion limits on $M_H$.  
 
\BC
\setlength{\unitlength}{1pt}
\begin{picture}(300, 80)
\ArrowLine(0,80)(60,40)
\ArrowLine(60,40)(0,0)
\Vertex(60,40){3}
\Photon(60,40)(110,40){3}{3.5}
\Vertex(110,40){3}
\DashCArc(135,40)(25,0,180){5}
\PhotonArc(135,40)(25,180,360){3}{5.5}
\Vertex(160,40){3}
\Photon(160,40)(210,40){3}{3.5}
\Vertex(210,40){3}
\ArrowLine(210,40)(270,80)
\ArrowLine(270,0)(210,40)
\put(125,75){{$H$}}
\end{picture}
\EC

There are many examples of EWPO where the Higgs particle contributes. For illustration, we choose here one of the most studied observables in the literature: The prediction for $M_W$ in terms of $\MZ$, $\al$, $G_F$ and ${\De r}$: 
$$
{\MW^2} \KL 1 - \frac{{\MW^2}}{\MZ^2} \KR = 
 \frac{\pi\,\al}{\wz\,G_F} \KL \frac{1}{1 - {\De r}} \KR.
$$
The parameter ${\De r}$ collects all the loop corrections and summarizes the deviations from the tree level relation: ${\MW^2} \KL 1 - \frac{{\MW^2}}{\MZ^2} \KR = 
 \frac{\pi\,\al}{\wz\,G_F}$. This $\De r$ can be evaluated, for instance,  from $\mu$ decay and from this one then gets $\MW$. The one-loop result for $\MW$ in the SM is well known (see, A.~Sirlin '80 and W.~Marciano, A.~Sirlin '80) and contains three relevant contributions given schematically by:
$$
\begin{array}{lccccc}
\De r_{1-\mathrm{loop}} = & \black{\De \al} & - &
 {\frac{\cw^2}{\sw^2} \De\rho} &
+ & { \De r_{\mathrm{rem}}(\MH)}\\[.5em]
& \black{\sim \log \frac{\MZ}{m_f}} && {\sim \Mt^2} && 
  \sim {\log(\MH/\MW)}\\[.5em]
& \black{\sim 6 \%} && {\sim 3.3 \%} && {\sim 1 \%}
\end{array}
$$
the contribution from the top quark is larger that the Higgs contribution, since the dependence with the top mass is quadratic. In contrast, the dependence on the Higgs mass is logarithmic, and indeed this dependence is quite general for many EWPO.  

This interesting exercise of comparing the SM prediction for $M_W$ as a function of the top mass and the Higgs mass with the data was done very often in the past, and more intensely in the era of LEP, SLD and TeVatron. These was done before the starting of the LHC and there were already some indications that the data seemed to prefer a light Higgs boson. We include below, on the left, one of these plots produced by the LEP Electroweak Working Group (LEPEWWG) in 2011. The oblique lines are the predictions for specific values of $M_H(\gev)$, 114, 300, 1000..and we can see clearly that the lines at the lowest values, i.e. those close to 114 GeV, fit better to the experimental measurement than the lines for heavier Higgs boson.     

\begin{minipage}[t]{9cm}
\hspace{-1cm}
\includegraphics[width=9cm]{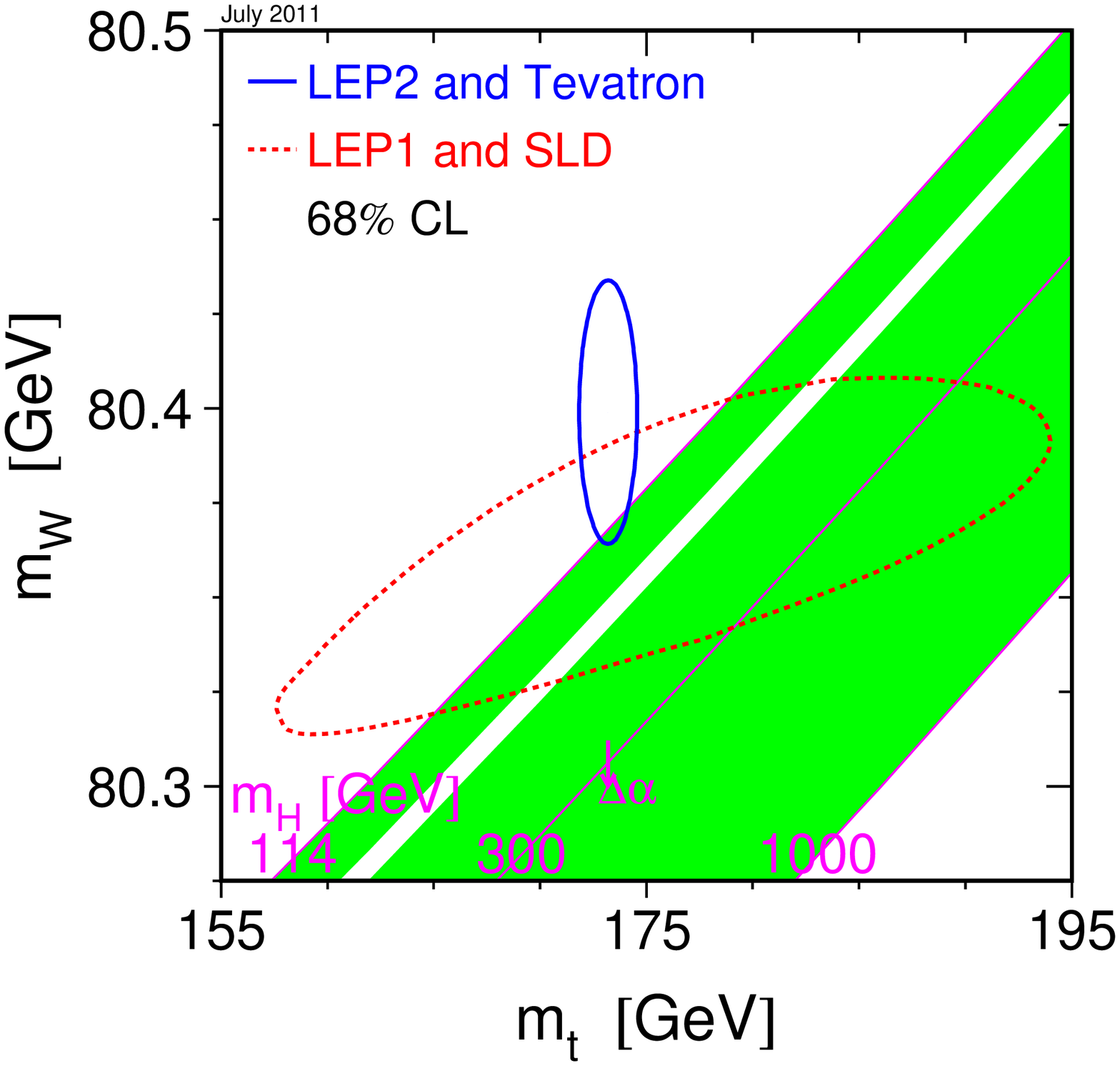}
\end{minipage}
\begin{minipage}[t]{9cm}
\hspace{-1cm}\includegraphics[height=8cm,width=8cm]{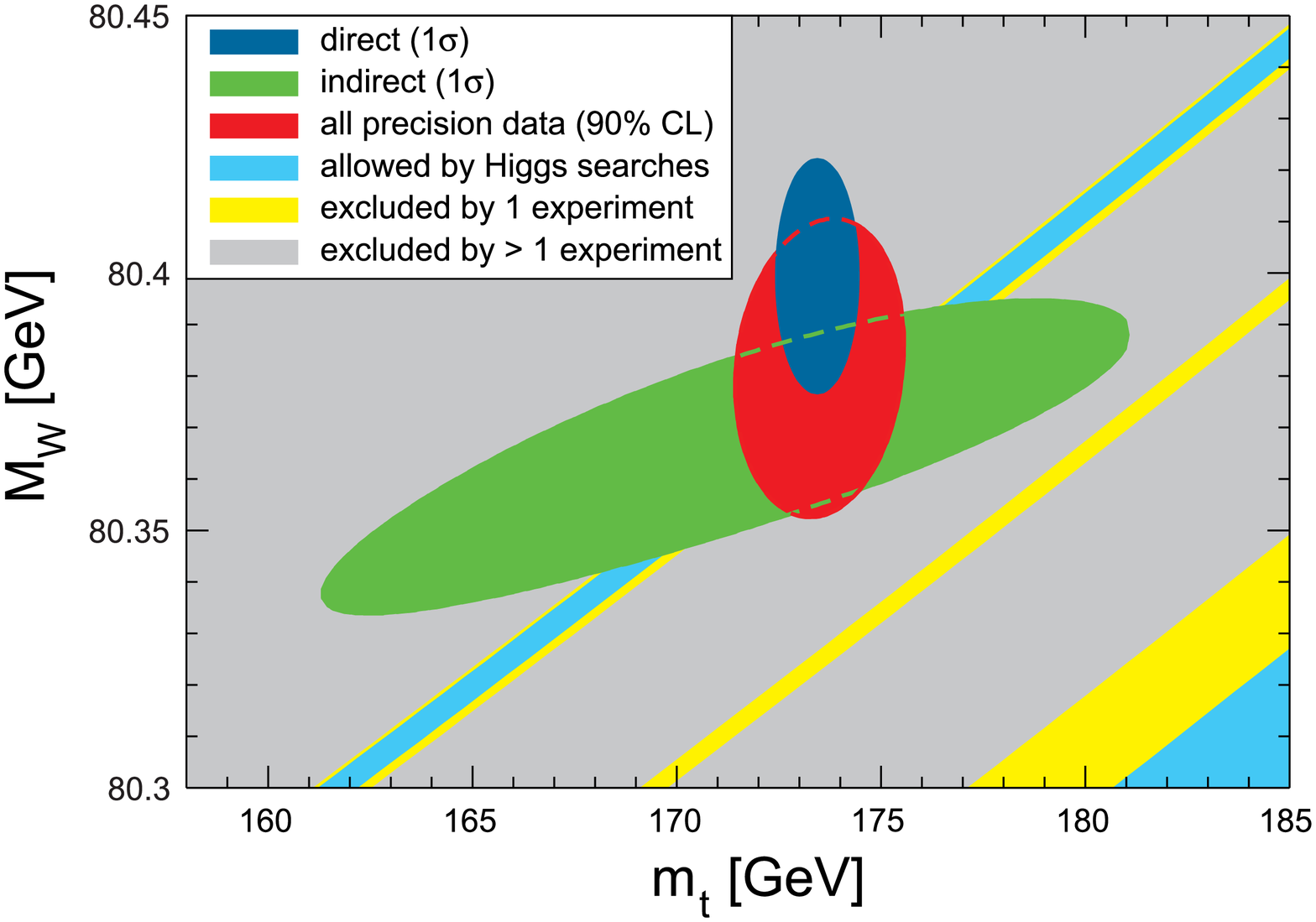}
 \end{minipage}
For comparison, we also include, on the right, the same kind of plot but corresponding to the situation after the LHC had started taking data. In this plot, the red area is allowed by all precision data at $90\% CL$ and the light blue oblique bands are the SM prediction for $M_W$ as a function of $m_t$, with $M_H$ allowed by Higgs searches at LHC, before the Higgs discovery: a) Central band:
$115.5 \gev < M_H < 127 \gev$, b) band at lower-right corner: $M_H>600 \gev$. Again we see that a light Higgs mass in this $115.5 \gev < M_H < 127 \gev$ window was preferred by all data before the Higgs discovery. It is certainly a good lesson to learn for the future searches of new physics beyond the SM, since the radiative corrections from this new physics could contribute to the precision observables in a relevant way and give us some clue on the scale where to look for this new physics.

\newpage
\blackl{The $\rho$ parameter and the custodial symmetry}\\[0.5em]
 
Another interesting parameter that measures the relevance of radiative corrections in the EW theory, and that is very sensitive to new physics beyond the SM is the so-called $\rho$ parameter.
This parameter is defined as the ratio of neutral to charged
current amplitudes at low energies:
\begin{equation}
\rho \equiv \frac{T_{NC}(q^2<<M_Z^2)}{T_{CC}(q^2<<M_W^2)}
\non
\end{equation}
>From $\nu$-scattering experiments and others there is a good measurement:
   $\rho_{\rm exp}=1.0008^{+0.0020}_{-0.0011}$(PDG 2012)\\
 The SM prediction at tree level is:
\begin{equation}
{\rho_{\rm tree}^{\rm SM}= \frac{M_W^{\rm 2\,\, tree}}
{M_Z^{\rm 2\,\, tree}\cos^2 \theta_W^{\rm tree}} =1 }
\non
\end{equation}
At one loop and keeping just the so-called 'oblique' corrections,
\begin{equation}
\rho =\frac{\rho_{\rm tree}}{1-\Delta \rho} \,\,;\,\,\Delta \rho  =	\frac{\Sigma^{\rm R}_Z (0)}{M_Z^2}
-  \frac{\Sigma^{\rm R}_W (0)}{M_W^2} \,\,\,\mbox{related\,\,to\,\,}T \mbox{\,\,\,parameter}
\non
\end{equation} 
For instance, the leading top and Higgs loop contributions give:
\begin{eqnarray}
(\Delta\rho)_t &=&\frac{g^2}{64\pi^2}N_C \frac{m_t^2}{M_W^2}+... 
\nonumber \\
(\Delta\rho)_H&=&-\frac{g^2}{64\pi^2}3
\tan^2\theta_W\log\frac{M_H^2}{M_W^2}+... \non
\end{eqnarray}
 The $\rho$ parameter being close to one is due to the so-called 
 custodial symmetry: a global symmetry of the SM Higgs sector in absence of gauge interactions.

In order to illustrate clearer this custodial symmetry it is convenient to use an alternative way 
of writing the (ungauged) Lagrangian of the SBS:
\begin{eqnarray}
{\cal L}_{\rm SBS}&=&\frac{1}{4} {\rm Tr}
\left [ (\partial_{\mu}M)^\dagger(\partial^{\mu}M)\right ] -V(M)\;; \non \\
V(M)&=&\frac{1}{4}\lambda \left [\frac{1}{2} {\rm Tr}(M^\dagger
M)+\frac{\mu^2}{\lambda}\right ]^2 \non
\end{eqnarray}
where $M$ is a $2\times 2$ matrix containing the four real scalar
fields of the $\Phi$ doublet:
\begin{eqnarray}
M &\equiv&
\sqrt{2}(\widetilde{\Phi}\Phi)=\sqrt{2}\left(\begin{array}{ll}
\phi_0^*&\phi^{\dagger}\\-\phi^-& \phi_0\end{array}\right)
\;;\; \nonumber  \\[-0.5em]
\Phi & = & \left(
\begin{array}{c}\phi^+ \\
 \phi_0 \end{array}\right)\;;\;    \nonumber \\[-0.5em]
\widetilde{\Phi}&=&i\tau_2\Phi^*=\left(\begin{array}{c}
\phi_0^*\\-\phi^-\end{array}\right)\nonumber
\end{eqnarray}
It is inmediate to check that  ${\cal L}_{\rm SBS}$ is invariant under the global
transformations:
\[M\rightarrow g_L M g_R^+ \;;\;\;g_L\subset SU(2)_L\;;\;\;g_R\subset SU(2)_R\]
This global symmetry $SU(2)_L \times SU(2)_R$ is called chiral EW symmetry (for analogy with QCD) and it is spontaneously broken down to the diagonal subgroup
{$SU(2)_{L+R} \equiv SU(2)_{\rm custodial}$}.\\
The pattern of global symmetry breaking is:
$$SU(2)_L \times SU(2)_R \to SU(2)_{\rm custodial}$$
Once $SU(2)_L \times U(1)_Y$ is gauged, the chiral symmetry (and the custodial) is explicitly broken.
 This custodial symmetry has many interesting implications for phenomenology, and it could provide some clue in the future searches of new physics beyond the Standard Model.

\section{Motivations for looking beyond the Standard Model}
In spite of the extraordinary success of the SM describing all the Particle Physics phenomena known so far in Nature, there is the general believe that the SM cannot be the ultimate theory of Fundamental Physics. There are several aspects where the SM does not provide a satisfactory answer. Here we list very briefly some of the issues that require going beyond the Standard Model:  

\begin{itemize} 
\item[-] The SM does not contain gravity. A fundamental theory including all the four known interactions: electromagnetic, weak, strong and gavitational, must go beyond SM.
\item[-] The SM does not provide gauge coupling unification. A fundamental theory that unifies all known gauge interactions must go beyond SM.
\item[-] The SM does not include neutrino masses nor intergenerational mixings for neutrinos. An explanation of the origin of non-vanishing neutrino masses and neutrino oscillations that are found in experiments require going beyond SM.
\item[-] The SM has no proper candidate for Dark Matter. The observations demonstrate that Dark Matter exists in Nature. Explaining the origin of Dark Matter requires going beyond SM. 
\item[-] The Higgs sector of the SM suffers of the so-called hierarchy problem. This will be shortly described below. Solving this problem
requires going beyond SM. 
\end{itemize}
 
\section{The hierarchy problem of the SM Higgs sector} 
This problem can be expressed as the instability of the value of the Higgs boson mass when radiative corrections are included in presence of a physical cut-off that is placed at energies far above the electroweak scale. It should be emphasized that it appears exclusively when the SM is considered as a low energy effective theory that is valid up to this given cut-off. But this assumption seems to be the most probable one, given the previous list of unsolved issues within the SM, one is lead to think that the SM is not a fundamental theory but rather a succesfull effective theory. 

Let us illustrate with a bit more detail how this hierarchy problem appears when computing the one-loop radiative corrections to the Higgs mass in presence of an ultraviolet cut-off $\Lambda$.
One starts with the Higgs progagator at the tree level, describing the free propagation, and then one adds the one-loop diagrams correcting this propagator, as the one shown below with the self-energy correction $\Si_H$ given by the fermionic contribution $\Si_H^f$, and this later being easily computable from the corresponding Feynman integral over the internal fermion momentum $k$. Schematically:\\[1em]
\black{Higgs propagator:}
\setlength{\unitlength}{1pt}
\begin{picture}(170, 20)
\DashLine(10,05)(160,05){5}
\Text(20,20)[]{$H$}
\Text(150,20)[]{$H$}
\end{picture}
\hspace{0.5cm}
inverse propagator: {$i (p^2 - \MH^2)$}\\[1em]
{Loop corrections:}
\setlength{\unitlength}{1pt}
\begin{picture}(170, 40)
\DashLine(10,00)(60,00){5}
\Text(20,20)[]{$H$}
\Vertex(60,00){3}
\ArrowArc(85,00)(25,0,180)
\ArrowArc(85,00)(25,180,360)
\Text(90,40)[]{$f$}
\Text(90,-40)[]{$\bar f$}
\Vertex(110,00){3}
\DashLine(110,00)(160,00){5}
\Text(150,20)[]{$H$}
\end{picture}
\hspace{0.5cm} 
inverse propagator: ${i (p^2 - \MH^2}{+ \Si_H^f}{)}$ \\[2em]
\setlength{\unitlength}{1pt}
\begin{picture}(550, 80)
\put(00,40){\rm{Fermion propagator:}}
\ArrowLine(120,40)(240,40)
\put(180,45){$k$}
\put(300,40){$\dfrac{i}{\slashed{k}-m_f}=\dfrac{i(\slashed{k}+m_f)}{k^2-m_f^2}$}
\end{picture}
\\[0.5em]
Integrating over all possible loop momenta $k$, considering $N_f$ degrees for fermion $f$ with mass $m_f$ and Yukawa coupling $\lambda_f$, and keeping just the dominant terms for large cut-off $\Lambda$, gives:  
 
\BEA
{\Si_H^f} &\sim&  N_f \; \lambda_{f}^2
\int d^4k \left(\frac{1}{k^2 - m^2_{f}} +
\frac{2 m^2_{f}}{(k^2 - m^2_{f})^2} \right)(-1)\non \\[0.5em]
{\rm for~} \Lambda \to \infty: \hspace{1cm}
{\Si_H^f} &\sim& N_f \; \lambda_f^2
\left(\;\; {\underbrace{\int \frac{d^4k}{k^2}}} \;\;  + \;\; 2 m_{\rm f}^2
{\underbrace{\int \frac{d^4k}{k^4}}} \;\; \right)(-1) \non
\EEA
\mbox{} \hspace{8.5cm} {$\sim \Lambda^2$} \hspace{1.5cm}
{$\sim \ln\Lambda$}\\[0.5em]
And from this, one finally gets the mass corrections as a function of $\Lambda$:
{$$\delta M_H^2=N_f \frac{\lambda_f^2}{16 \pi^2}\left(-2\Lambda^2+6m_{\rm f}^2
\log \frac{\Lambda}{m_f}+\dots\right).$$} 
\\ 
There are two dominant contributions at high values of the cut-off: the largest one that grows quadratically as $\Lambda^2$ and the other one that grows logarithmically as $\log \Lambda$. 
Thus, for instance, if one takes the cut-off at the Planck energy scale: $\La = M_{\rm Pl}$:
$$
 \de \MH^2 \sim M_{\rm Pl}^2
{\quad \RA \quad \de \MH^2 \approx 10^{30} \, \MH^2},
$$
i.e., one finds an unacceptable huge correction that is 30 orders of magnitude larger than the starting tree level squared Higgs mass (for $\MH \lsim 1 \tev$). 

Another popular example is the case of Grand Unified Theories (GUT), where the physical cut-off is at $\Lambda=M_{\rm GUT}\sim 10^{16} \gev$ and one also gets huge corrections given by $\de \MH^2 \approx M_{\rm GUT}^2$. 

In summary, the hierarchy problem is the instability of the small Higgs mass to
large corrections in a context where the SM is a low energy remnant of a more fundamental theory  with a large mass scale in addition to the
weak scale. Furthermore, this instability occurs because:\\[0.5em]
$-$ there is no additional symmetry for $\MH = 0$,\\[.3em]
$-$ and in consequence, there is no protection against large corrections. 

\section{Two main avenues to solve the hierarchy problem}
There are two qualitatively different proposals to solve the hierarchy problem of the SM. Generically: One avenue assumes new symmetries and the Higgs boson is an elementary particle; The other avenue assumes new interactions and the Higgs boson is a composite particle. At present there is not yet any experimental evidence in favor of none of these two possibilities, therefore the issue of the elementarity/compositeness of the observed Higgs particle is still an open question. The generic features of these two main avenues are summarized schematically below:\\[0.5em]
\begin{minipage}{17em}
{\bf Elementary Higgs}\\
{$\star$}There should exist an extra symmetry
(at least) and new particles with couplings
dictated by this symmetry such that 
the most problematic quadratic sensitivity to the high scale cancels.\\[.3em]
{$\star$}  The typical example is 
{\it Supersymmetry} where
the sparticle partner cancels the quadratic divergence 
generated by the particle.\\ 
{$\star$} The soft SUSY breaking scale acts as
a cutoff of divergences \\
{$\star$}   The Higgs boson is weakly
interacting\\
{$\star$} The Higgs self-coupling is related to the  EW gauge
coupling\\
{$\star$}  The Higgs boson mass is close to the EW scale\\
{$\star$} Typically a bunch of new elementary particles appear in the spectrum:\\
{$\star$} Including several Higgs particles, besides the SM Higgs-like boson.


 \end{minipage}
\hspace{1.5em}
\begin{minipage}{17em}
{\bf Composite Higgs}\\
{$\star$} At some scale the Higgs
dissolves \\
and the theory of constituents is at work \\
{$\star$} Similar to QCD where the pions
dissolve into quarks \\
{$\star$} The compositeness scale acts as
a cutoff of quadratic divergences \\
{$\star$} The typical example is 
{\it Technicolor} Theories where \\
the Higgs boson is strongly interacting and the 
Higgs mass is at TeV scale\\
{$\star$} Modern theories of compositeness 
involve an {\it extra dimension} through
the AdS/CFT correspondence.
The Higgs mass value and the size of its couplings are very model dependent.\\
{$\star$} The smallness of $m_H$, close to the EW scale, can be explained if $H$ is a Pseudo-Goldstone boson. \\ 
{$\star$} Typically, new composite resonances appear in the spectrum.
\end{minipage}
 

\section{Supersymmetry} 

The existence of one (or more) new symmetry relating fermions and bosons is the most popular proposal to solve the hierarchy problem of the SM Higgs sector. This new symmetry is called supersymmetry (SUSY) and generically acts as:

\BEA
Q | \mbox{boson} \rangle & = & | \mbox{fermion} \rangle \non \\
Q | \mbox{fermion} \rangle & = & | \mbox{boson} \rangle \non
\EEA
which, effectively, produces an enlargement with respect to the SM spectrum: the SM particles have SUSY partners that share their same quantum numbers but differ by one half unit in their spin. Thus, the SM left-handed and right-handed fermions, $f_{L,R}$, have their corresponding SUSY partners, named sfermions, $\Sferm_{L,R}$, that are scalar particles, the SM gauge bosons have their SUSY partners named gauginos that are fermions, and the Higgs bosons have their SUSY partners named higgssinos that are also fermions.

SUSY then solves the hierarchy problem by the additional contributions from sfermions. 
When computing the one-loop radiative corrections to the Higgs boson propagator in SUSY theories one has to add new contributions from the scalar fermion partners that are given by the two graphs below. As previously done, one then focus on the behavior of these new contributions at large values of the cut-off $\Lambda$:\\[1em] 

\BC
\setlength{\unitlength}{1pt}
\begin{picture}(500, 80)
\DashLine(10,40)(60,40){5}
\Text(20,60)[]{$H$}
\Vertex(60,40){3}
\DashArrowArc(85,40)(25,0,180){5}
\DashArrowArc(85,40)(25,180,360){5}
\Text(90,80)[]{$\Sferm_{L,R}$}
\Text(90,0)[]{$\bar{\Sferm}_{L,R}$}
\Vertex(110,40){3}
\DashLine(110,40)(160,40){5}
\Text(150,60)[]{$H$}
\DashLine(200,40)(350,40){5}
\Vertex(275,40){3}
\DashArrowArc(275,65)(25,90,270){5}
\DashArrowArc(275,65)(25,270,90){5}
\Text(275,105)[]{$\Sferm_{L,R}$}
\Text(210,60)[]{$H$}
\Text(340,60)[]{$H$}
\end{picture}
\EC

\vspace{-1.0em}
$$
\Si_H^{\Sferm} \sim \; \mbox{first\;diagram\;}(\sim \log \Lambda)\;+ \; \;\;N_{\Sferm} \; \la_{\Sferm}
\int d^4k \left(\frac{1}{k^2 - m^2_{\tilde f_L}} +
\frac{1}{k^2 - m^2_{\tilde f_R}} \right) 
$$

for $\La \to \infty$: \quad
$\delta M_H^2=2N_{\Sferm}\frac{\la_{\Sferm}}{16 \pi^2}\left(\Lambda^2-2 m^2_{\tilde f} \log \frac{\Lambda}{m_{\tilde f}}\right)+ \dots$
 \\[1em]
where, for simplicity, $m_{\tilde f_L}=m_{\tilde f_R}=m_{\tilde f}$ is assumed. $N_{\tilde f}$ is the number of sfermion modes and $\la_{\Sferm}$ is the sfermions coupling to Higgs bosons involved in the second diagram.  

 From the previous result, it is clear that when adding these sfermion corrections to the previous fermionic corrections one finds that the quadratic contributions, ${\cal O}(\Lambda^2)$, cancel in the total Higgs boson mass squared if the following equations are satisfied:
\BEA
N_{\sfl} = N_{\sfr}= N_{\tilde f}&=& N_f \non \\[.5em]
\la_{\Sferm}&=& \la_f^2 \non
\EEA
and these are precisely the conditions imposed by SUSY, namely, the identity in the number of bosonic and fermionic degrees of freedom, and the specific relations between their couplings. 
Notice also that the total dominant corrections, including ${\cal O}(\Lambda^2)$ and ${\cal O}(\log \Lambda)$, vanish if furthermore
$$
m_{\Sferm} = m_f
$$
i.e. for exact degeneracy between fermions and sfermions.

Generically, one may then characterize the SUSY breaking by the mass splitting between fermions and their sfermion partners. Namely, if  $m_{\Sferm}^2 = m_f^2 + \De^2 $ and $\la_{\Sferm} = \la_f^2$ then one gets a total correction given by:
$$
\Si_H^{f + \Sferm} \sim  N_f \; \lambda_f^2 \;
\De^2 + \ldots
$$
and this correction stays acceptably small if the mass splitting $\De$ is small, say not much heavier than  the weak scale.\\[.7em]
One then concludes that the stability of the Higgs boson mass corrections is realized if the mass scale of the SUSY partners is not very far above the weak scale, or simply:
$$
\msusy \lsim {\cal O}(1 \TeV)
$$
Therefore, SUSY at the TeV scale provides an attractive solution to the
hierarchy problem. But this implies finding SUSY at these scales, which has not happened so far in the experiments. Setting $\msusy$ above these values, say  $\msusy$ larger than a few TeV, leads to the so-called Split SUSY models, but all these suffer of some kind of instabilities due to reintroduction of large Higgs mass corrections.    
 

{}

\subsection{SUSY-breaking Models}
As we have seen, exact SUSY requires mass degeneracy between particles and sparticles: $m_f = m_{\tilde f}$, \ldots, etc. However  in a realistic model SUSY must be broken somehow, since the SUSY partners with such masses have not been found in Nature.

On the other hand, it is known that satisfactory models of SUSY breaking must proceed via spontaneous SUSY breaking at some high energy scale.  
Specific SUSY-breaking schemes (see below) in general yield an effective
Lagrangian at low energies that is supersymmetric except for some explicit 
{\it soft} SUSY-breaking terms. 
These soft SUSY-breaking terms have the interesting property of 
not altering, via the radiative corrections, the dimensionless couplings of the theory. Therefore, they preserve the nice property of the unbroken SUSY case with no quadratic divergences, as required by the SUSY solution to the hierarchy problem. In fact this is true in all orders of perturbation
theory and the specific types of soft SUSY-breaking terms are well known and classified in the literature. For the case of minimal SUSY particle content and two Higgs doublets, $H_d(=H_1)$ and $H_u(=H_2)$, the soft SUSY-breaking Lagrangian can be written as,
\beqar
{\cal L}_{\rm soft}  &=&
 - \frac{1}{2} \Bigl({M_1} \widetilde{B}\widetilde{B} + {M_2}
\widetilde{W}\widetilde{W} + {M_3} \tilde{g}\tilde{g} \Bigr) +
{\rm h.c.}  \non \\
&& {} - ({m_{H_u}^2 + |\mu|^2}) H_u^+ H_u 
      - ({m_{H_d}^2 + |\mu|^2}) H_d^+ H_d -
 ( {B} H_u H_d + {\rm h.c.}) \non \\
&& {} - \Bigl( \tilde{u}_R {{A_u}} \widetilde{Q} H_u -
 \tilde{d}_R {{ A_d}} \widetilde{Q} H_d -
 \tilde{e}_R {{ A_e}} \widetilde{L} H_d \Bigr) +
{\rm h.c.} \non \\
&& {} - \widetilde{Q}^+  {{ m_{\widetilde{Q}}^2}} \widetilde{Q}
 - \widetilde{L}^+  {{ m_{\widetilde{L}}^2}} \widetilde{L}
 - \tilde{u}_R {{ m_{\widetilde{u}}^2}} \tilde{u}_R^*
 - \tilde{d}_R {{ m_{\widetilde{d}}^2}} \tilde{d}_R^*
 - \tilde{e}_R {{ m_{\widetilde{e}}^2}} \tilde{e}_R^* .
\non
\eeqar
This is the most general parameterization of SUSY-breaking terms that keeps relations
between dimensionless couplings unchanged; hence not generating quadratic divergences. It includes mass terms for the gauginos, $M_i$ ($i=1,2,3$), for the scalars $m_S^2$, a Higgs bilinear $B$ term and also trilinear couplings between the Higgs bosons and the sfermions $A_f$. Notice that $ m_{\tilde f}^2$ and $A_f$ are $3\times 3$ matrices in family space, therefore, in general, they introduce many new parameters.
Most of the SUSY-breaking models assume that the soft SUSY-breaking mass scales involved are not far above the TeV scale, {$\msusy \lsim 1 \tev$},  such that they all avoid the hierarchy problem. 

Generically, these SUSY-breaking Models can be classified in two big groups: Unconstrained Models and Constrained Models, according to the following general features:\\[1em]
\blackl{Unconstrained models (MSSM,..)}\\[.3em]
$\star$ These are agnostic about how SUSY breaking is achieved and
no particular SUSY breaking mechanism is assumed.\\[.3em]
$\star$ They implement instead a general parameterization of all
possible soft SUSY-breaking terms.\\[.3em]
$\star$ The relations between dimensionless couplings are unchanged\\
and, therefore, no quadratic divergences are re-introduced by the SUSY breaking.\\[0.3em]
$\star$ The simplest and most popular of these models is the Minimal Supersymmetric Standard Model (MSSM). It is called minimal because it is based in the minimal number of Supersymmetries and in the minimal particle content. In spite of being the simplest case, the MSSM still has plenty of parameters. In  
the most general case there are 
105 new parameters, including couplings, masses, mixing angles and phases.\\[0.5em]

 \blackl{Constrained models (mSUGRA, \ldots):}\\[0.3em]
$\star$ In these models there are specific assumptions on the scenario that achieves spontaneous SUSY breaking.\\[0.3em]
$\star$ Therefore, they provide specific predictions for the soft SUSY-breaking terms in terms of a smaller set of parameters.\\[0.3em] 
$\star$ An experimental determination of the SUSY parameters would imply setting the 
patterns of SUSY breaking.\\[0.3em]
$\star$ All constrained models are special versions of the MSSM.\\[0.3em]
$\star$ There are different kinds of Constrained models, mainly according to the origin of the SUSY breaking and the way it is transmitted from the so-called Hidden sector" to the "Visible sector".  
For illustration, we include below, in an schematic way, some examples:  
\BC
\begin{tabular}{ccc}
\black{\mbox{} \hspace{-2.5em} ``Hidden sector'': } &
\hspace{.7em} {$\longrightarrow$ } \hspace{.7em} &
\black{``Visible sector``: }\\
\black{\mbox{} \hspace{-2.5em} SUSY breaking } & & \black{MSSM }
\end{tabular}
\EC


\BC

``Gravity-mediated'': CMSSM/mSUGRA\\
``Gauge-mediated'':   GMSB\\
``Anomaly-mediated'': AMSB\\
``Gaugino-mediated''\\
\ldots

\EC
For instance, in two of the most popular ones CMSSM (Constrained MSSM) and mSUGRA (Minimal Supergravity), the mediating interactions are gravitational. In contrast in the GMSB (Gauge Mediated SUSY Breaking) the mediating interactions are gauge interactions, etc..
 
Since all models are specific versions of the MSSM, we will focus next in the general features of the MSSM spectrum and in particular of the MSSM Higgs sector. 
 
%



\subsection{MSSM spectrum}
We summarize in the table below the particle content within the MSSM. Besides the SM particles, there are their corresponding SUSY partners and the extra Higgs boson particles that correspond to the enlarged Higgs sector of the MSSM with two Higgs doublets. The SM interaction eigenstates are also specified in the table. The interaction eigenstates that have the same quantum numbers mix and give rise to the physical mass eigenstates which are also specified in the table. The MSSM physical states include the squarks, sleptons, sneutrinos, gluinos, charginos, neutralinos and the physical Higgs bosons, these particles and all these particles are being searched for at the present experiments.
\BC
{\tiny 
\begin{tabular}{|c|c|c|c|c|} 
\hline \hline 
\cline{1-2} 
& \multicolumn{4}{|c|}{\textcolor{black}{SUSY particles}}{}{}\\ 
\hline 
\cline{1-3} 
{Extended Standard}&\multicolumn{2}{|c|}{$SU(3)_C \times SU(2)_L \times U(1)_Y$}& 
\multicolumn{2}{|c|}{Mass eigenstates}\\  
{Model spectrum}&\multicolumn{2}{|c|}{interaction eigenstates}& 
\multicolumn{2}{|c|}{}\\  
\cline{1-5} 
& \textcolor{black}{Notation} & \textcolor{black}{Name} & \textcolor{black}{Notation} & \textcolor{black}{Name}\\ 
\cline{1-5} 
&&&&\\  
$q=u,d,s,c,b,t$ &  
\textcolor{black}{$\tilde q_L,\tilde q_R$} & squarks &\textcolor{black}{$\tilde q_1,\tilde q_2$}  & squarks\\ 
$l=e,\mu,\tau$ & \textcolor{black}{$\tilde l_L,\tilde l_R$} & sleptons & 
\textcolor{black}{$\tilde l_1,\tilde l_2$} & sleptons\\  
$\nu =\nu_e, \nu_{\mu}, \nu_{\tau}$ &  \textcolor{black}{$\tilde \nu$}&  
sneutrino &\textcolor{black}{$\tilde \nu$}& sneutrino\\  
&&&&\\  
\cline{1-5} 
&&&&\\ 
g &\textcolor{black}{$\tilde g$} & gluino &\textcolor{black}{$\tilde g$} & gluino\\ 
&&&&\\  
\cline{1-5} 
&&&&\\  
$W^{\pm}$ & \textcolor{black}{$\tilde W^{\pm}$} & wino & &\\ 
$H_1^{+} \supset H^+$ &\textcolor{black}{$\tilde H_1^{+}$} & higgsino & \textcolor{black}{$\tilde \chi^{\pm}_{i}\, 
{\scriptstyle {(i=1,2)}}$} & 
charginos\\ 
$H_2^{-} \supset H^-$ & \textcolor{black}{$\tilde H_2^{-}$} & higgsino & &\\ 
&&&&\\  
\cline{1-5} 
&&&&\\  
$\gamma$ & \textcolor{black}{$\tilde \gamma$}& photino & &\\ 
$Z$ & \textcolor{black}{$\tilde Z$} & zino & &\\ 
$H_1^{o} \supset h^0,\,H^0,\,A^0 $ &\textcolor{black}{$\tilde H_1^{o}$} & higgsino & \textcolor{black}{$\tilde \chi^{o}_{j}\, 
{\scriptstyle {(j=1,\ldots,4)}}$} & neutralinos\\ 
$H_2^{o} \supset h^0,\,H^0,\,A^0$ & \textcolor{black}{$\tilde H_2^{o}$} & higgsino & &\\ 
$W^3$ & \textcolor{black}{$\tilde W^3$} & wino & &\\ 
$B$ & \textcolor{black}{$\tilde B$} & bino & &\\ 
&&&&\\ 
\hline \hline 
\end{tabular}
}
\EC 
{}  
 

{}

\subsection{Enlarged Higgs sector of the MSSM versus SM}  
In the MSSM two Higgs doublets $H_d$ (=$H_1$) and $H_u$ (=$H_2$) are needed to give masses to down- and up-type fermions. 
This is in contrast to the SM where with just one Higgs doublet both $u$ and $d$ type masses can be generated via the Higgs Mechanism. For instance, for the quarks of the first generation: 
$$
\lag_{\rm SM} = {\underbrace{m_d \bar Q_L \Phi d_R}} + 
{\underbrace{m_u \bar Q_L \tilde \Phi u_R}}
$$
\mbox{} \hspace{6cm} {d-quark mass} \hspace{0.5cm} {u-quark mass}
$$
Q_L = \VL u \\ d \VR_L, \quad \tilde \Phi = i \si_2 \Phi^*, \quad
\Phi \to \VL 0 \\ v \VR, \quad \tilde \Phi \to \VL v \\ 0 \VR
$$
There are two main reasons for two Higgs doublets in the MSSM and not jut one:\\[.5em] 
On one had, Supersymmetry implies that the Superpotential must be an holomorphic function of the chiral superfields, i.e. it
depends only on $\varphi_i$, not on $\varphi_i^*$. Therefore a term like $\bar Q_L \Phi^*$ is not allowed. \\[.5em]
On the other hand, two Higgs doublets are also needed in SUSY for cancellation of anomalies. The fermionic partners of these Higgs scalars would otherwise contribute to non-vanishing anomalies.
 
The specific components of these two Higgs doublets are given by:
\\[-1em]
\BEA
\black{H_1} &=& \VL H_1^1 \\ H_1^2 \VR = \VL {v_1}
                                         + (\phi_1 + i\chi_1)
                                 /\sqrt2 \\ \phi_1^- \VR \non \\
\black{H_2} &=& \VL H_2^1 \\ H_2^2 \VR =  
                           \VL \phi_2^+ \\ {v_2}
                             + (\phi_2 + i\chi_2)/\sqrt2 \VR 
                  \phantom{ \, e^{i\magenta{\xi}}} \non
\EEA
where $v_1$ and $v_2$ are the respective vacuum expectation values of the two neutral scalar bosons, and their ratio defines a very relevant MSSM parameter: $\tan \beta = v_2/v_1$. Notice that these two complex doublets imply the introduction of eight degrees of freedom, but only three of them are really needed to provide masses to the three weak bosons. Therefore, the implementation of the Higgs Mechanism in this case will lead to five physical scalars remaining in the spectrum.

The Higgs potential of the MSSM is given in terms of these two Higgs doublets by:
\BEA
V &=& m_1^2 H_1\bar{H}_1 + m_2^2 H_2\bar{H}_2 
                         - m_{12}^2 (\epsilon_{ab}
      H_1^aH_2^b + \hc)  \non \\[.5em]
   && {} + {\frac{{g'}^2 + g^2}{8}}\,
      (H_1\bar{H}_1 - H_2\bar{H}_2)^2
      +{\frac{g^2}{2}}\, |H_1\bar{H}_2|^2 \non 
\EEA
One remarkable feature of this potential is that the Higgs self-couplings are given in terms of the EW gauge couplings, in contrast to the
      SM potential where the self-couplings were given by the unknown $\lambda$.\\[1em]
After the spontaneous EW symmetry breaking, one finds the announced five physical Higgs particles, 
{$h^0, H^0, A^0, H^{\pm}$}, which appear as 'oscillations' around the asymmetric vacuum state.\\[.5em]
The would-be Goldstone bosons dissapear in the unitary gauge and one gets the wanted gauge boson masses:
$$
{\MW^2} = \frac{1}{2} g^{\prime 2} (v_1^2 + v_2^2), \quad
{\MZ^2} = \frac{1}{2} (g^2 + g^{\prime 2}) (v_1^2 + v_2^2), \quad
{M_\ga} = 0
$$
All the predictions in the MSSM Higgs sector can then be expressed in terms of two input parameters, $\tan \beta$  and the mass of the CP-odd Higgs boson $\MA$:
$$
{\Tb} = \frac{{v_2}}{{v_1}}, 
\qquad {\MA^2} = -m_{12}^2 (\Tb + \CTb) 
$$
 
\blackl{The masses of the MSSM Higgs bosons} \\[0.5em]
The tree-level prediction for the Higgs boson masses in the MSSM are:
$$
m^2_{H, h} = 
 \frac{1}{2} \KKL \MA^2 + \MZ^2
         \pm \sqrt{(\MA^2 + \MZ^2)^2 - 4 \MZ^2 \MA^2 \CQZb} \KKR , 
$$ 
and 
$$m^2_{H^\pm}=M_A^2+M_W^2.$$
This implies an upper mass bound for the lightest Higgs mass $\mh$: \\[0.5em]
$\RA$ $\mh^2 \leq \MZ^2 \cos^2 2 \beta$ (the equality holds for $\MA>>\MZ$).\\[0.5em]
Therefore, at tree level: $\mh < \MZ$, what is clearly in contradiction with the experiments that 
have not found such a light Higgs particle.

However, when radiative corrections are included, the Higgs masses get shifted respect to the tree level values and, in particular, the lightest Higgs corrected mass $\mh$ (we use here the same notation for the tree and corrected masses, for shortness) can be considerably enhanced in some regions of the MSSM parameter space. In fact there are large corrections from the Yukawa couplings of the third generation quarks, being the corrections from the top quark the largest ones. For instance, the dominant 1-loop corrections from the top-stop sector can be written (for $\MA \gsim 150 \gev$) approximately as:
$${\mh^2} \simeq \MZ^2 \cos^2 2 \beta + \frac{3 g^2}{8 \pi^2}\frac{{\Mt^4 }}{\MW^2}
\black{\log \left(\frac{{\mst^2}}{\mt^2}\right)},$$
which clearly indicates the large size of the correction, since it goes with a large factor $m_t^4$ in front,  and we also see the logarithmic growing of this correction with the relevant SUSY scale, here the stop mass. 

The approximate behavior of the corrected $\mh$ as a function of the relevant SUSY mass for several values of 
$\tan \beta$ is illustrated in the plot below. 
In this plot we see a very relevant increase of $\mh$
with $\Delta_{\rm S}=\mst$ and with $\tan \beta$ that lead to Higgs mass values that are compatible with data.\\
The present status of the predictions for $\mh$ in the MSSM is that a complete one-loop result and `almost complete' two-loop result are available. In particular, one can find some regions of the MSSM parameter space where the predicted $\mh$ is compatible with the Higgs mass value that has been recently measured at LHC. Generically, one can conclude that for soft SUSY masses at or slightly below a few TeV the lightest Higgs boson of the MSSM is indeed light, with $m_h$ of order $\sim {\cal O}(100 \gev)$, and it could perfectly fit the LHC measured Higgs mass value.
\BC
\includegraphics[width=11.5cm,angle=0]{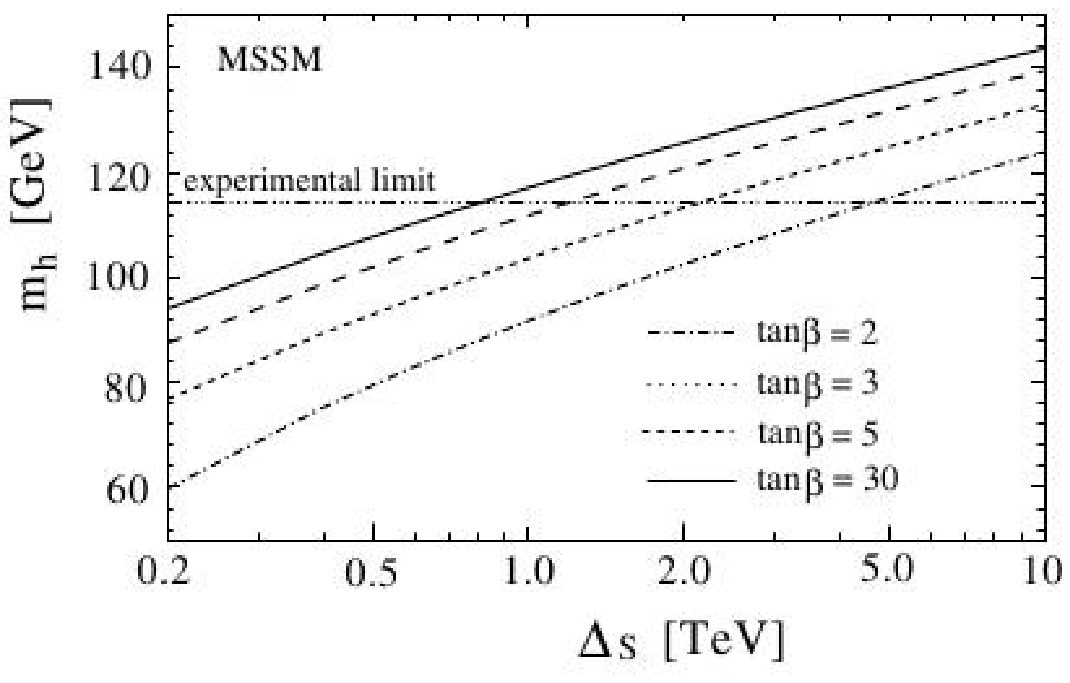}\\
(Plot taken from \cite{Wells:2009kq})
\EC
 
 
\blackl{The couplings of the MSSM Higgs bosons:}\\[0.5em]
The MSSM (neutral) Higgs boson couplings to weak gauge bosons and fermions are predicted to be at tree level as follows:
\beqar
g_{hVV} &=& {\sin(\be - \al) } \; g_{HVV}^{\rm SM}, \quad
V = W^{\pm}, Z \non \\[.5em]
g_{HVV} &=& {\cos(\be - \al) } \; g_{HVV}^{\rm SM} \non \\[.5em]
g_{hAZ} &=& {\cos(\be - \al) } \; \frac{g'}{2 \cos\theta_{W}} 
\non \\[0.5em]
g_{h b\bar b}, g_{h \tau^+\tau^-} &=& 
 {- \frac{\sin\al}{\cos\be} } \; 
   g_{H b\bar b, H \tau^+\tau^-}^{\rm SM} \non \\[.5em]
g_{H b\bar b}, g_{H \tau^+\tau^-} &=& 
 {\frac{\cos\al}{\cos\be} } \; 
   g_{H b\bar b, H \tau^+\tau^-}^{\rm SM} \non \\[.5em]   
g_{h t\bar t} &=& {\frac{\cos\al}{\sin\be} } \; 
g_{H t\bar t}^{\rm SM} \non \\[.5em]
g_{H t\bar t} &=& {\frac{\sin\al}{\sin\be} } \; 
g_{H t\bar t}^{\rm SM} \non \\[.5em]
g_{A b\bar b}, g_{A \tau^+\tau^-} &=& 
{\gamma_5\tan\be} \; g_{H b\bar b, H \tau^+\tau^-}^{\rm SM}\non \\[.5em]
g_{A t\bar t} &=& 
{\gamma_5\cot\be} \; g_{H t\bar t}^{\rm SM} \non \\[.2em]
&& \non
\eeqar
where $\alpha$ is the mixing angle in the CP-even Higgs sector that is given at tree level by:
\BE 
\alpha=\arctan\left[\frac{-(m_A^2+M_Z^2)\sin\beta\cos\beta}{M_Z^2\cos^2\beta+m_A^2\sin^2\beta-m_h^2}\right]\ ,\ -\frac\pi2<\alpha<0\ \non,
\EE 
and the $g^{\rm SM}$ couplings are the corresponding Higgs boson couplings of the SM. 

The first obvious conclusion from the formulas above is that the lightest MSSM Higgs boson couplings to weak gauge bosons are always smaller than the corresponding SM Higgs boson couplings, 
\black{$g_{hVV} \leq g_{HVV}^{\rm SM}, \quad V = W^{\pm}, Z $},
whereas, the couplings to $b$ quarks ($t$ quarks) and to $\tau$ leptons 
 $g_{h b\bar b}, g_{h \tau^+\tau^-}$ ($g_{h t\bar t}$) can get a significant
 enhancement (suppression) with respect to the corresponding SM coupling at large $\tan \beta$. This enhancement (suppression) is illustrated in the plot below, where the ratio $C_{qqh}=g_{h q\bar q}/g_{H q\bar q}^{\rm SM}$ is shown as a function of $M_A$ for two values of $\tan \beta$.\\[-9em]
\BC
\includegraphics[width=70mm]{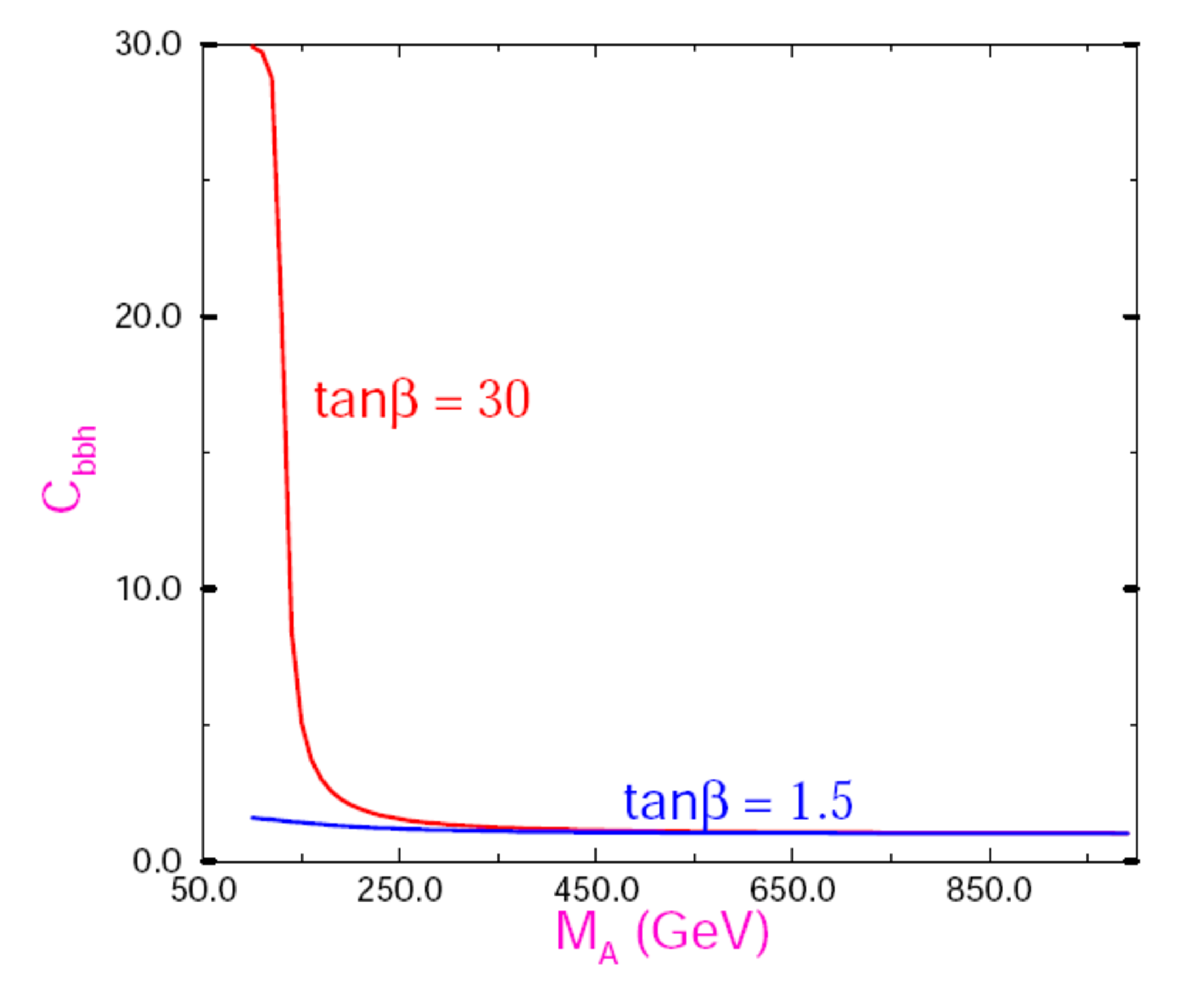}
\includegraphics[width=70mm]{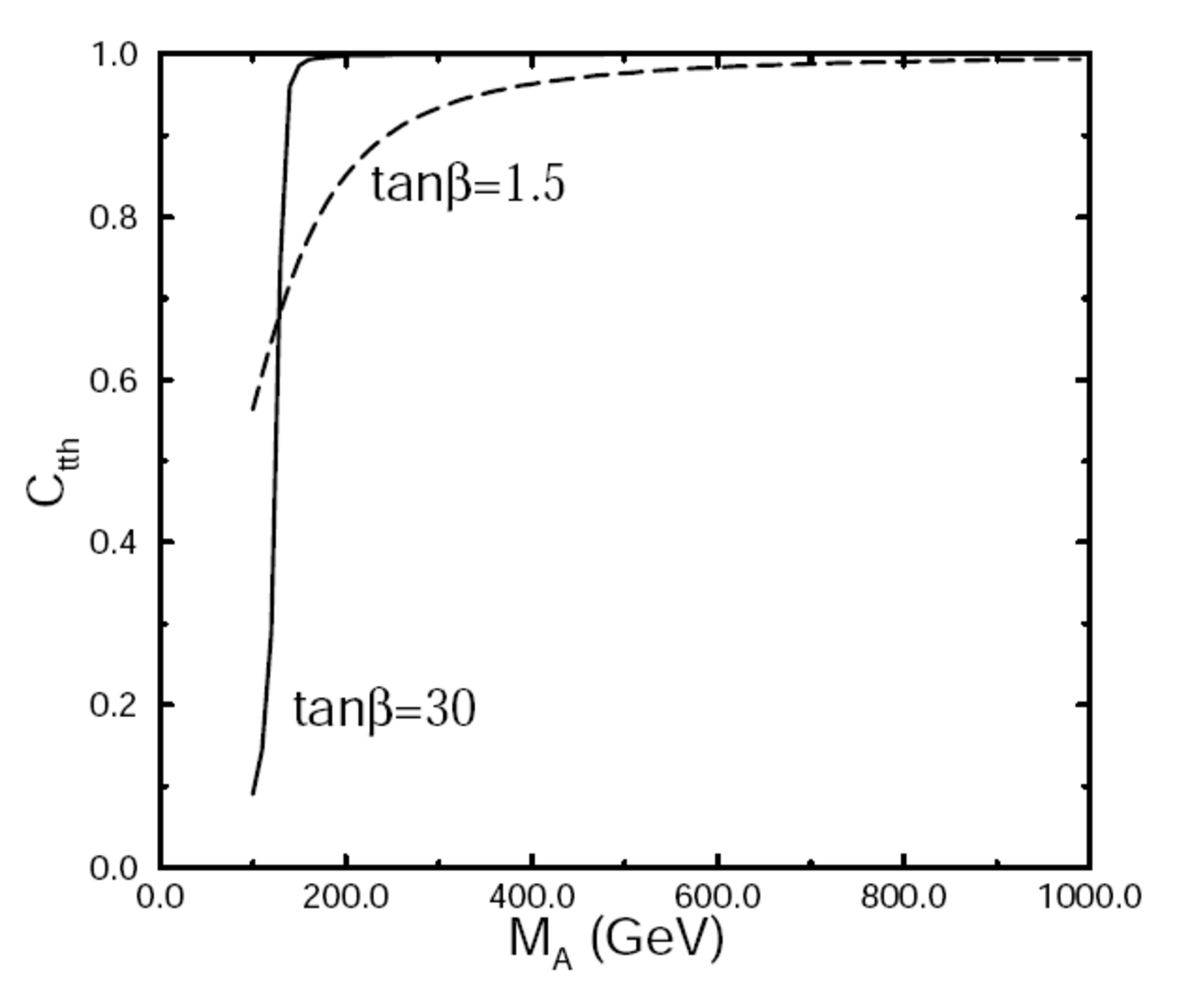}

(Plots taken from \cite{Dawson:1997tz})
\EC
As the Higgs boson masses, all the couplings above talso ge corrected by loops when going beyond the tree level and in some regions of the MSSM parameter space these radiative corrections can be sizable.

\blackl{The decoupling limit} 

The so-called decoupling limit corresponds to taking large values of the input mass $M_A$,
$\MA >> \MZ$, and it has the interesting peculiarity of getting the MSSM Higgs sector converging to the SM Higgs sector. More specifically, the lightest MSSM Higgs boson behaves as the SM Higgs boson, whereas the other MSSM Higgs bosons get heavy, close to $M_A$, and they decouple from the low energy physics. 
In particular, one can check that for $\MA >> \MZ$ the following limits are obtained:\\[1em]
$m_h^{{\rm tree}} \rightarrow \MZ |\cos 2 \beta |$ \\[1em]
$-\frac{\sin\ \alpha}{\cos \beta}\rightarrow 1 $,
$\frac{\cos\ \alpha}{\sin \beta}\rightarrow 1 $,
$\sin(\beta-\alpha)\rightarrow 1$
$\Rightarrow$ $g_{hVV} \rightarrow g_{HVV}^{\rm SM}$,
$g_{h f\bar f} \rightarrow g_{H f\bar f}^{\rm SM}$\\[1em]
{$\MA \approx \MH \approx \MHp$}\\[1em]
In fact, as can be seen in the plot below, this decoupling limit is 
already effective at {$\MA \gsim 150 \gev$:}\\[1em]
\BC
\includegraphics[width=9cm,angle=0]{MAmh01.cl.eps}

(Plot generated with FeynHiggs2.2)
\EC
And one can conclude that the SM is the resulting low energy effective theory of the MSSM once the heavy Higgs bosons (and the heavy SUSY partners) are decoupled.

\subsection{Another interesting properties of SUSY}
Among the most interesting properties of SUSY theories that are specially relevant for phenomenology, there are the following: 1) Coupling constant unification, 2) Radiative Electroweak Symmetry Breaking and 3) R-parity.\\[1em]
\blackl{Coupling constant unification}\\[0.5em]

Let us consider the running of the three coupling constants $\alpha_1$,
$\alpha_2$ and $\alpha_3$ of the 
$SU(3)_C \times SU(2)_L \times U(1)_Y$ gauge theory. This running is dictated by the Renormalization Group Equations (RGE's) that connect parameters at different energy scales and by the particle content of the theory. In particular, if we use the RGE's to evolve the gauge coupling constants from the electroweak (EW) scale to the Gran Unification Theory (GUT) scale,  
$$\al_i({Q_{\rm EW}}) {\to} \al_i({Q_{\rm GUT}})$$
one gets different answers depending if one uses just the SM or the SUSY enlarged models. For instance, comparing the SM with the MSSM, one gets that the three coupling constants do not meet in the SM case, whereas they do unify in the MSSM at $Q=Q_{\rm GUT} \simeq 10^{16} \gev$. This unification is illustrated in the next plot where the inverse of the coupling constants are shown as  functions of $\log Q$, with $Q$ expressed in GeV. Again, this is under the assumption that the soft SUSY breaking masses are not far above the TeV scale, 
$M_{\rm SUSY} \lsim 1 \tev$.    
\BC   
\includegraphics[width=16.5cm]{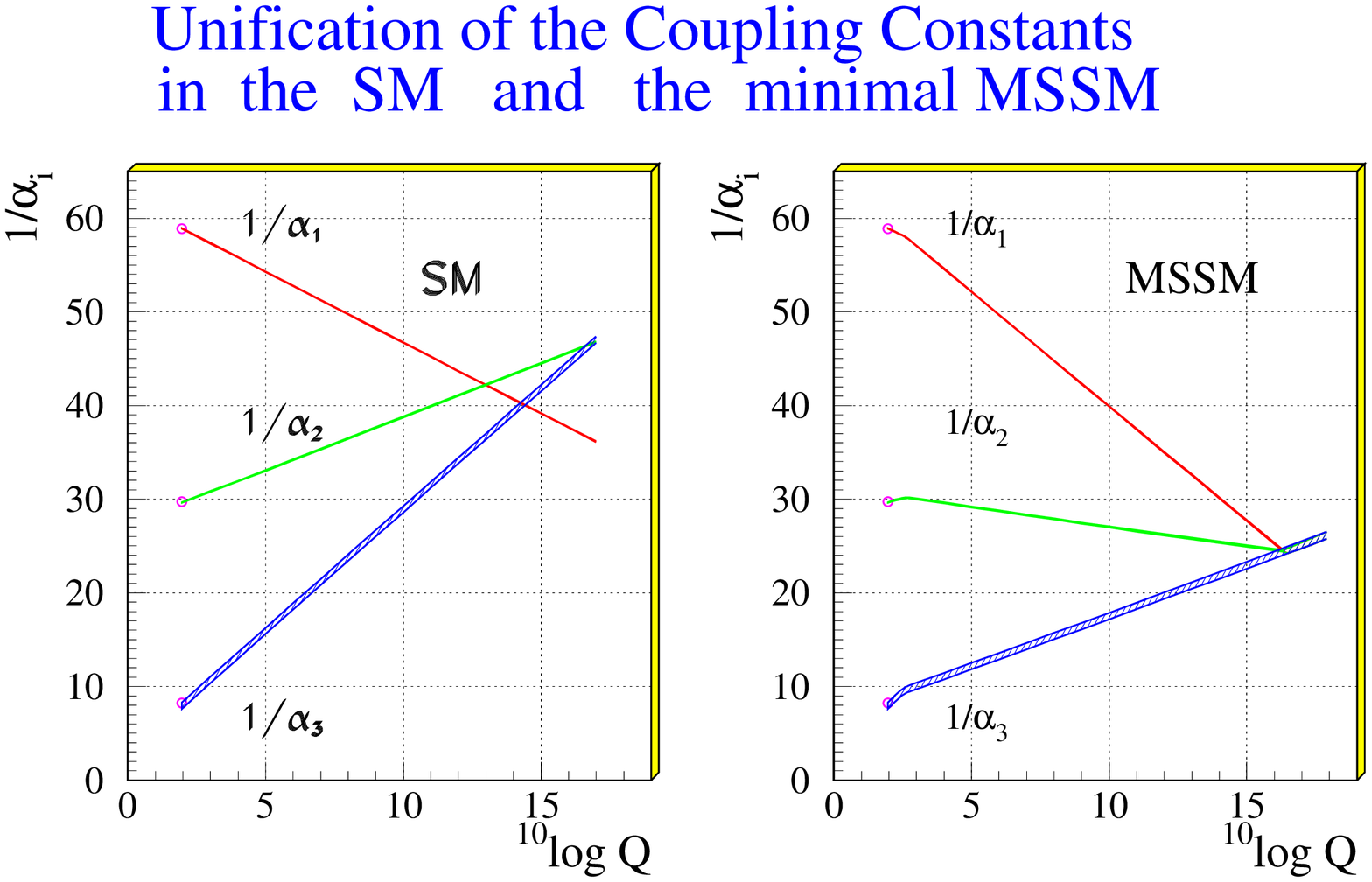}

\vspace{-0.7cm}
\EC
 

\blackl{Electroweak Radiative Symmetry Breaking}\\[-0.5cm]

\BC 
\includegraphics[width=10cm]{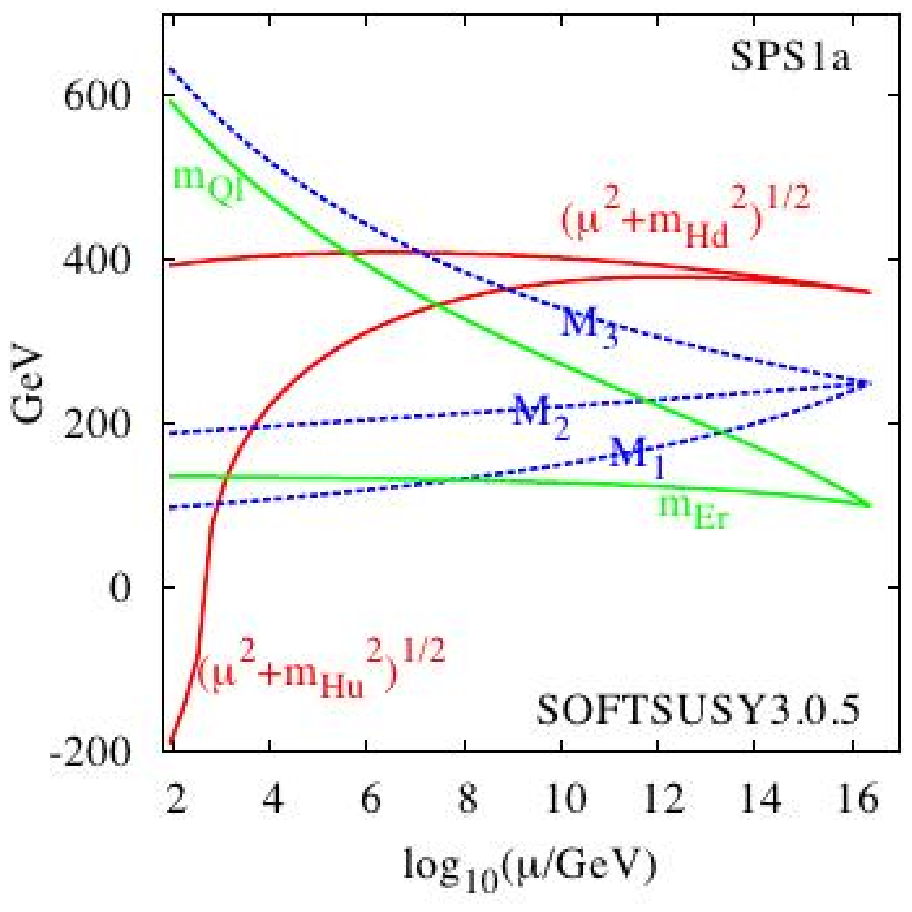}\\
(Plot generated with SOFTSUSY3.0.5)
\EC 

\vspace{-0.1cm}
In SUSY theories, one can get the Electroweak Symmetry Breaking by means of radiative corrections that produce the needed negative squared scalar mass, as required by the Higgs Mechanism. This can be clearly illustrated by means of the running of the relevant parameters from the high energy scale down to the EW scale as provided by the RGE's. For instance, if one starts with a set of universal conditions at the GUT scale for the soft masses and run them down to the EW scale, one finds out that one scalar mass parameter gets negative, and it is precisely the needed one to produce the wanted $SU(2)_L \times U(1)_Y \to U(1)_{\rm em}$ breaking. This is illustrated in the included plot where the $(\mu^2+m^2_{H_u})^{1/2}$ mass runs into negative values. And it turns out that it only works properly if again $M_{\rm SUSY} \lsim 1 \tev$.\\[0.6cm] 
\blackl{R-parity}\\[0.6cm]
The MSSM and other SUSY models have an extra symmetry, called the ``R-parity'' that implies the conservation of a new multiplicative quantum number defined for each particle as, \\[.5em]
$$P_R=(-1)^{3(B-L)+2s},$$
where $B$, $L$ and $s$ are the baryon number, the lepton number and the spin of the particle respectively.  
It happens that all SM-particles and Higgs bosons have even R-parity, $P_R = +1$, whereas, 
all superpartners have odd R-parity, $P_R = -1$. 

This symmetry has  very important consequences for phenomenology. 
First, the SUSY particles appear only in pairs. For instance, the production of neutralinos in electron-positron collisions occur in pairs, like, e.g.:
 $e^+ e^- \to \tchi^+_1 \tchi^-_1$.
 
It also has very important consequences for Dark Matter Physics, since it provides a natural particle candidate for explaining the Dark Matter: the lightest SUSY particle (LSP) that, due R-parity, is stable. 
 More specifically, the lightest neutralino is usually the most popular 
 candidate for (Cold) Dark Matter in the MSSM and other SUSY models. 
 
Another interesting and very relevant consequence of R-parity for collider phenomenology is that, since the LSP is neutral and uncolored, it leaves no traces in collider 
detectors and, therefore, the typical SUSY signatures are events with unbalanced energy, i.e with apparent ``missing energy''. 

Another proposals as: the gravitino, the axino and others have also been considered as Dark Matter candidates in the literature. \\[.3em]
 
\section{Compositness}
Another different avenue is to think of the Higgs boson 
as a composite particle instead of a fundamental one which was the avenue in the previous SUSY models. The hypothesis of a composite Higgs is perfectly compatible with present data, therefore why not to consider it. Another important qualitative difference with respect to the previous SUSY case is that typically the interactions leading to a composite state must be strong. So the composite hypothesis for the Higgs system leads generically to new strong interactions beyond the Standard Model, in constrast to the SUSY hypothesis that does not introduce generically new interactions but instead leads to new fundamental particles.

Once one assumes a composite Higgs, then one has to make a particular assumption on what is the strong dynamics that is responsible for the production of such a composite state if one is interested in studying further phenomenological consequences of its existence in Nature. This implies to set a specific underlying Quantum Field Theory describing these new strong interactions beyond the Standard Model and, therefore, the predictions will be obviously model dependent. There are many of such strongly interacting models and we will not go through all of them. However, one can learn some interesting general features by just looking for singularities with some well known physical examples where composite particles appear as a result of strong dynamics. First we know that in a theory with strong dynamics it is common to generate a populated spectra with a dense tower of resonances built from the more fundamental objects that interact strongly. The mass of these resonances are in general related to the typical scale where the strong dynamics condensates,
$\Lambda_{\rm strong}$. Apart from resonances there are another interesting composite systems that appear in strong interacting theories having a global symmetry that is spontaneously broken. According to the Goldstone Theorem, in these cases there may appear in the spectrum new scalar or pseudoscalar composite states, the Goldstone Bosons, that should be massless if the global symmetry of the Lagrangian was exact, or they may get a small mass in the case that this is not exact but an approximate global symmetry. The small breaking in this later case is usually accounted for by some small explicit breaking masses in the Lagrangian, given generically by a new scale, $\Lambda_{\rm break}$ that may be or not related with $\Lambda_{\rm strong}$.  Thus, under the generic hypothesis of the Higgs boson being a composite/dynamical state of a strongly interacting gauge theory, still there are two generic possibilities. Either it emerges as a resonance, and its mass is related to the condensation scale, $\Lambda_{\rm strong}$, or it is a Pseudo-Goldstone boson and its mass emerges as a consequence of the 'small' breaking of the global symmetry, therefore it is related with $\Lambda_{\rm break}$.    
In the following we will look into particular examples that are illustrative of what could happen in case that the Higgs boson is a composite particle. The most popular example is to think that the Higgs system of the SM relies in a strongly theory that is a copy of QCD, but operating at higher energies.\\[1em]

\subsection{Electroweak Chiral Symmetry and Composite Higgs} 
In order to understand the possible similarities of the Higgs system with QCD, it is more convenient to use the alternative parametrization of the SM Higgs field that were commented in a previous section of these lectures.
 
Let us again consider here this alternative way of writing the (ungauged) Lagrangian for the Symmetry Breaking Sector of the SM, given by:
\begin{eqnarray}
{\cal L}_{\rm SBS}&=&\frac{1}{4} {\rm Tr}
\left [ (\partial_{\mu}M)^\dagger(\partial^{\mu}M)\right ] -V(M)\;; \non \\
V(M)&=&\frac{1}{4}\lambda \left [\frac{1}{2} {\rm Tr}(M^\dagger
M)+\frac{\mu^2}{\lambda}\right ]^2 \non
\end{eqnarray}
where $M$ is a $2\times 2$ matrix containing the four real scalar
fields of the doublet $\Phi$:
\begin{eqnarray}
M &\equiv&
\sqrt{2}(\widetilde{\Phi}\Phi)=\sqrt{2}\left(\begin{array}{ll}
\phi_0^*&\phi^{\dagger}\\-\phi^-& \phi_0\end{array}\right)
\;;\; \nonumber  \\[-0.5em]
\Phi & = & \left(
\begin{array}{c}\phi^+ \\
 \phi_0 \end{array}\right)\;;\;    
\widetilde{\Phi}=i\tau_2\Phi^*=\left(\begin{array}{c}
\phi_0^*\\-\phi^-\end{array}\right)\nonumber
\end{eqnarray}
Written in this way it is easier to see the existence of an extra global symmetry of this SBS Lagrangian. More specifically, ${\cal L}_{\rm SBS}$ is invariant under the global and 
independent transformations $g_L$ and $g_R$ acting on $M$:
\[M\rightarrow g_L M g_R^+ \;;\;\;g_L\subset SU(2)_L\;;\;\;g_R\subset SU(2)_R\]
This global symmetry $SU(2)_L \times SU(2)_R$ is called {\it Electroweak Chiral Symmetry} due to the obvious analogy with the well known Chiral Symmetry of QCD. 
One can also check that this Electroweak Chiral Symmetry is spontaneously broken down to the diagonal subgroup, usually called the custodial symmetry group,
{$SU(2)_{L+R} \equiv SU(2)_{\rm custodial}$}.\\
The pattern of the global symmetry breaking in the SM Higgs system is therefore:
$$SU(2)_L \times SU(2)_R \to SU(2)_{\rm custodial}$$
This breaking, according to the Goldstone Theorem, would give rise to the appearance in the spectrum of three massles Goldstone Bosons, which in this case are pseudo scalars  (i.e. with negative parity) since it is the $SU(2)_{L-R}$ symmetry the one that is not a symmetry of the vacuum. Once the gauge interactions are incorporated as usual,  by the gauge principle associated to the $SU(2)_L \times U(1)_Y$ gauge symmetry, then these three GBs dissapear from the spectrum and the three corresponding gauge bosons, $W^{\pm}$ and $Z$, get the proper masses. On the other hand, within the full SM, there are symmetry breaking terms in the Lagrangian that indeed break the two global symmetries $SU(2)_L \times SU(2)_R$ and $SU(2)_{\rm custodial}$, and this is why the associated GBs of the EW Chiral symmetry breaking are not strictly massless and they are usually referred to as pseudo-GBs with a 'small' associated mass.    

 Some immediate possibilities arise for a composite Higgs from the above way of thinking:  It could  be that the Higgs is a
scalar resonance emerging from some new strong interactions among new fermions, as it happens in
QCD where many resonances (like $\sigma$, $\rho$, etc) emerge from the strong
interactions among quarks. Or it could happen that the Higgs particle emerges as a pseudo-Goldstone boson associated to a spontaneous symmetry breaking of a larger symmetry group, containing the previous 'minimal breaking pattern' given by $SU(2)_L \times SU(2)_R \to SU(2)_{\rm custodial}$.\\[1.5em] 
\subsection{Chiral Symmetry and the Chiral Lagrangian of QCD} 
 We have seen in the previous lecture that in QCD with two massless quarks, $u$ and $d$, there is an extra global symmetry, the Chiral Symmetry of QCD,  that is spontaneously broken down to the isospin group: 
 $$SU(2)_L \times SU(2)_R \to SU(2)_V,$$
 by the non-vanishing quark condensate, $<0|\bar{q} q|0> \neq 0$,
 and this breaking explains the smallness of the pion masses as compared to the typical resonance masses of QCD. The three pions, $\pi^{\pm}$ and $\pi^0$ are the three pseudo-GBs associated to this breaking whereas the other hadrons emerge in the QCD spectrum as resonances made up by quarks and antiquarks. In this case, the masses of the pions are associated to the scale of the Chiral Symmetry breaking in QCD, $m_\pi \sim \Lambda^{\rm QCD}_{\rm break}$ whereas the masses of the other hadrons are related to the scale of the QCD strong interactions, $\Lambda^{\rm QCD}_{\rm strong}$ (usually called $\Lambda_{\rm QCD}$ in short). If these two scales of QCD are or are not related is still under debate and we will not go further into this point here.
      
  Regarding the dynamics of pions in  QCD to low energies, it is well described by an
effective Lagrangian, the so-called Chiral Lagrangian of QCD, that is invariant under the Chiral Symmetry. The Effective Quantum Field Theory that is built from this Chiral Symmetry is called Chiral Perturbation Theory (ChPT).

  The Chiral Lagrangian of QCD is usually written in terms of a {non-linear representation of the GBs}:
$$U(x)=\exp \left(\frac{i}{f_\pi}\pi_a(x)\sigma^a \right)\,\,\,
\mbox{with}\,\,\,\sigma^a (a=1,2,3)=\mbox{Pauli matrices},$$
where $f_\pi$ is the pion decay constant that is measured, for instance, from the $\pi^+ \to \mu^+ \nu_\mu$ decay:\\[-.1em]
$$<0|J^{+\mu}|\pi^-(p)>= \frac{if_\pi}{\sqrt{2}}p^\mu\,\,,\,\,\,f_\pi=94\,{\rm MeV}$$
Under a chiral transformation the $U(x)$ transforms linearly (but $\pi$ transform non-linearly):
$$U(x)\rightarrow g_LU(x)g^+_R\,\,\,
\mbox{with}\,\,\,g_L\in SU(2)_L\,\,\,,\,\,\,g_R\in SU(2)_R$$
 
The most general chiral invariant Lagrangian is a sum of an
infinite number of
terms with increasing number of derivatives in the $U(x)$
and the $U^+(x)$ fields and with an infinite number of arbitrary
parameters. This provides a systematic expansion in powers of momenta and also in powers of $m_\pi$ if the explicit Chiral symmetry breaking terms are included into the Lagrangian. 

Thus in ChPT to lowest order, ${\cal O}(p^2)$, and by neglecting the explicit chiral symmetry breaking terms, there is just one term in the Chiral Lagrangian, given by:\\[.7em]
$${\cal L}_0=\frac{f_\pi^2}{4}Tr(\partial_{\mu}U\partial^{\mu}U^+)\,\,\,,$$
and from this one gets the well known expressions for the the pion-pion scattering amplitudes, which were refered to previously as the Low Energy Theorems (LET's) of QCD:
$$  T(\pi^+\pi^-\rightarrow \pi^+\pi^-)=-\frac{u}{f_{\pi}^2}\,\,\,,\,\,\,
T(\pi^+\pi^-\rightarrow \pi^0\pi^0)=\frac{s}{f_{\pi}^2}\,\,\,.$$ 

Going to next to leading order in ChPT (we are still ignoring the explicit chiral symmetry breaking terms to make this presentation here as simpler as possible) amounts to consider the next order Lagrangian ${\cal L}_1$  with  all the possible chiral invariant terms with four derivatives, i.e. ${\cal O}(p^4)$ terms, with the corresponding chiral parameters in front, usually called $L_i$ parameters in QCD. Some of these terms are, for instance:
 \begin{equation} 
{\cal L}_1=  
L_1 \left[ Tr\left( (\partial_\mu U) U^\dagger(\partial^\mu U) U^\dagger \right)\right]^2 +
L_2  \left[ Tr\left( (\partial_\mu U) U^\dagger (\partial_\nu U) U^\dagger \right)\right]^2+...   \dots
\non 
\end{equation}
In practical terms, and in order to perform a one-loop computation of a given observable with the Chiral Lagrangian (CL) of QCD,  
$$ {\cal L}_{\rm CL} =  {\cal L}_0+{\cal L}_1+...,$$
 by means of Feynman diagrams, one has to compute all the contributing tree-level  diagrams from ${\cal L}_0$ and from 
${\cal L}_1$, add the one-loop contributing diagrams generated with the Feynman rules of ${\cal L}_0$, and finally perform the renormalization of all the entering chiral parameters $L_i$.
  All together provide well defined predictions for low energy observables  like, for instance, the pion-pion scattering amplitudes, which are given in terms of a finite set of renormalized chiral parameters, usually called $L^r_i$. These predictions are usually compared with data, and from this comparison one gets the preferred by data values for these $L^r_i$. This procedure can be done to higher orders in ChPT, i.e to ${\cal O}(p^6)$ with two-loop contributions included, etc, and it can also be generalized to include all the proper  explicit chiral symmetry breaking terms into the Lagrangian.   The typical dimensionless parameter of the low energy expansion provided by ChPT, once loop contributions are incorporated, is given by $p/(4\pi f_\pi)$. Therefore, one expects a good convergence of this expansion for energies
well below $4\pi f_\pi \sim 1200$ MeV. ChPT has been checked to work pretty well in comparison with data for many years.
 
In addition, by going to higher orders in ChPT, i.e ${\cal O}(p^4)$ and above, and using either unitarization methods or dispersion relations, the emerging {resonances} can also be implemented. These are seen as resonant peaks in $\pi \pi \to \pi \pi$ scattering.
For instance, the $\rho$ vector meson appears clearly in the phase shift 
$\delta_{IJ}$ for $I=J=1$.  The next figures illustrate $\delta_{11}(\sqrt{s})$ as a function of the energy $\sqrt{s}$ and we see that $\delta_{11} \simeq 90^o$ when 
$\sqrt{s}=775\,\,{\rm MeV}$, signaling clearly the emergent $\rho$ resonance with mass at $m_\rho=775\,\,{\rm MeV}$. \\ 
\begin{minipage}[h]{9cm}
\includegraphics[width=8cm]{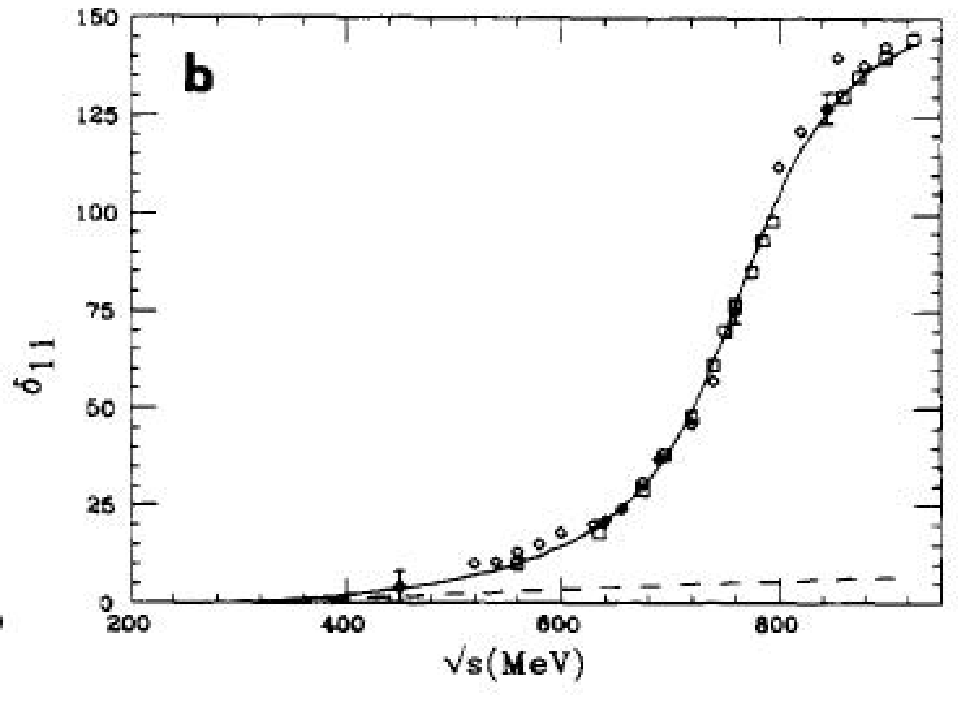} \\[-1cm] 
\BC
(Plot taken from \cite{Dobado:1989qm})
\EC
\end{minipage}
\begin{minipage}[h]{9cm}
\includegraphics[width=8cm]{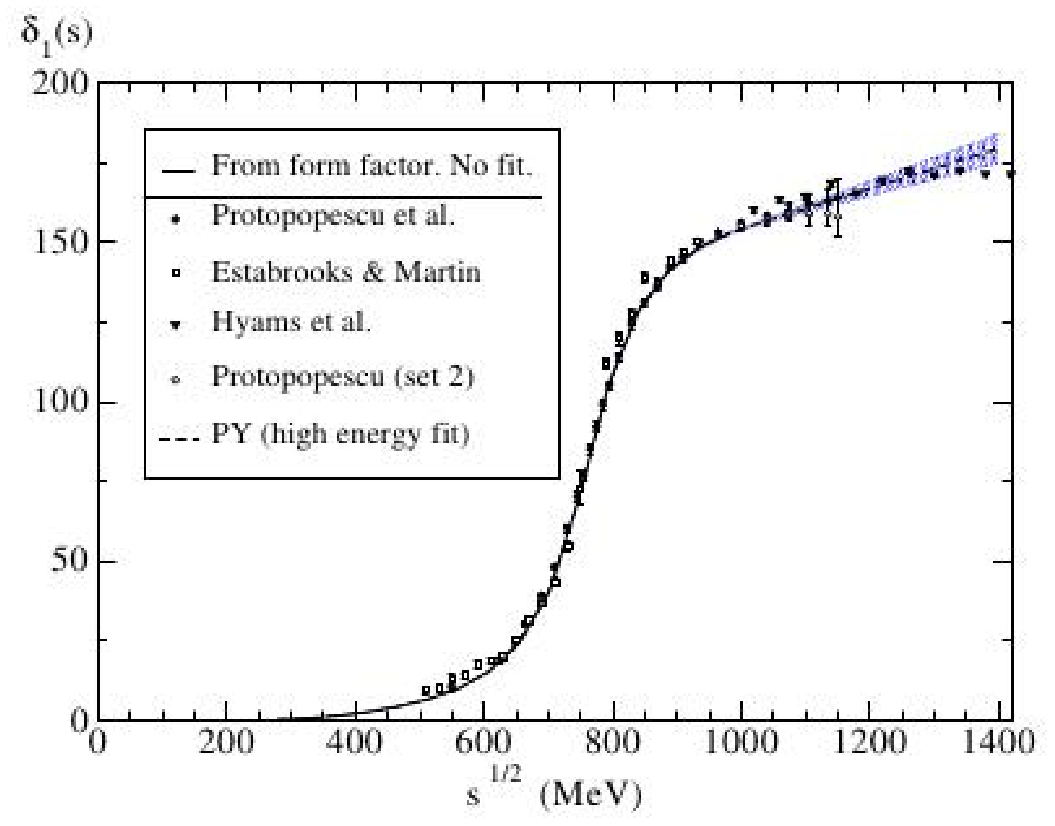} \\[-1cm] 
\BC
(Plot taken from \cite{Pelaez:2004vs})
\EC
\end{minipage} 
 
\subsection{From QCD to Technicolor Theories} 
The most popular examples of strongly interacting theories that can provide the needed framework for a composite Higgs boson are the Technicolor Models. Generically, these models 
assume a new $SU(N_{TC})$ gauge symmetry describing the dynamics of new strong interactions in analogy to the usual $SU(3)_C$ gauge interactions of QCD. Also following the guide of QCD, Technicolor assumes the existence of new fermionic constituents of matter, and new gauge bosons that are the intermediate bosons of the new strong interactions among these new constituents.  By analogy to QCD these are named respectively as\\[0.7em]
$\star$ {New constituents}: Techniquarks $q_{TC}$ \\[0.7em]
$\star$ {New gauge bosons}: Technigluons $g_{TC}$ \\[0.7em]
And the number of Technicolors is given by $N_{TC}$. 

Regarding the global symmetries, the simplest Technicolor Models also assume that the Lagrangian has the global chiral symmetry of the Electroweak Theory, i.e. the Electroweak Chiral Symmetry,  and that it is broken spontaneously to the custodial symmetry group by the techniquark condensate:\\ 
$$<0|\bar{q}_{TC}q_{TC}|0> \neq 0 \RA SU(2)_L \times SU(2)_R \to  SU(2)_{L+R}.$$ 
Then, following again the guide of QCD,  the 3 resulting Goldstone bosons are identified with
the 3 Technipions: $\pi_{TC}^\pm$ and $\pi_{TC}^0$. 
 When the subgroup $SU(2)_L \times U(1)_Y$ is gauged: the three GBs, $\pi_{TC}^\pm$ and 
 $\pi_{TC}^0$, dissapear from the spectrum and they are replaced by the longitudinal gauge bosons, $W^{\pm}_L$, $Z_L$. The EW bosons then get the proper mass by means of the Higgs Mechanism, but in this case without the appearance in the spectrum of an elementary Higgs boson. Notice that preserving the degrees of freedom here does not require the introduction of any extra scalars other than the needed three GBs. Notice also that, as in the SM, the gauging of the EW symmetry breaks explicitly the global EW Chiral symmetry and therefore the would-be GBs are not massless but they acquire masses ($M_W$ and $M_Z$)  that are typically smaller than the masses of the other resonances or composite particles in these Technicolor Models that typically appear at or above 1 TeV.

Regarding the dynamics of these technipions to low energies, one also follows the guide of QCD and uses effective Lagrangians that are based in the Electroweak Chiral Symmetry but applied to the technipions case. Similarly, 
 the coupling of the technipions to the weak
current (in analogy to $f_\pi$) is given by:
\begin{equation}
<0|J^{+\mu}_L|\pi^-_{TC}(p)>=\frac{iF_{\pi}^{TC}}{\sqrt{2}}p^{\mu} \,\,\,
\mbox{with}\,\,\,F_{\pi}^{TC}=v=246\;GeV
\end{equation}
The spectrum of $SU(N_{TC})$ appear as a replica of the QCD spectrum but with all the masses shifted upwards. Thus, there are
Technipions $(\pi_{TC}^{\pm},\pi_{TC}^0)$, Technirhos
$(\rho_{TC}^{\pm}, \rho_{TC}^0)$, Techniomegas, Technietas, etc.. 

By using large $N$ techniques one can re-scale QCD quantities to the 
Technicolor ones. For instance, the ratio between the technimeson mass and the meson mass can be estimated as:
\begin{equation}
\frac{m_{\rm Tmeson}}{m_{\rm meson}} \sim
\frac{F_{\pi}^{TC}}{f_{\pi}}\cdot \sqrt{\frac{N_C}{N_{TC}}}\,\,\, \mbox{with} \,\,\,\frac{F_{\pi}^{TC}}{f_{\pi}}=\frac{246\;GeV}{0.094\;GeV}\sim 2700
\non
\end{equation}
Thus, the first expected resonance is the technirho with a mass and a total width given by:
\begin{equation}
m_{\rho_{TC}}=
\frac{F_{\pi}^{TC}}{f_{\pi}} \sqrt{\frac{N_C}{N_{TC}}}
m_{\rho},\non
\end{equation}
\begin{equation}
\Gamma_{\rho_{TC}}=\frac{N_C}{N_{TC}}
\frac{m_{\rho_{TC}}}{m_{\rho}}\Gamma_{\rho}.\non
\end{equation}
For example, taking $N_C=3$, $N_{TC}=4$, $m_{\rho}=760\;MeV$,  and $\Gamma_{\rho}=151\;MeV$ lead to:
$ m_{\rho_{TC}}=1.8\;TeV$ and $\Gamma_{\rho_{TC}}= 260\; GeV$. 

In Technicolor Theories the Higgs particle does not appear as an elementary particle but as a composite scalar resonance. Therefore, as any other resonance of Technicolor, the mass of the Higgs boson resonance should be at the $O(1\;TeV)$ energy scale. Correspondingly, the effective
cut-off of Technicolor Theory where the new physics sets in is at:
\[ \Lambda_{TC}^{\rm eff} \sim O(1\;TeV) \] 
and therefore there is not hierarchy problem in Technicolor Theories. However, this avenue to solve the hierarchy problem having the simplest implementation of a composite Higgs boson as an emerging  resonance in Technicolor Models at $O(1\;TeV)$ is not anymore aceptable, if one identifies this boson with the   recently discovered Higgs particle that has a relative low mass at $m_H=125.6$ GeV. There could be however, different implementations of the leading ideas of Technicolor Models that could lead to more compatible with data predictions, even though we do not have yet a quite satisfactory Theory of Technicolor.

Following the sensibilities with QCD,  the resonances of Technicolor would then appear in $V_LV_L$ scattering ($V=W,Z$), as the $\rho$ of QCD appears in $\pi\pi$ scattering.  
In order to describe these scattering processes and other interesting observables that could be measured it is common to use the effective Lagrangians technique. In particular, Technicolor and other Strongly Interacting theories of EWSB can be described generically with effective Electroweak Chiral Lagrangians and with the equivalent to ChPT in this other context.\\[0.7em]
\blackl{Present bounds on Technicolor}\\[0.7em]
Although the general motivation for Techicolor Theories is very appealing, it turns out that these theories are very much constrained by past and present data.

 On one hand, Technicolor models when connecting quarks with  techniquarks tend to produce too much Flavor Changing Neutral Currents (FCNC). The absence of FCNC in data sets very strong constraints on these models. However,  these bounds are very model dependent. 
 
On the other hand, the Electroweak Precision Observables (EWPO) also set very restrictive bounds in these and other models. In particular, the so-called oblique parameters $S$ and $T$ (Peskin, Takeuchi,1990) that measure possible deviations from the SM predictions due to new physics in self-energies $\Pi_{XY}$ of  EW gauge bosons are the most constraining ones. These are defined by: 
$$\hat{\alpha}(M_Z) T \equiv \frac{\Pi^{\rm new}_{WW}(0)}{M_W^2}- 
\frac{\Pi^{\rm new}_{ZZ}(0)}{M_Z^2},$$
$$\frac{\hat{\alpha}(M_Z)}{4 {\hat s}^2_Z {\hat c}^2_Z}  S \equiv  
\frac{\Pi^{\rm new}_{ZZ}(M_Z^2)-\Pi^{\rm new}_{ZZ}(0)}{M_Z^2}. $$
 Indeed, the present experimental bounds on $S$ and $T$ exclude already many Technicolor models, as can be seen in the next plot.\\[1em]

\begin{minipage}[h]{10cm}
\BC
\includegraphics[width=9cm]{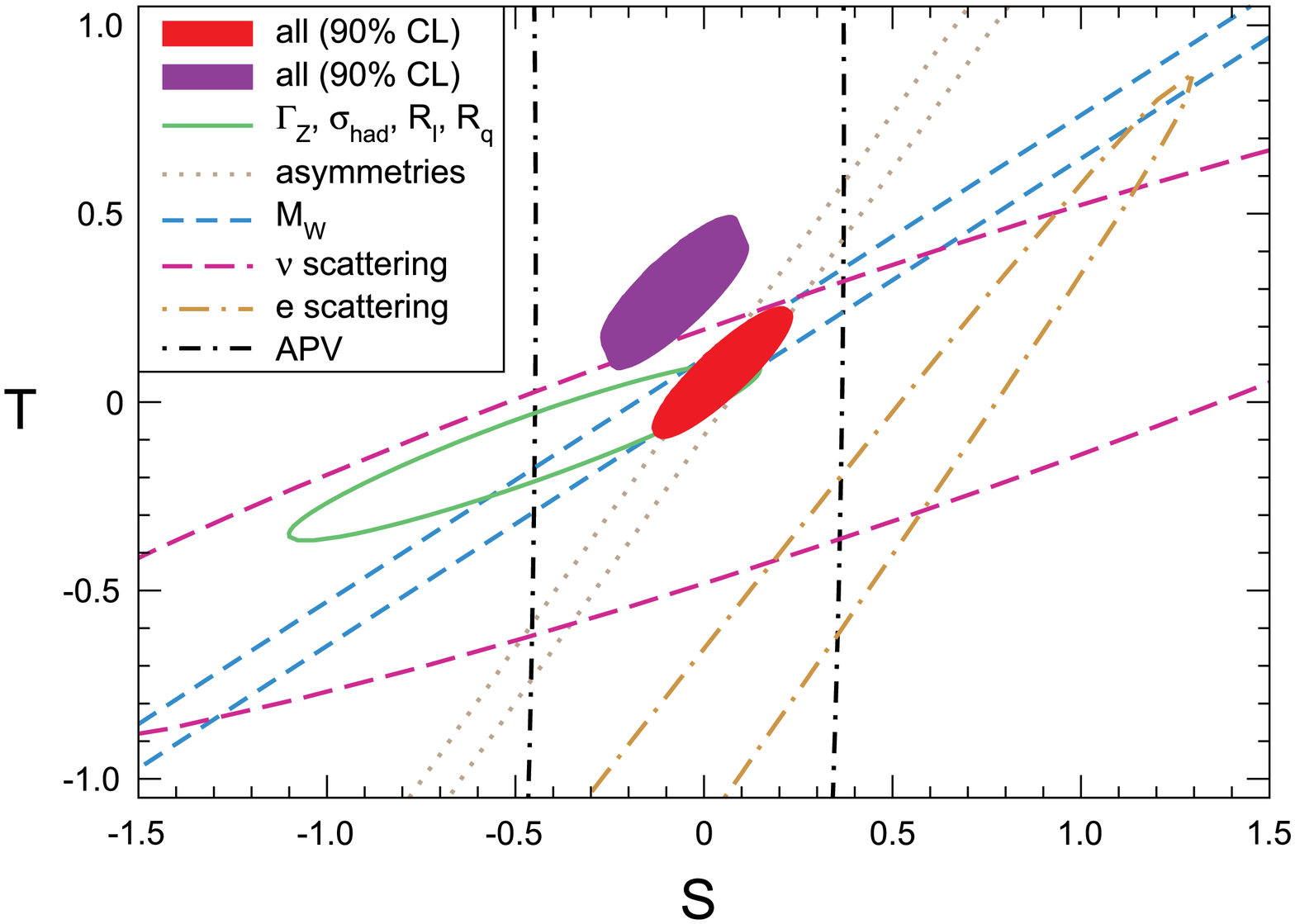}
\EC
\end{minipage}
\begin{minipage}[h]{8cm}
\BC
PDG July 2012~\cite{pdg} \\
1 $\sigma$ $(39.35 \%)$ constraints \\
$S_{\rm exp}=0.04 \pm 0.09$ \\
$T_{\rm exp}=0.07 \pm 0.08$\\
\purple{violet}  $ 600 \gev  < M_H < 1000 \gev$\\
\red{red}  $115.5\gev < M_H < 127 \gev$ 
\EC
\end{minipage}
Particularly, the models that are based on simple scaling from QCD, are already excluded by these data, since each technicolor and each technidoblet contributes to $S$ and one has:  $S_{TC} \propto N_{TC} N_{D}$, for $N_{TC}$ technicolors and $N_{D}$ technidoblets. This leads to 
$S_{TC} \sim 0.45$ for $N_{TC}=4$ and $N_D=1$. If this is compared 
with ($1\sigma$, $39.35\%$): $S_{\rm exp}=0.04\pm 0.09$, we conclude that this simple model is indeed  many sigmas away from data!.   

Finally, one can also get very restrictive bounds from present colliders data. As we can see in the next plots, both experiments at LHC, ATLAS and CMS,   exclude light $\rho_{TC}$ masses from direct searches and from its couplings to standard fermions. Concretely, from the last PDG2012 one gets the exclusion region: $m_{\rho_{TC}} < 260-480 GeV$ (depending on channels).

\vspace{1cm} \hspace{-1cm}
\begin{minipage}[h]{9cm}
\includegraphics[width=9cm]{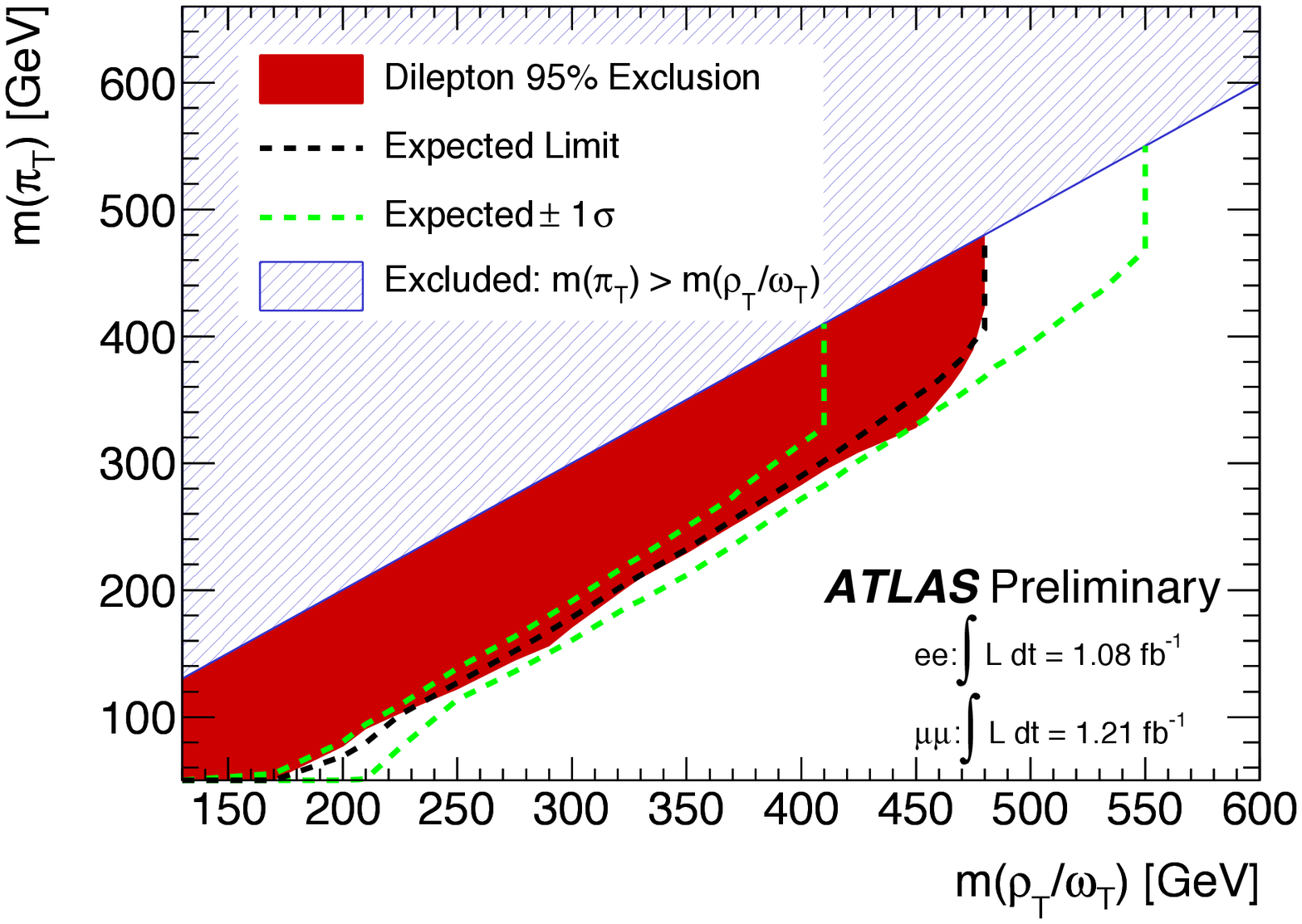} 
\end{minipage}  
\begin{minipage}[h]{9cm}
\includegraphics[width=8cm]{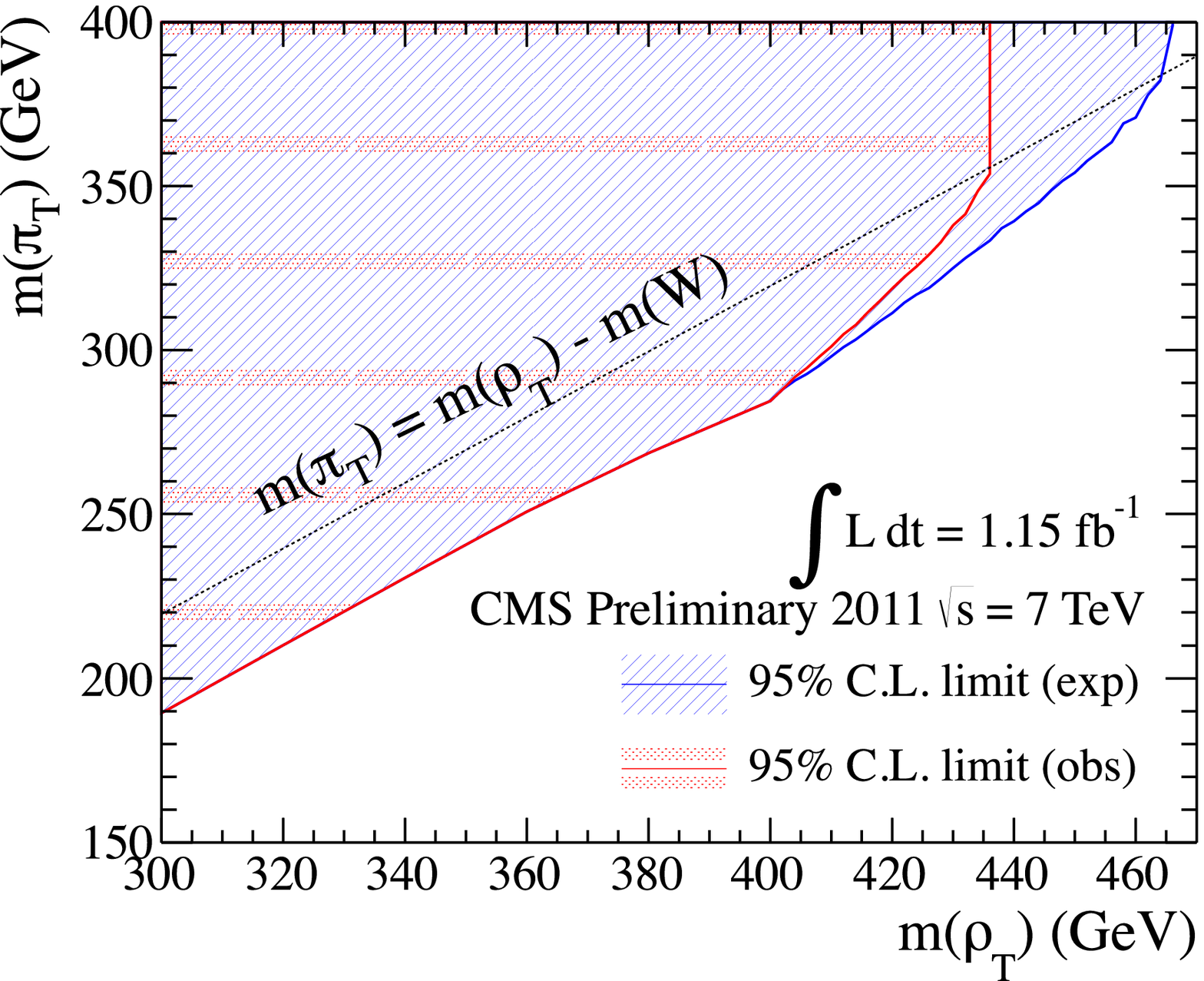}
\end{minipage}\\ 
\BC
(Plots taken from \cite{pdg})
\EC
 
 
\subsection{Composite Higgs in Extra Dimensions} 
Here we comment shortly on the main ideas underlying some proposals for a composite Higgs boson within the context of theoretical models with extra dimensions. Usually the number of total dimensions is chosen to be five.
  
To get a ligth Higgs boson ${\cal O}(100 \gev)$ 
in theories with extra dimensions, one interesting possibility is to assume that the Higgs boson field is the scalar component of a new gauge field living in five dimensions. 
Another interesting feature of these models in extra dimensions is that the mass of the Higgs is then protected by gauge symmetry (this happens in Gauge-Higgs Unification Models). More specifically,  the Higgs mass is zero at the tree level and a non-zero mass value is generated radiatively at one-loop, in a similar way as in the well known Coleman-Weinberg Model.

The connection of a light Higgs boson with the physics of strongly interacting theories comes by means of the famous ${\rm AdS}_5/{\rm CFT}_4$ correspondence that relates weakly coupled theories of gravity (Anti-de Sitter, AdS) in 5 dimensions (5D) with strongly coupled Conformal Field Theories (CFT) in 4 dimensions (4D). More specifically,  
the breaking of the bulk (5D) gauge group by boundary conditions on the InfraRed (IR) brane 
is described in the CFT (4D) as the Spontaneous Symmetry Breaking of a global symmetry $G$ into a subgroup $H$, and this breaking $G \to H$  occurs by the strong dynamics at the TeV scale. 
The Higgs boson in 4D is then identified with one of the associated GBs of this Spontaneous Symmetry Breaking (similar to Little Higgs Models).  

When the 5th dimension $y$ is compactified and the geometry is warped (Randall Sundrum (RS) Models) the small ratio between the infrared brane at the TeV scale  and the Ultraviolet brane at the Planck energy scale, $1 {\rm TeV}/M_{\rm Pl}$, is explained in terms of the exponential
suppression produced  by the 'warp' factor $e^{-ky}$, with $k$ being the ${\rm AdS}_5$ curvature $\sim {\cal O}(M_{\rm Pl})$.\\  
\begin{center}
\includegraphics[width=8cm]{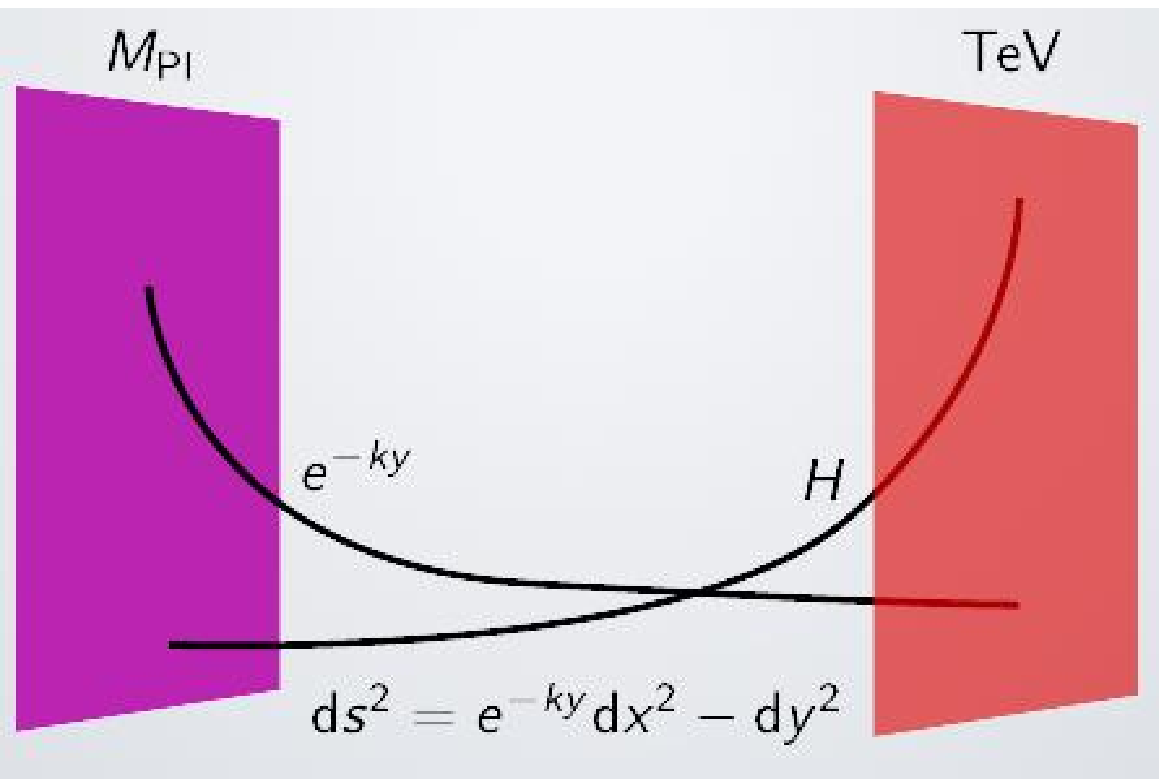}
\end{center}

The main problem of all these models with extra-dimensions is that they are strongly constrained by the EWPO data. It turns out that the typical excitation modes in these models, the so-called
Kaluza-Klein (KK) modes that appear due to the compact extra 5th dimension, contribute dangerously to $S$ and/or $T$ parameters and therefore  very stringent lower mass bounds are found leading to very heavy KK masses. The specific mass bounds are model dependent. In the plot shown next: $m_{KK}> {\cal O}(10-5\,\,\,{\rm TeV})$ for models (a) and (b) with RS metric, and $m_{KK}> {\cal O}(5-2 \,\,\,{\rm TeV})$ for models (c) and (d) with RS-deformed metric. Indeed, these later models are the only ones that allow for a light Higgs having a compatible with data mass value. Generically, the most important restrictions  of extra dimension models come from the breaking of custodial symmetry that is very common in these models. Therefore, usually the most realistic models include an additional symmetry in 5D leading to the needed custodial symmetry protection in 4D.\\[-0.1cm]
\begin{center} 
\includegraphics[width=9cm]{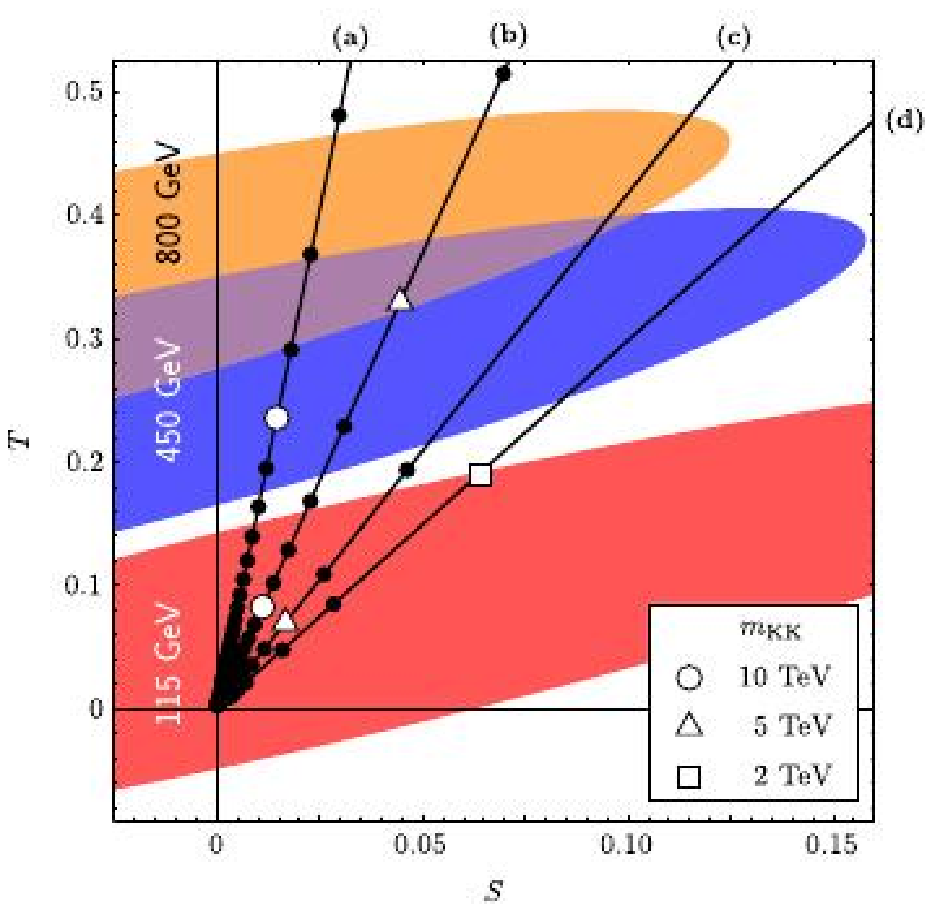} \\[-0.5cm]
(Plot taken from \cite{Cabrer:2011vu})
\end{center}
\section{Electroweak Chiral Lagrangians} 
Electroweak Chiral Lagrangians are based on the existence of the electroweak  chiral symmetry, the accidental symmetry of the symmetry breaking sector of the SM that has been introduced in these lectures before. The building of these  effective field theories rely on specific non-linear implementations of the
global symmetry breaking pattern of the electroweak chiral symmetry down to the custodial symmetry:
$$SU(2)_L \times SU(2)_R \to SU(2)_{\rm custodial}.$$
And non-linear here means that the three associated GBs of these breaking transform non-linearly under the global $SU(2)_L \times SU(2)_R$ transformations.
Once the subgroup $SU(2)_L \times U(1)_Y$ is gauged in order to include the EW gauge interactions by means of the gauge principle, then both the electroweak chiral symmetry and the custodial symmetry are explicitly broken by the hypercharge gauge interactions with coupling $g'$. In addition, when including fermions into the EW Theory, also the mass differences between the two fermion components of the $SU(2)_L$ doublets break explicitly the custodial symmetry. 

There are many models that one can build with Electroweak Chiral Lagrangians. Here we do not review them, but just for illustration of some of their generic properties, we simply classify them into two main cualitative different classes of models:\\[0.2cm]
$\star$ Electroweak Chiral Lagrangians without a light Higgs (ECL) \\[0.2cm]
$\star$ Electroweak Chiral Lagrangians with a light Higgs (ECLh) \\[0.2cm]
Obviously, the first ones are not so much interesting at present, however they were proposed first in the literature (before the discovery of the Higgs boson) and they provided already many of the most relevant features of these Non-linear Effective Theories. Indeed, some of these features are common with the other class of models, therefore, it is worth to remind here some of their most relevant properties.
\subsection{Electroweak Chiral Lagrangians without a light Higgs (ECL)} 
The simplest Electroweak Chiral Lagrangian (ECL) is similar to the lowest order Chiral Lagrangian for QCD, but with the proper gauging for $SU(2)_L\times U(1)_Y$:  
\begin{equation}
{\cal L}^0_{\rm ECL}  =  \frac{v^2}{4}\; Tr\left[ D_\mu U^\dagger D^\mu U \right]
+ {\cal L}_{\rm YM}
\label{NLL} \non
\end{equation} 
 where the first term is usually called the gauged non-linear sigma model Lagrangian and the second one, ${\cal L}_{\rm YM}$, is the usual Yang Mills Lagrangian for the electroweak gauge fields.
 The unitary matrix containing the three GBs of the EWSB, $w^+, w^-, w^0$, and its covariant derivative are usually written as:
\begin{eqnarray}
U & \equiv & \displaystyle{\exp\left( {i \;
\frac{\vec{\tau}\cdot\vec{w}}{v}}\right)},\;\;\;
v  = 246 \;{\rm GeV}, \;\;\; \vec{w} = (w^1,w^2,w^3)
\non \\
D_\mu U & \equiv & \partial_\mu
U +i  \frac{g}{2}\vec{W}_\mu \cdot \vec{\tau}  U -i  \frac{g'}{2}U  B_\mu \;
\tau^3 \non
\end{eqnarray}
where $\tau_i, i=1,2,3$ are the $2 \times 2$ Pauli matrices.

When going to next to leading order, as in the case of QCD, one adds to the previous Lagrangian, ${\cal L}^0_{\rm ECL}$, all possible $SU(2)_L \times U(1)_Y$ gauge invariant terms  with higher dimension (i.e. with dimension 4 in this case).  These new terms must also be invariant under the global $SU(2)_L\times SU(2)_R$ symmetry (and the custodial symmetry) in the same way as in the SM case, that is if the $g'$ and
the mass differences between the two components of the fermion $SU(2)_L$   doublets are set to zero. 
For instance, in the next to leading order Lagrangian, ${\cal L}^1_{\rm ECL}$, one has the following two terms, among others, that are of ${\cal O}(p^4)$ in momentum space:  
\begin{equation}
{\cal L}^1_{\rm ECL}= a_4  \left[ Tr\left( (D_\mu U) U^\dagger (D_\nu U) U^\dagger \right)\right]^2 \,\,+\,\,
 a_5  \left[ Tr\left( (D_\mu U) U^\dagger(D^\mu U) U^\dagger \right)\right]^2 +\dots
\non 
\end{equation}
Notice, that these particular $a_4$ and $a_5$ would correspond to $L_2$ and $L_1$ respectively of the Chiral Lagrangian for low energy QCD.

Then, for a specific computation of an observable to next to leading order, one includes the tree level contributions from ${\cal L}^0_{\rm ECL}$ and ${\cal L}^1_{\rm ECL}$ and adds the one loop contributions using the Feynman Rules of the lowest order Lagrangian ${\cal L}^0_{\rm ECL}$. Then, finally one defines properly the renormalized ECL parameters, $a^r_i$, such that all the predictions at the one-loop level are finite and well defined, following a simmilar effective field theory procedure as in the ChPT of QCD. Previous to the discovery of the Higgs particle at LHC, there were indeed many works in the literature studying phenomenological implications of these kind of Electroweak Chiral Lagrangians and some comparisons with data were also done. By using these effective field theory techniques it is possible, for instance, to make well defined predictions to one-loop level for the scattering amplitudes of longitudinal electroweak gauge bosons in terms of these renormalized $a^r_i$ parameters and then compare the expected events from these processes at high energy colliders, both hadronic ones like LHC etc, and also future $e^+e^-$ colliders like ILC etc for various specific scenarios corresponding to different settings for the numerical values of these $a^r_i$ parameters. The final goal would be obviously to determine which is the underlying fundamental dynamics in the SBS that explains these particular values. 

Besides, and also following similar techniques as in the QCD case, one can incorporate the description of resonances with the ECL.  
It turns out that the two particular previous ECL parameters, $a_4$ and $a_5$, have important consequences for the phenomenology of $V_LV_L$ scattering, in particular for the appearance of a resonant behavior in the scattering amplitudes like, for instance, $W^+_LW^-_L \to W^+_LW^-_L$,  $W^+_LW^-_L \to Z_L Z_L$, and $W^+_L Z_L \to W^+_L Z_L$. In the next plot we include some examples of possible resonances that could show up in the LHC when the initial quarks radiate these longitudinal electroweak gauge bosons and they re-scatter with amplitudes as computed to one-loop level from the previous ECL with some particular
values of $a_4$ and $a_5$ and once the proper dispersion relations (or unitarization requirements) have been implemented. One can see in these plots that several possibilities could arise, and several kinds of resonances, scalar, vector, etc, could be seen at LHC with masses around  ${\cal O}(1\,\, {\rm TeV})$.   
\\[0.7em]
\begin{minipage}[h]{8cm}
\blackl{Some examples of resonances:}\\[1em]
(Plots taken from \cite{Dobado:1999xb})\\
\includegraphics[width=6cm]{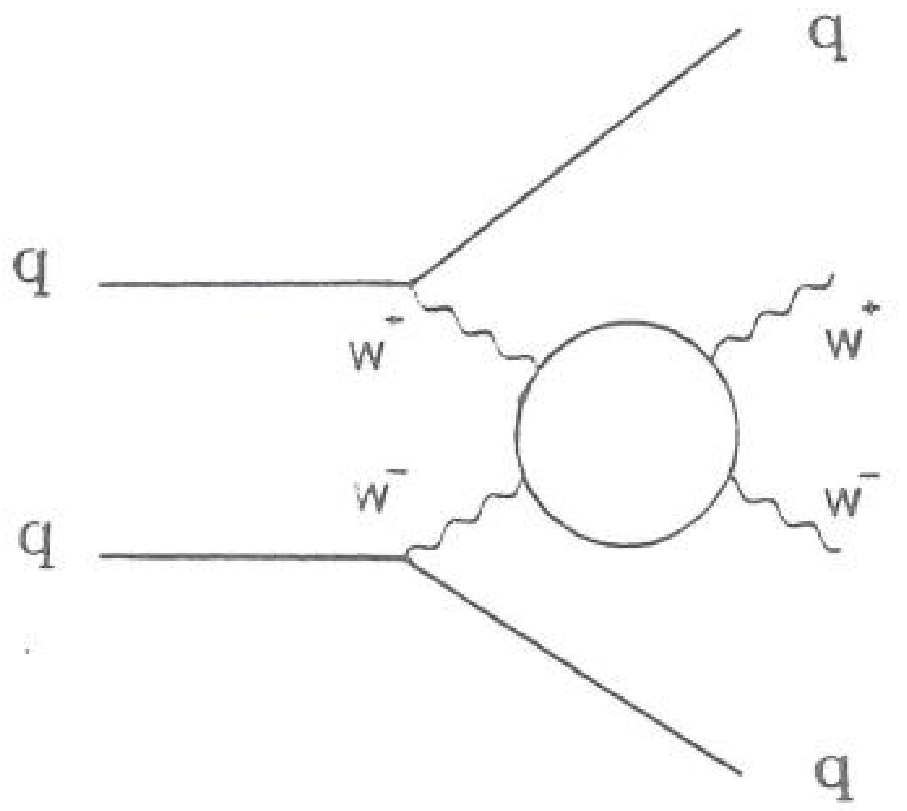} \\
The resonances should show clearly in WW scattering\\
Looking at peaks in invariant mass of WW, WZ, ZZ, pairs\\
Depending on the particular model (i.e. values of $a_i$)
There could be: scalar, vector,...both.  
So far.......not seen any
\end{minipage}  
\begin{minipage}[h]{10cm}
\includegraphics[width=10cm]{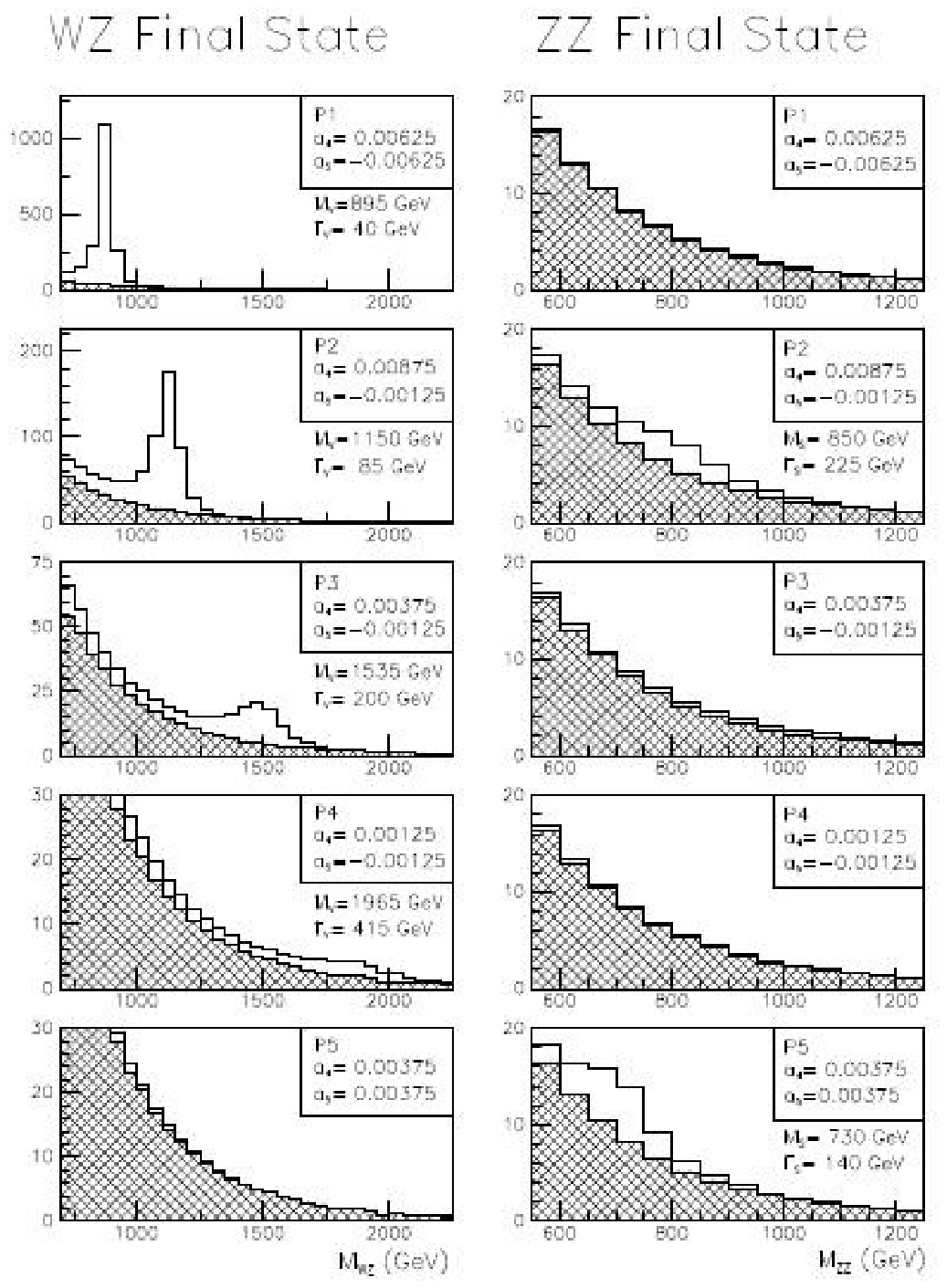} 
\end{minipage}\\
\subsection{Electroweak Chiral Lagrangians with a light Higgs (ECLh)}
After the discovery of the Higgs particle at the LHC, new Electroweak Chiral Lagrangians have been proposed that include a dynamical Higgs-like particle into the formulation of the EFT. Following a similar procedure as explained in the previous case, one can build non-linear effective Lagrangians that are $SU(2)_L \times U(1)_L$ gauge invariant and that have, in addition to the three GBs, transforming non-linearly under the Electroweak Chiral Symmetry, a new dynamical field $h(x)$ that is a singlet under this symmetry. This field then appears explicitly in the effective Lagrangian interaction terms by means of factor functions $(1+k_1 (h(x)/v)^1+k_2 (h(x)/v)^2+..)$ containing a series of powers of the dimensionless field ($h(x)/v$).

If one assumes again that the custodial symmetry breaking is as in the SM, and that no large CP violation nor  Flavor Violation occur beyond SM, then the simplest Electroweak Chiral Lagrangian with a light Higgs, keeping for instance just up to ${\cal O}(h(x)/v)$ terms in these series expansions, and including also fermions, is given by (we use here the first generation quarks notation):
\begin{eqnarray}
{\cal L}_{\rm ECLh} & = &  \frac{v^2}{4}\; Tr\left[ D_\mu U^\dagger D^\mu U \right] \left(1+2 a \frac{h}{v}\right) -\frac{v}{\sqrt{2}} (\bar{u}_L \bar{d}_L) U \left(1+c^{u,d}\frac{h}{v}\right) (y^u u_R \,\,\,y^d d_R)^T\,+\,h.c. \non \\
&+& {\cal L}_{\rm YM}+ \frac{1}{2}(\partial_\mu h)(\partial^\mu h)
\non
\end{eqnarray} 
where,  ${\cal L}_{\rm YM}$ is again the Yang Mills Lagrangian including all the kinetic terms for the  
$SU(2)_L \times U(1)_Y$ gauge fields, and the coefficients $a$ and $c^{u,d}$,   are  the relevant parameters of the Electroweak Chiral Lagrangian at this leading order of the non-linear effective field theory. These ECLh parameters are usually treated as phenomenological parameters, but the final 
goal with this approach would be to have a list of predictions for these parameters from different underlying theories. Then from the comparison with data, that will set experimentally preferred values for these ECLh parameters  one expects to be able to conclude on the preferred fundamental theory underlying the Higgs System. As in the previous ECL case, one can go beyond leading order and include higher dimensional terms also in the ECLh. 

There are recent studies with this Lagrangian ${\cal L}_{\rm ECLh}$ and also comparisons of their predictions at the tree level with present data at LHC and they are already providing some constraints on the values of the $c$ and $a$ parameters that are compatible with data. For instance, a global fit analysis, combining  several channels, assuming universality in the fermion parameters, $c_f=c$, and also assuming that no non-SM particles contribute to potential anomalous effective vertices like $(h/v)G_{a\mu\nu}G^{a\mu\nu}$ or $(h/v)F_{\mu\nu}F^{\mu\nu}$ leads to constraints on the $(a,c)$ plane as those shown in the next plot~\cite{Ellis:2013lra}. Here, the more likely regions of this parameter space have lighter shading, and the 68\%, 95 and 99\% CL contours are indicated by dotted, dashed and solid lines respectively. The yellow lines are the predictions in various models. 

We see clearly in this  plot, that the space left by data in the $(a,c)$ plane for potential deviations from the SM values, $a_{\rm SM}= c_{SM}=1$, is being reduced considerably with the increasing statistics at LHC, and this kind of analysis indeed sets already constraints on the ECLh parameters.

As in the previous ECL case, one can also include here higher dimensional operators, and in addition there are also terms with higher powers of $(h(x)/v)$. All these have been considered in the literature, but the total number of invariant terms increases considerably, as well as the number of parameters involved in the ECLh and the phenomenological analysis of these models with so many parameters gets very involved.  

Besides, as in any of the previously commented non-linear Chiral Effective Field theories , when doing predictions with the ECLh beyond leading order, one has to include not only the the tree level contributions from the previous Lagrangian, but also the one-loop generated contributions from the lowest order terms, and then finally one has to renormalize all the ECLh parameters to provide finite predictions for the observables at one-loop. 

\BC
\includegraphics[width=10cm]{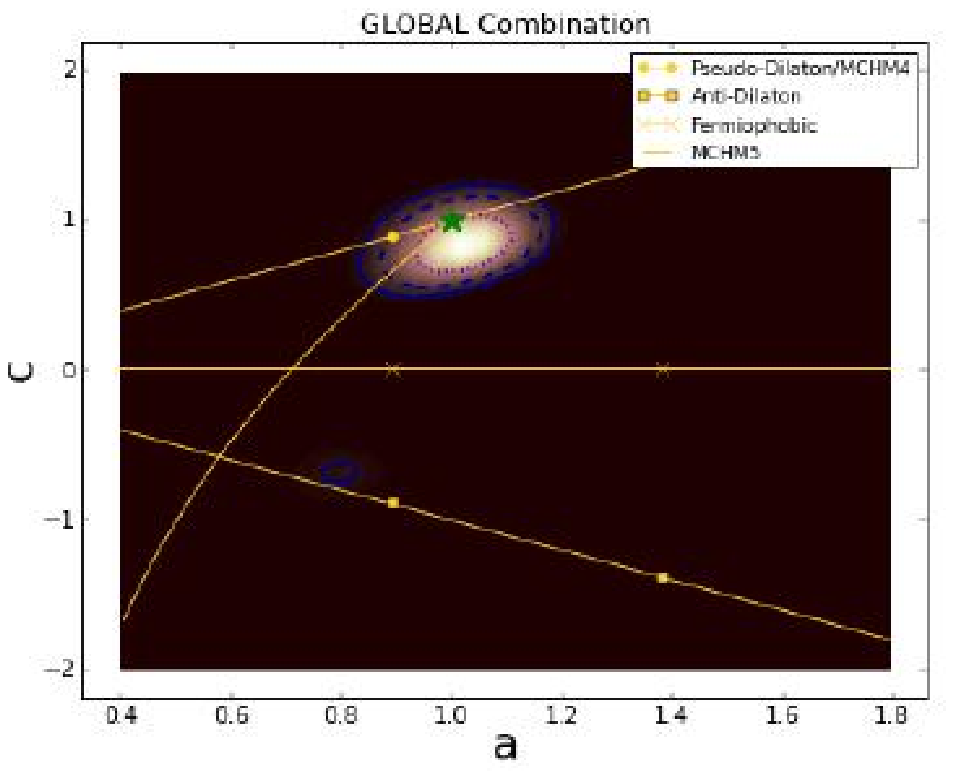}\\ 
(Plot taken from \cite{Ellis:2013lra})
\EC

This interesting program of building consistent and well defined effective quantum field theories for the Electroweak Theory with the help of Chiral Lagrangians is very active nowdays and seems promising. In fact there are already some predictions in the literature showing that, not only in the ECL but also in the ECLh case, there could appear new resonances in the scattering of longitudinal $W$ and $Z$ gauge bosons close to the TeV scale, and these could be seen at LHC in the next stage.   

\section*{Acknowledgments}
I wish to thank Carlos Merino for inviting me to give these lectures at the 2013 IDPASC school
at Santiago de Compostela, for the very efficient organization and for contributing to create the kind atmosphere with the students. I also acknowledge Sven Heinemeyer and Mariano Quiros for interesting discussions and for lending me their slides of previous Lectures on related subjects to some considered here. This work was partially supported by the European Union FP7 ITN
INVISIBLES (Marie Curie Actions, PITN- GA-2011- 289442), by the CICYT through the
project FPA2012-31880 and by the CM (Comunidad Autonoma de Madrid) through the project HEPHACOS S2009/ESP-1473. The work is also supported in part by the Spanish Consolider-Ingenio 2010 Programme CPAN (CSD2007-00042). The author also acknowledges the support of the Spanish MINECO's "Centro de Excelencia Severo Ochoa" Programme uder grant SEV-2012-0249.

\end{document}